\documentclass[aps,prd,superscriptaddress,preprint,floatfix,tightenlines,showpacs,preprintnumbers,nofootinbib,eqsecnum]{revtex4-1}

\pdfoutput=1
\usepackage[T1]{fontenc} % if needed
\usepackage[utf8]{inputenc}

\usepackage{url}
\usepackage{epsfig}
\usepackage{amsmath}
\usepackage{amssymb}
\usepackage{amsfonts}
\usepackage{bbm}
\usepackage{graphicx}
\usepackage{mathrsfs}
\usepackage{color}
\usepackage[dvipsnames]{xcolor}
\usepackage{comment}
\usepackage{mathtools}
\usepackage{braket}
\usepackage{float}
\usepackage{subfig}
\usepackage{subfloat}
\usepackage[normalem]{ulem}

\usepackage{todonotes}

\usepackage{hyperref}
\hypersetup{colorlinks=true, linkcolor=teal, urlcolor=blue, citecolor=red}
\usepackage{multirow}
\usepackage{relsize}
\usepackage{fullpage}
\usepackage{makecell}
\usepackage{enumitem,xcolor}

\usepackage{xspace}
\usepackage{bm}         % for bold math characters in a math einvironment
\usepackage{array}
\usepackage{booktabs}   % for fancy looking tables
\usepackage{mleftright}

\usepackage{bbold}
\usepackage{amscd}
\usepackage{soul}
\usepackage{tikz-cd}
\usepackage{empheq}
\usepackage[most]{tcolorbox}
\usepackage{slashed}
\usepackage{dsfont}
\usepackage[makeroom]{cancel}

\usepackage[capitalise]{cleveref}

\crefname{section}{Sec.}{Secs.}
\crefname{table}{Tab.}{Tabs.}
\crefname{figure}{Fig.}{Figs.}
\crefname{equation}{Eq.}{Eqs.}
\crefname{appendix}{Appendix\ }{Appendix\ }

\newcommand{\U}[1]{\mathrm{U}(1)_{\mathrm{#1}}}			% Use this for U(1) groups
\newcommand{\SU}[2]{\mathrm{SU}(#1)_{\mathrm{#2}}}		% Use this for SU(N) groups
\newcommand{\SO}[2]{\mathrm{SO}(#1)_{\mathrm{#2}}}		% Use this for SO(N) groups
\newcommand{\OO}[1]{\mathrm{O}(#1)}	% Use this for O(N) groups
\newcommand{\E}[1]{\mathrm{E}_{#1}}	% Use this for Exeptional En groups
\newcommand{\T}[1]{\mathrm{T}^{#1}}
\newcommand{\F}[1]{\mathrm{F}^{#1}}

\newcommand{\ZZ}{\mathbb{Z}^\mathrm{F}_3 \times \mathbb{Z}^\mathrm{C}_3}
\newcommand{\ZC}{\mathbb{Z}^\mathrm{C}_3}
\newcommand{\ZF}{\mathbb{Z}^\mathrm{F}_3}
\newcommand{\ZCF}{\mathbb{Z}^\mathrm{C,F}_3}

\colorlet{PLUM}{Plum}

\begin{document}

\title{Sculpting the Standard Model from low-scale Gauge-Higgs-Matter 
$\E{8}$ Grand Unification in ten dimensions}
\vspace{1cm}

\author{\vspace{1cm} Alfredo Aranda}
\email{fefo@ucol.mx}
\affiliation{Facultad de Ciencias-CUICBAS, Universidad de Colima, 
C.P.28045, Colima, M\'exico 01000, M\'exico}
\affiliation{Dual CP Institute of High Energy Physics, 
C.P. 28045, Colima, M\'exico}

\author{Francisco~J.~de~Anda}
\email{fran@tepaits.mx}
\affiliation{Tepatitl{\'a}n's Institute for Theoretical Studies, 
C.P. 47600, Jalisco, M{\'e}xico}
\affiliation{Dual CP Institute of High Energy Physics, 
C.P. 28045, Colima, M\'exico}

\author{Ant{\'o}nio~P.~Morais}
\email{aapmorais@ua.pt}
\affiliation{Departamento de F\'isica, Universidade de Aveiro, Campus de Santiago, 
3810-183 Aveiro, Portugal}
\affiliation{Centre  for  Research  and  Development  in  Mathematics  and  Applications (CIDMA), Campus de Santiago, 
3810-183 Aveiro, Portugal}
\affiliation{Department of Astronomy and Theoretical Physics, 
Lund University, 221 00 Lund, Sweden\vspace{1cm}}

\author{Roman~Pasechnik \vspace{1cm}}
\email{Roman.Pasechnik@thep.lu.se}
\affiliation{Department of Astronomy and Theoretical Physics, 
Lund University, 221 00 Lund, Sweden\vspace{1cm}}

\begin{abstract}
\vspace{0.5cm}
The construction and general implications of a model with complete supersymmetric unification of the Standard Model matter content, interactions and families' replication into a single $\E{8}$ gauge superfield in ten dimensions is presented. The gauge and extended Poincar\'e symmetries are broken through compactification of the $\mathbb{T}^6/(\mathbb{Z}_3\times \mathbb{Z}_3)$ orbifold with Wilson lines, which reduces the original symmetry and matter content into those of the Standard Model plus additional heavier states. Proton decay can be suppressed automatically while the compactification scale may be as low as $10^9~\rm{GeV}$, so that the corresponding GUT-scale physics may be potentially accessible and testable by future collider measurements.
\end{abstract}
\maketitle

%%%%%%%%%%%%%%%%%%%%%%%%%%%%%%
\section*{Introduction}
\label{sec:Intro}
%%%%%%%%%%%%%%%%%%%%%%%%%%%%%%

The Standard Model (SM) is currently the best theory at disposal to accurately describe three of the four fundamental interactions in nature. It is based upon modern Quantum Field Theory (QFT) and its predictions have matched some of the most stringent tests~\cite{ATLAS:2012yve,CMS:2012qbp,UA1:1983crd,GargamelleNeutrino:1973jyy,Hasert:1973cr,CDF:1995wbb,Parker:2018vye}. Despite its success, there are clear phenomenological indications that the SM is not the ultimate picture, such as neutrinos being massive particles \cite{Super-Kamiokande:1998kpq}, the existence of Dark Matter (DM) \cite{Bertone:2004pz} or the unexplained structure of the SM fermion families. Furthermore, both its gauge structure, mathematically described by the symmetry $\SU{3}{C}\times \SU{2}{L}\times \U{Y}$, as well as its matter content, are purely phenomenologically driven and a consensual first-principles explanation for their origin is still lacking. 

In fact, it is a remarkable theoretical challenge to explain and even eventually to derive the basic properties of the SM from a unified concept at high-energy scales. Typically, Grand Unified Theories (GUTs) feature unification of the SM gauge couplings into a single one by means of embedding the SM symmetry group into a larger simple gauge group at a certain energy scale, commonly dubbed the GUT scale. In this way, the SM Higgs, matter and gauge fields, denoted by green, blue and pink labels, respectively, in \cref{tab:SM}, can be a part of larger representations of the GUT symmetry group.
%%%%%%%%%%
	\begin{table}[htb]
		\begin{center}
			\begin{tabular}{c|c}
				\toprule
				$\mathrm{T}^4 \ltimes \SO{3,1}{}$&                 
				$\SU{3}{C} \times \SU{2}{L} \times \U{Y}$ \\
				\midrule
				$(\bm{1})$ & $\textcolor{OliveGreen}{(\bm{1},\bm{2},3)}$ \\
				$(\bm{2})$ & $3\times\textcolor{blue}{(\bm{1},\bm{2},-3)} + 3\times\textcolor{blue}{(\bm{1},\bm{1},6)} + 3\times\textcolor{blue}{(\bar{\bm{3}},\bm{1},-4)} + 3\times\textcolor{blue}{(\bar{\bm{3}},\bm{1},2)} + 3\times\textcolor{blue}{(\bm{3},\bm{2},1)}$ \\
				$(\bm{4})$ & $\textcolor{magenta}{(\bm{8},\bm{1},0)} + \textcolor{magenta}{(\bm{1},\bm{3},0)} + \textcolor{magenta}{(\bm{1},\bm{1},0)}$ \\
				\bottomrule
			\end{tabular}
		\end{center}
		\caption{The transformation properties of the SM fields under the gauge and Poincar\'e symmetries. $(\bm{1})$, $(\bm{2})$ and $(\bm{4})$ denote scalars, spin-$1/2$ Weyl fermions and spin-$1$ vectors in four dimensions, respectively.}
		\label{tab:SM}
	\end{table}
%%%%%%%%%%%	

Historically, the first and simplest way to achieve such a gauge unification is by utilising the $\SU{5}{}$ gauge symmetry at the GUT scale \cite{Georgi:1974sy}. The three gauge interactions are a remnant of a more fundamental one where the SM vector bosons are different components of a single $\bm{24}$-dimensional representation as shown in \cref{tab:SU5}. However, the matter sector is only partially unified where the three fermionic generations belong to three distinct $\bm{\bar{5}} + \bm{10}$ spin-$1/2$ copies. The Higgs sector, on the other hand, is enlarged to a spinless $\bm{\overline{5}}$ irreducible representation (irrep), bringing along new coloured scalars. It is well known that the latter need to be rather heavy, typically with masses above $10^{16}~\mathrm{GeV}$ in order to prevent the proton from rapidly decaying, while the doublet component is required to be at the TeV scale. The need for such a remarkable hierarchy is known as the \textit{doublet-triplet splitting problem} \cite{Sakai:1981gr} and is recognized as one of the theoretical challenges of minimal four-dimensional $\SU{5}{}$ GUTs.
%%%%%%%%%%
		\begin{table}[htb!]
			\begin{center}
					\begin{tabular}{c|c}
						\toprule
						$\mathrm{T}^4 \ltimes \SO{3,1}{}$&                 
						$\SU{5}{}$ \\
						\midrule
						$(\bm{1})$ & $\textcolor{OliveGreen}{(\bar{\bm{5}})}$ \\
						$(\bm{2})$ & $3\times\textcolor{blue}{(\bar{\bm{5}})} + 3\times\textcolor{blue}{(\bm{10})}$ \\
						$(\bm{4})$ & $\textcolor{magenta}{(\bm{24})} $ \\
						\bottomrule
					\end{tabular}
			\end{center}
			\caption{The SM field content embedded in $\SU{5}{}$ with transformation properties under the gauge and Poincar\'e symmetries.}
		\label{tab:SU5}
		\end{table}
%%%%%%%%%%

Another possibility came with the $\SO{10}{}$ GUT \cite{Fritzsch:1974nn} where, besides gauge field unification, one can also unify all the SM fermions of an entire matter family into a single $16$-dimensional $\SO{10}{}$ representation as shown in \cref{tab:SO10}.
%%%%%%%%%%
		\begin{table}[htb]
			\begin{center}
				\begin{tabular}{c|c}
					\toprule
					$\mathrm{T}^4 \ltimes \SO{3,1}{}$&                 
					$\SO{10}{}$ \\
					\midrule
					$(\bm{1})$ & $\textcolor{OliveGreen}{(\bm{10})}$ \\
					$(\bm{2})$ & $3\times\textcolor{blue}{(\bm{16})}$ \\
					$(\bm{4})$ & $\textcolor{magenta}{(\bm{45})} $ \\
					\bottomrule
				\end{tabular}
			\end{center}
			\caption{The SM field content embedded in $\SO{10}{}$ with its transformation properties under the gauge and Poincar\'e symmetries.}
		\label{tab:SO10}
		\end{table}
%%%%%%%%%%
Furthermore, the three matter $\bm{16}$-plets contain a SM-singlet each offering three generations of right-handed neutrinos and the possibility for explaining the observed active neutrino masses via a seesaw mechanism. In this context, the seesaw scale can be attributed to the $\SO{10}{}$ breaking scale such that the smallness of neutrino masses results from a large hierarchy between GUT and electroweak (EW) scales.

A particularly attractive option is the $\E{6}$ GUT \cite{Gursey:1975ki}, where matter unification is realized in three copies of the fundamental $\bm{27}$ irrep. However, in its minimal form, the SM Higgs doublet is also part of a scalar $\bm{27}$-plet suggesting a higher degree of unification if both the fermion and scalar counterparts are related by means of simple ${\cal N}=1$ supersymmetry (SUSY) transformations \cite{King:2005my}. A SUSY $\E{6}$ GUT is then the minimal theory featuring Higgs and matter unification in three distinct copies of $\bm{27}$ supermultiplets as summarized in \cref{tab:E6}.
%%%%%%%%%%
		\begin{table}[htb]
			\begin{center}
				\begin{tabular}{c|c}
					\toprule
					$S^1 \ltimes [\mathrm{T}^4 \ltimes \SO{3,1}{}]$&                 
					$\mathrm{E}_6$ \\
					\midrule
					$(2 \times \bm{1} + \bm{2})$ & $3 \times \textcolor{ForestGreen}{(\bm{27})}$ \\
					$(\bm{4} + \bm{2})$ & $\textcolor{magenta}{(\bm{78})} $ \\
					\bottomrule
				\end{tabular}
			\end{center}
			\caption{The SM field content embedded in SUSY $\E{6}$ with the corresponding transformation properties under the gauge and super-Poincar\'e symmetries. $S^1$ denotes simple ${\cal N} = 1$ SUSY.}
		\label{tab:E6}
		\end{table}
%%%%%%%%%%
		
Note that the unification of matter fields both in $\SO{10}{}$ and $\E{6}$ GUTs is realized independently for each generation without offering any fundamental reason to explain the family replication observed in nature. This can be accomplished at the level of a SUSY $\E{7}$ GUT if the entire SM fermion and Higgs sectors are embedded in a single real $\bm{912}$ supermultiplet. The challenge that such a picture poses is on how to obtain chiral matter from a real representation. While this is not realizable in four space-time dimensions, the presence of orbifolded extra dimensions (EDs) can leave massless chiral sectors in a low-energy SM-like effective field theory (EFT). While this can be achieved with one ED, SUSY representations become simpler in the case of two compact EDs. This enlarges the super-Poincaré symmetry to $\T{6} \ltimes \SO{5,1}{}$ with the minimal field content of \cref{tab:E7}.
%%%%%%%%%%
		\begin{table}[htb]
			\begin{center}
				\begin{tabular}{c|c}
					\toprule
					$S^1 \ltimes [\mathrm{T}^6 \ltimes \SO{5,1}{}]$&                 
					$\mathrm{E}_7$ \\
					\midrule
					$(4 \times \bm{1} + \bm{4})$ & $\textcolor{ForestGreen}{(\bm{912})}$ \\
					$(\bm{6} + \bm{4})$ & $\textcolor{magenta}{(\bm{133})} $ \\
					\bottomrule
				\end{tabular}
			\end{center}
			\caption{The SM field content embedded in SUSY $\E{7}$ with the corresponding transformation properties under the gauge and super-Poincar\'e symmetries. $S^1$ denotes simple ${\cal N} = 1$ SUSY.}
		\label{tab:E7}
		\end{table}
%%%%%%%%%%
Such a picture offers a possible framework for both gauge unification and Higgs-matter-family unification where the three fermion families can be realised together with other observed properties of the SM particle spectra and interactions.

The pattern of enlarging of the GUT symmetry can be seen through the Dynkin diagrams and follows the exceptional chain \cite{Buchmuller:1985rc,Koca:1982zi},
\begin{equation}
 \SU{3}{C}\times \SU{2}{L}\times \U{Y} \subset \SU{5}{} \subset \SO{10}{} \subset \E{6} 
 \subset \E{7} \subset \E{8} \,. \label{Eq:chain}
\end{equation}
The largest group in this chain, $\E{8}$, has an unique property that its adjoint representation $\textbf{248}$ coincides with its fundamental one \cite{Slansky:1981yr}. This suggests that all the SM gauge and matter fields can be, in principle, unified into a single gauge $\textbf{248}$ representation provided that a maximal $\mathcal{N}=4$ SUSY is realised. Furthermore, it also provides an $\SU{3}{}$ family symmetry as a coset of $\E{8}$ to $\E{6}$ reduction as has been thoroughly explored earlier in Refs.~\cite{Camargo-Molina:2016yqm,Camargo-Molina:2017kxd,Morais:2020ypd,Morais:2020odg}.

There is a plethora of models that aim to build GUTs including family symmetries \cite{King:2001uz,King:2017guk,Hagedorn:2010th,Antusch:2014poa,Bjorkeroth:2015ora,Bjorkeroth:2015uou,Bjorkeroth:2017ybg,deAnda:2017yeb,CarcamoHernandez:2020owa,Morais:2020ypd,Morais:2020odg,Camargo-Molina:2016yqm,Camargo-Molina:2017kxd,Camargo-Molina:2016yqm,Camargo-Molina:2016bwm} and EDs \cite{Reig:2017nrz,Altarelli:2008bg,Burrows:2009pi,Burrows:2010wz,deAnda:2018oik,Altarelli:2006kg,Adulpravitchai:2010na,Adulpravitchai:2009id,Asaka:2001eh,deAnda:2019jxw,deAnda:2018ecu,deAnda:2018yfp}. These models require various independent groups and a large number of fields. The $\E{8}$ group is a good bet for a complete unification of SM vectors, fermions and scalars, as well as both the gauge and Yukawa interactions, and has indeed been widely studied both in the context of string theory \cite{Ibanez:1987pj,Parr:2020oar,Manolakos:2020cco} and within the framework of QFT \cite{Adler:2002yg,Adler:2004uj,Garibaldi:2016zgm,Thomas:1985be,Konshtein:1980km,Baaklini:1980fv,Baaklini:1980uq,Barr:1987pu,Bars:1980mb,Koca:1981xd,Mahapatra:1988gc,Ong:1984ej,Camargo-Molina:2016yqm,Olive:1982ai}.

Similarly to $\E{7}$, the main challenge posed by $\E{8}$ is that it is a real group -- the same as the extended ${\cal N}=4$ SUSY theory -- while the SM requires chiral representations. With orbifolded EDs, the compactification procedure could leave massless chiral sectors in a low-energy SM-like EFT. Furthermore, extended ${\cal N}=4$ SUSY in four dimensions can be obtained from a ${\cal N}=1$ SUSY 10d theory \cite{Arkani-Hamed:2001vvu,Brink:1976bc}. The orbifolding mechanism provides a natural way to geometrically break a GUT symmetry group and to generate large masses for some or most of the unobserved states. However, a consistent way of getting the correct flavour structure and interactions in a SM-like low-energy EFT from the $\E{8}$ GUT, that is compatible with all existing phenomenological constraints, has not yet been developed in the literature.

In this paper, we propose a particular realisation of a GUT based upon the $\E{8}$ gauge symmetry and $\mathcal{N}=1$ SUSY in ten spacetime dimensions (10d) where the full SM matter, Higgs and gauge field content is unified into a single $\E{8}$ gauge superfield transforming as in \cref{tab:E8}.
%%%%%%%%%%
		\begin{table}[htb!]
			\begin{center}
				\begin{tabular}{c|c}
					\toprule
					$S^1 \ltimes [\mathrm{T}^{10} \ltimes \SO{9,1}{}]$&                 
					$\mathrm{E}_8$ \\
					\midrule
					$(\bm{10} + \bm{8})$ & $\textcolor{magenta}{(\bm{248})} $ \\
					\bottomrule
				\end{tabular}
			\end{center}
			\caption{Full unification of the SM field content in a single adjoint $\bm{248}$ supermultiplet composed by a 10d real vector and a 10d Weyl/Majorana fermion.}
		\label{tab:E8}
		\end{table}
%%%%%%%%%%
The $\mathbb{T}^6/(\mathbb{Z}_3\times \mathbb{Z}_3)$ orbifold compactification breaks SUSY and the original gauge symmetry {\it directly} to that of the SM, $\E{8}\to \SU{3}{C}\times \SU{2}{L} \times \U{Y}$ (i.e. without any intermediate, typically anomalous, symmetry groups), leaving only the SM gauge and fermion fields massless. The operators responsible for the proton decay naturally appear to be strongly suppressed alleviating the need for a very strong hierarchy between the ED compactification scale and the EW one. While the Higgs doublets acquire a positive mass at one of the Wilson-line breaking scales, via the Renormalisation Group (RG) evolution, one of the Higgs doublet squared mass terms runs negative at a TeV scale, thus, triggering the EW symmetry breaking radiatively. This is known as the radiative EW symmetry breaking (REWSB) mechanism (for its particular realisation in the framework of non-SUSY trinification model, see Ref.~\cite{Camargo-Molina:2016bwm}).

The high-energy theory only has one arbitrary parameter, the $\E{8}$ gauge coupling, while the details of the orbifold compactification procedure considered in this article provide two real and twelve complex additional arbitrary parameters coming from two radii, six complex Wilson-line effective vacuum expectation values (VEVs) and six complex Higgs VEVs. These parameters fully determine the properties of the SM-like EFT emergent at low energies and all the related phenomenology at the EW scale and beyond.

The layout of the paper is as follows. In \cref{sec:structure}, the key basics and definitions of the proposed $\E{8}$ GUT in 10d are given. In \cref{sec:t6z3z3}, a specific $\mathbb{T}^6/(\mathbb{Z}_3\times \mathbb{Z}_3)$ orbifold is introduced and its compactification mechanism is studied in detail. In \cref{sec:anom}, anomaly cancellation in the considered model is discussed. In \cref{sec:trilag}, the low-energy EFT Lagrangian of the $\E{8}$ GUT is presented, and its fermion and scalar mass spectra are derived and discussed in detail. In \cref{sec:prodec} the processes that generate highly suppressed proton decay are analyzed. In \cref{sec:gcu}, the RG evolution of the gauge couplings is presented as well as the relevant phenomenology it entails is reviewed. An outlook is given in \cref{sec:string-diff} elaborating on basic differences of the proposed framework from string theory. Finally, in \cref{sec:conclusion} the main conclusions of this work are presented. 

%%%%%%%%%%%%%%%%%%%%%%%%%%%%%%%%%%%%%%%%%%%%
\section{Ten-dimensional $\E{8}$ GUT}
\label{sec:model}
%%%%%%%%%%%%%%%%%%%%%%%%%%%%%%%%%%%%%%%%%%%%

%%%%%%%%%%%%%%%%%%%%%%%%%%%%%%%%%%%%%%%%%%%%
\subsection{Model structure}
\label{sec:structure}
%%%%%%%%%%%%%%%%%%%%%%%%%%%%%%%%%%%%%%%%%%%%

In this article, the $\mathcal{N}=1$ SUSY model of \cref{tab:E8} based on an $\E{8}$ GUT in 10d is introduced. The corresponding spacetime contains six compact EDs factorizable as $\mathbb{R}^4\times(\mathbb{T}^2)^3$, with $\mathbb{T}^2$ being a 2-torus, which are denoted by three complex coordinates $z_i$ in what follows. The non-compact coordinates of the physical four-dimensional (4d) spacetime are then labeled as $x^\mu \equiv x$. This GUT is a pure super Yang-Mills (SYM) theory that contains a single 10d vector superfield $\mathcal{V}_{\textbf{248}}(x,z_i)$ in the adjoint (which is also the fundamental) representation of $\E{8}$. The 10d $\mathcal{N}=1$ SUSY then decomposes into $\mathcal{N}=4$ SUSY in 4d spacetime \cite{Arkani-Hamed:2001vvu,Brink:1976bc} and needs to be broken through the orbifold boundary conditions to a simple $\mathcal{N} = 1$ SUSY theory. In such a scenario, the single 10d vector superfield $\mathcal{V}$ decomposes into a 4d vector superfield $V$, and a sector with three 4d chiral superfields $\phi_i$, one per complex ED, supplemented with an infinite tower of Kaluza–Klein (KK) states for each of these superfields. A thorough description of the decomposition of the ED Poincar\'e representations in the current setup is shown in~\cref{app:edpoinc} while further details about the procedure of a generic 10d orbifolding is reviewed in~\cref{app:orb}. The Lagrangian of the model and its decomposition are detailed in \cref{app:lag}.

Upon orbifolding, the super-Poincar\'e group on the first column of \cref{tab:E8} becomes incomplete and the full symmetry of the model becomes
\begin{equation}
    S^1 \ltimes [(\mathrm{T}^{4} \times \mathrm{T}^6 /{\Gamma}) \ltimes (\SO{9,1}{}/{\mathrm{F}})] \ltimes  \E{8}\,,
    \label{eq:incSP}
\end{equation}
where $S^1$ denotes $\mathcal{N}=1$ SUSY, the $\mathrm{T}$ represents the translation group and $\SO{9,1}{}$ is the 10d Lorentz group. Here, the discrete subgroup of ED translations $\Gamma\simeq \mathbb{Z}^6\subset \mathrm{T}^6$ is also called the lattice group\footnote{This is a direct product of six times the group of integers} and compactifies the EDs, while $\mathrm{F}$ is a discrete subgroup of ED rotations $\mathrm{F}\subset \mathrm{SO}(6)$, which defines the orbifolding. In this article, the case $\mathrm{F}\simeq\ZZ$ is explored for the first time (here and below, the superscripts F and C stand for family and colour, respectively).

Both the lattice group $\Gamma$ and the orbifolding group $\mathrm{F}$ transformations are accompanied by a gauge transformation, so that both $\Gamma,\mathrm{F}\subset \E{8}$. Therefore, the spacetime symmetry no longer commutes with the gauge one, i.e.
\begin{equation}
\exists \ \ \ \alpha\in \Gamma, \beta\in \mathrm{F}, \gamma\in \mathrm{E}_8 \ \ \ | \ \ \
    [\alpha, \gamma] \neq 0,\  \ [\beta, \gamma] \neq 0\,,
    \label{eq:E8GF}
\end{equation}
such that the gauge sector becomes sensitive to geometrical effects leading to its breaking. 

In the considered model, the EDs are orbifolded as $\mathbb{T}^6/(\ZZ)$ with
\begin{equation}\begin{split}
\ZF:\ (x,z_1,z_2,z_3)&\sim (x,\omega^2 z_1,\omega^2 z_2,\omega^2 z_3),\ \ \ \mathcal{V}\to e^{-2i\pi q_{8}^{\mathrm{F}}/3}\ \mathcal{V}\,, \\
\ZC:\ (x,z_1,z_2,z_3)&\sim (x, \omega^3z_1,\omega z_2,\omega^2 z_3),\ \ \ \ \mathcal{V}\to e^{-2i\pi( q_{8}^{\mathrm{C}}+q_{8}^{\mathrm{F}}-q_{3}^{\mathrm{F}})/3}\ \mathcal{V}\,,
\label{eq:modrot}
\end{split}\end{equation}
where $\omega = e^{2i \pi /3}$ denotes the complex cubic root of unity and $q_8^\mathrm{C,F}$ are the Abelian charges under the $T^{3,8}$ generators of $\SU{3}{C,F}$ symmetry groups emerging via the following decomposition
\cite{Slansky:1981yr}
\begin{equation}
    \E{8}  \supset \SU{3}{C} \times \SU{3}{L} \times \SU{3}{R} \times \SU{3}{F} \,,
\end{equation}
respectively. The lattice of the $\mathbb{T}^6$ tori is defined by the coordinate transformations
\begin{equation}
 z_i\sim z_i+\tau^r_i \quad \text{with} \quad \tau_i^1 = 2\pi R_i \quad \text{and} \quad  \tau_i^2 = 2\pi e^{i\pi/3} R_i \,,
\label{eq:modtra}
\end{equation}
where the $R_i$ are the radii of the different tori. Note that each torus is built from two circles of equal radii. Each lattice translation is accompanied by a gauge transformation $U^r_{i=1,2,3}$ which is called a Wilson line
\begin{equation}
\mathcal{V}(x,z_i)=U^r_i \mathcal{V}(x,z_i+\tau^r_i) \,.
\label{eq:vwl}
\end{equation}
A non-trivial $U^r_i$ gauge transformation, i.e.~a Wilson line, can be seen as an effective VEV in the chiral superfields \cite{Candelas:1985en}. For consistency with the orbifolding group, it must be aligned with a representation that has a zero mode in its KK tower. One must note that it is not an usual VEV, as it does not come from the minimization of a potential. Instead, it comes from the ED profiles of the scalar fields and makes the same impact on the theory as an ordinary VEV in the Higgs mechanism.

Even though the effective (Wilson-line) VEVs are extracted from the properties of the orbifold, they acquire radiative corrections. By integrating out all the fields, except the scalars getting the effective VEVs, one obtains an effective potential that drives the magnitude of such VEVs. This is the essence of the Hosotani mechanism
\cite{Hosotani:1983xw,Hosotani:1983vn,Hosotani:2004wv,Hosotani:2004ka,Haba:2004qf,Haba:2002py}. One must bear in mind that the direction in the space of effective VEVs must be chosen a priori, as it comes from the translational boundary conditions. Therefore, the symmetry is already broken by the orbifold while the VEV magnitude is determined radiatively to be of the required size from the Hosotani mechanism  (as the tadpoles are absent) and is proportional to the radii of the extra dimensions.

In what follows, a continuous Wilson line will be associated to $U^r_{i}$, whose gauge transformation is consistent with \cref{eq:modrot}. Choosing the three different $R_i$ and the arbitrary dimensionless parameters associated to the Wilson line in $U^r_{1}$, one completely defines the parameter space of the considered GUT. As described in detail in the next section, the model is based on an orbifold and Wilson-line symmetry breaking mechanism that reduce $\E{8}$ directly down to $\SU{3}{C} \times \SU{2}{L} \times \U{Y}$, where the EW symmetry can further be broken radiatively (see below).

%%%%%%%%%%%%%%%%%%%%%%%%%%%%%%%%%%%%%%%%%%%%%%%%%%%%%%%%%%%%%%%%%
\subsection{Orbifolding the EDs with $\mathbb{T}^6/(\ZZ)$}
\label{sec:t6z3z3}
%%%%%%%%%%%%%%%%%%%%%%%%%%%%%%%%%%%%%%%%%%%%%%%%%%%%%%%%%%%%%%%%%

The model proposed in this article is built upon the orbifold defined by \cref{eq:modrot}. The first $\mathbb{Z}_3$, denoted as $\ZF$ in \cref{eq:modrot}, breaks $\E{8}\to \E{6}\times \SU{3}{F}$ while the second one, $\ZC$, breaks $\E{8}\to \SU{9}{}$ \cite{Aranda:2020noz,deAnda:2020prd}. When $\ZF$ and $\ZC$ act together, they effectively break $\E{8}$ into an intersection of $\E{6}\times \SU{3}{F}$ and $\SU{9}{}$ symmetries containing trinification and family symmetries  as
\begin{eqnarray}
\E{8} \to (\E{6}\times \SU{3}{F})\cap \SU{9}{}=\SU{3}{C}\times \SU{3}{L}\times \SU{3}{R}\times \U{F3}\times \U{F8} \,, 
\end{eqnarray}
with labels resembling the symmetries in the low-energy SM-like EFT limit. Although this is the breaking associated to the rotational boundary conditions only, it is useful to use this intersection group as a basis for describing the $\textbf{248}$ in terms of the orbifold charges as presented in \cref{tab:pe3}.

The lattice of $\mathbb{T}^6$ is factorized into three tori. Each torus is built from two circles which must have equal radii. Different tori can have different radii defined in \cref{eq:modtra}, where $R_i$ denotes the three different radii. The ED lattice must be invariant under the orbifold action to be consistent. One can see that the 6d lattice is built from three copies of the 2d lattice generated by the basis $(1,e^{i\pi/3})$ which is just a normalization of \cref{eq:modtra}. The orbifold action is generated by multiples of $\omega=e^{2i\pi/3}$ acting on the lattice as
\begin{equation}
e^{2i\pi/3}(1,e^{i\pi/3})=(e^{i\pi/3}-1,-1) \,,
\label{eq:basis}
\end{equation}
which is known as the modular transformation (i.e. the transformed basis is just an integer linear combination of the original basis). Therefore, indeed, the considered orbifold is consistent.

One of the features of orbifolds is the existence of singular points which imply certain boundary conditions responsible for the symmetry breaking. The singular points are the ones invariant under the orbifold transformations and are found as solutions of the following equations
\begin{equation}
    \mathcal{Z}^{\rm C,F}(z_i)=z_i \,, \ \ \   
    \mathcal{Z}^{\rm C,F}(z_i)=z_i+\tau^r_i\,,
    \label{eq:ZCF}
\end{equation}
where $\mathcal{Z}^{\rm C,F}$ denotes the action of the orbifold group $\ZCF$ on the coordinates.
It defines the fixed points which in this case are
\begin{equation}
\left(x,\left\{0,\frac{1+e^{i\pi/3}}{3},\frac{2+2e^{i\pi/3}}{3}\right\},
\left\{0,\frac{1+e^{i\pi/3}}{3},\frac{2+2e^{i\pi/3}}{3}\right\},
\left\{0,\frac{1+e^{i\pi/3}}{3},\frac{2+2e^{i\pi/3}}{3}\right\}\right) \,.
\label{eq:fp}
\end{equation}

\begin{figure}[htb!]
    \includegraphics[scale=0.4]{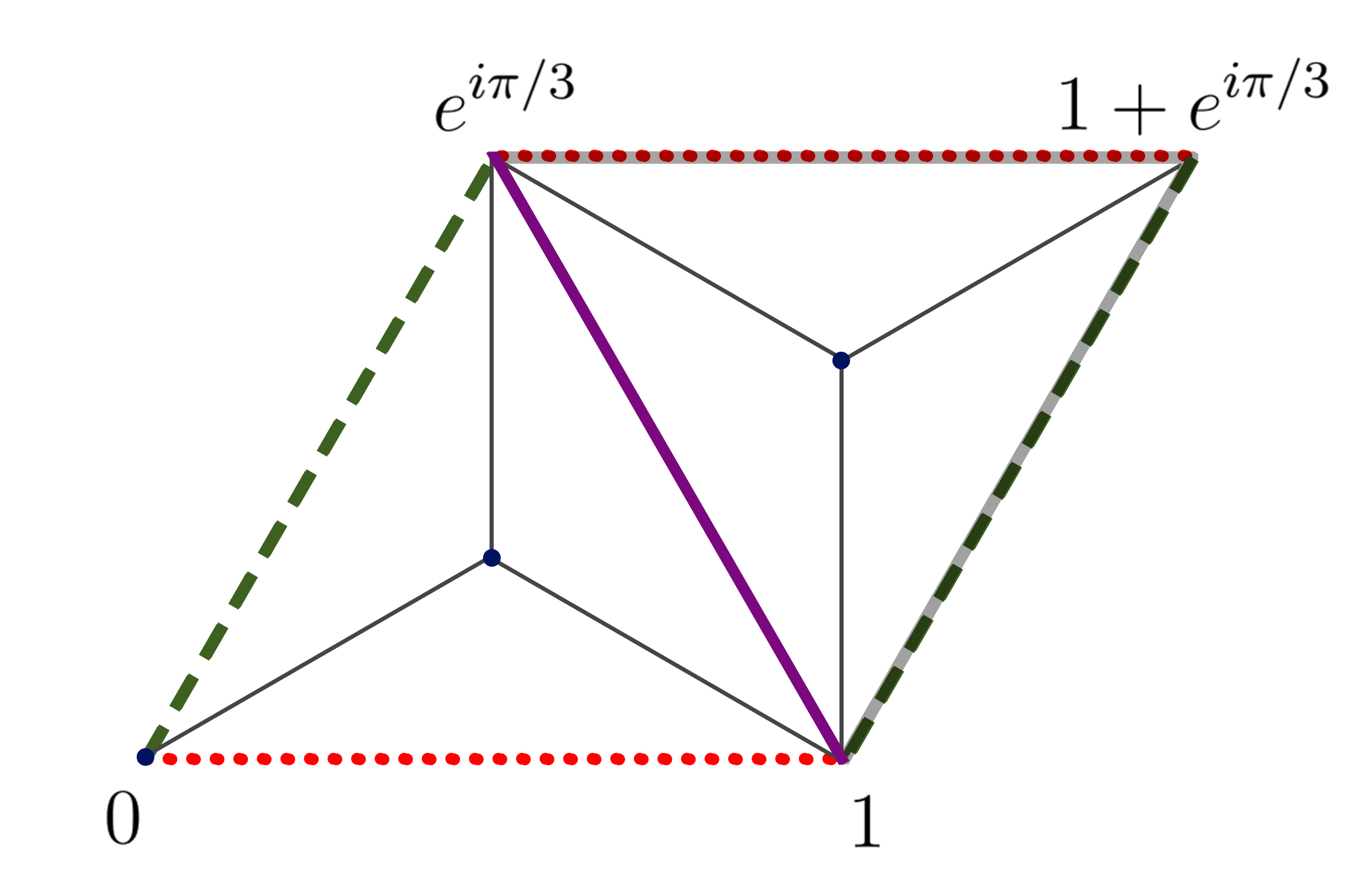}
    \includegraphics[scale=0.4]{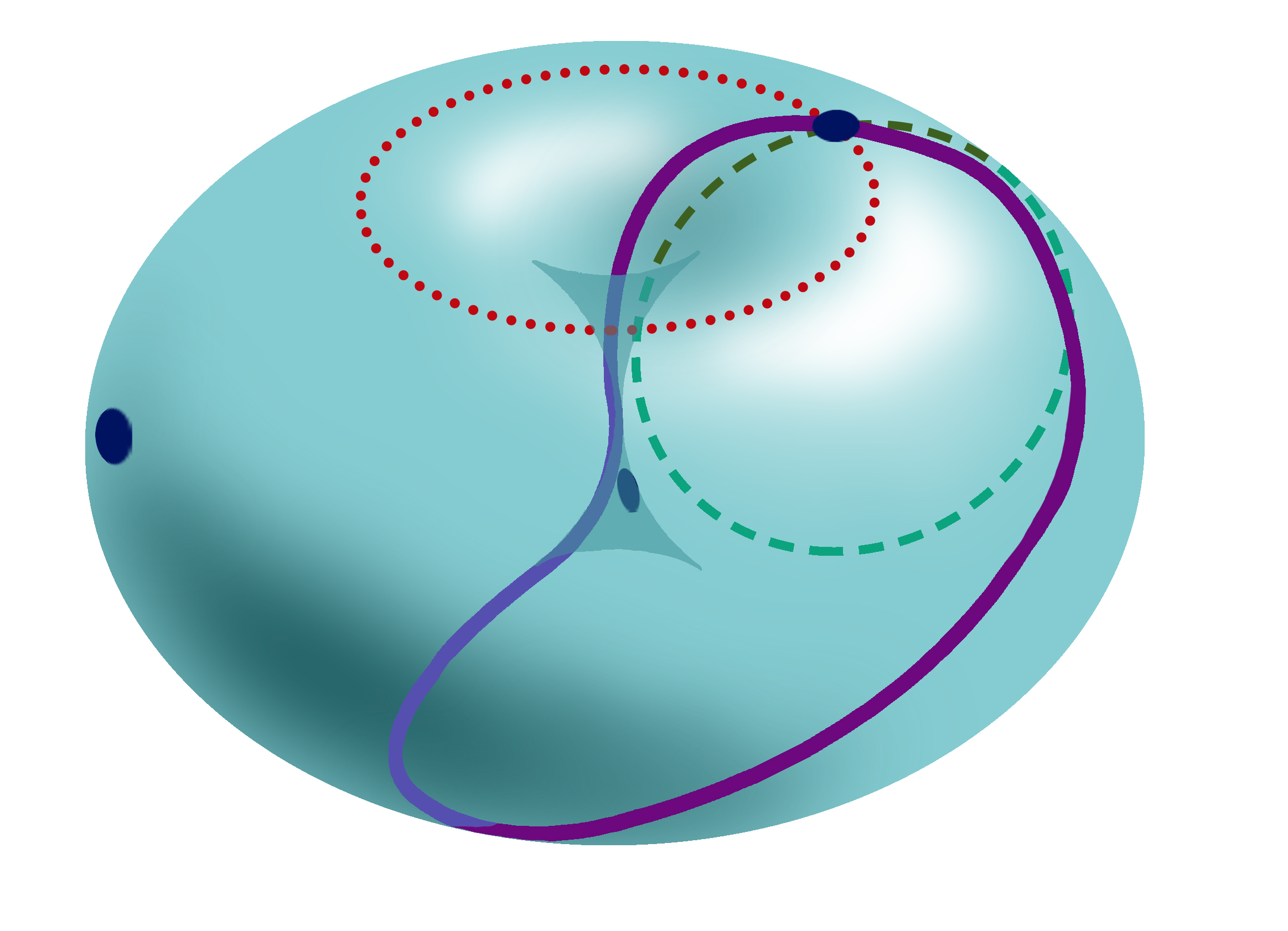}
    \begin{minipage}[t]{.5\linewidth}
    \centering
    \caption*{(a) The unfolded compactified ED\\ space of a single coordinate $z_i$.} \label{fig:edspace}
    \end{minipage}%
    \begin{minipage}[t]{.5\linewidth}
    \centering
    \caption*{(b) $\mathbb{T}^2$ torus visualization.\\ Colours are as in (a).}
    \end{minipage}
    \includegraphics[scale=0.27]{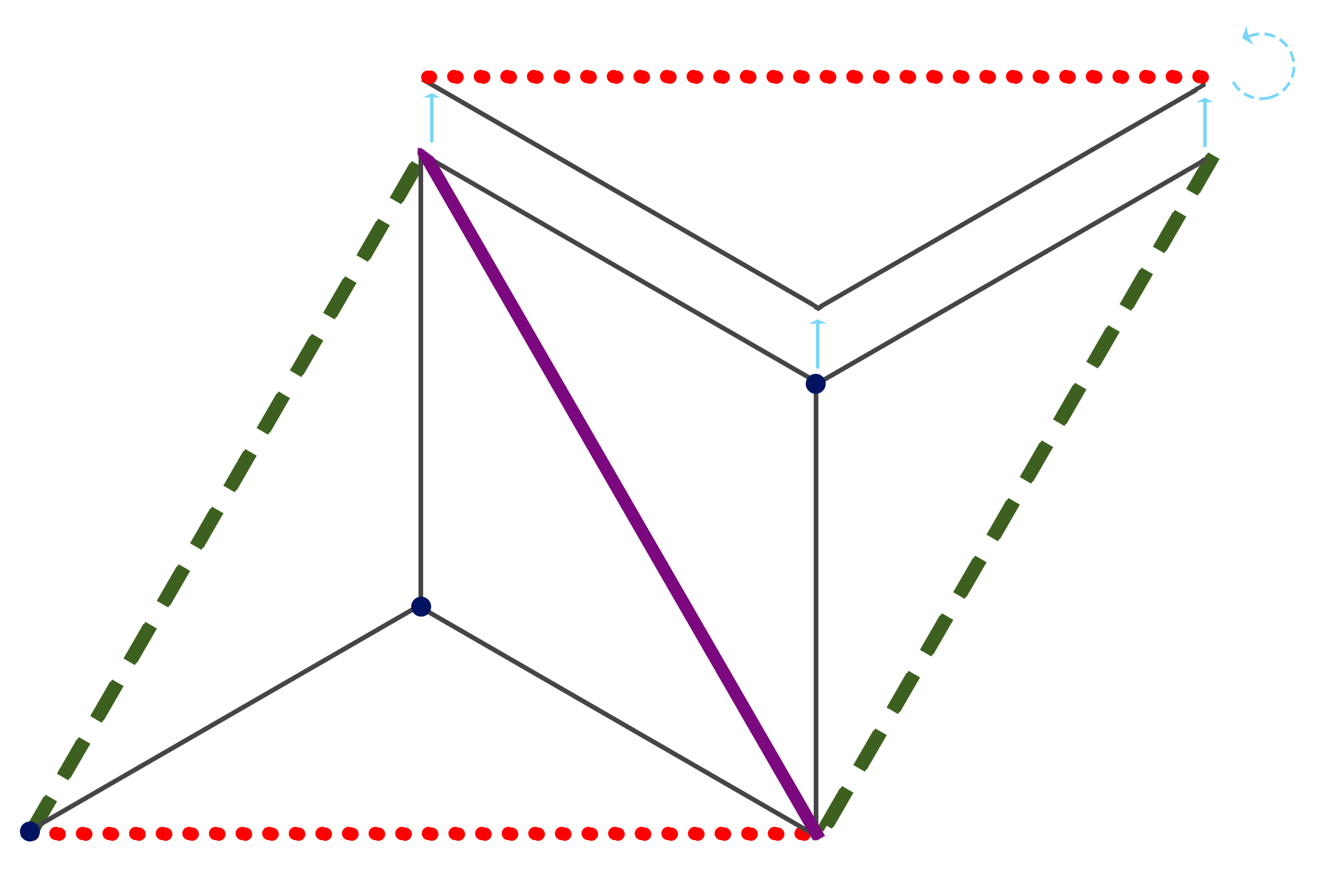}
	\includegraphics[scale=0.27]{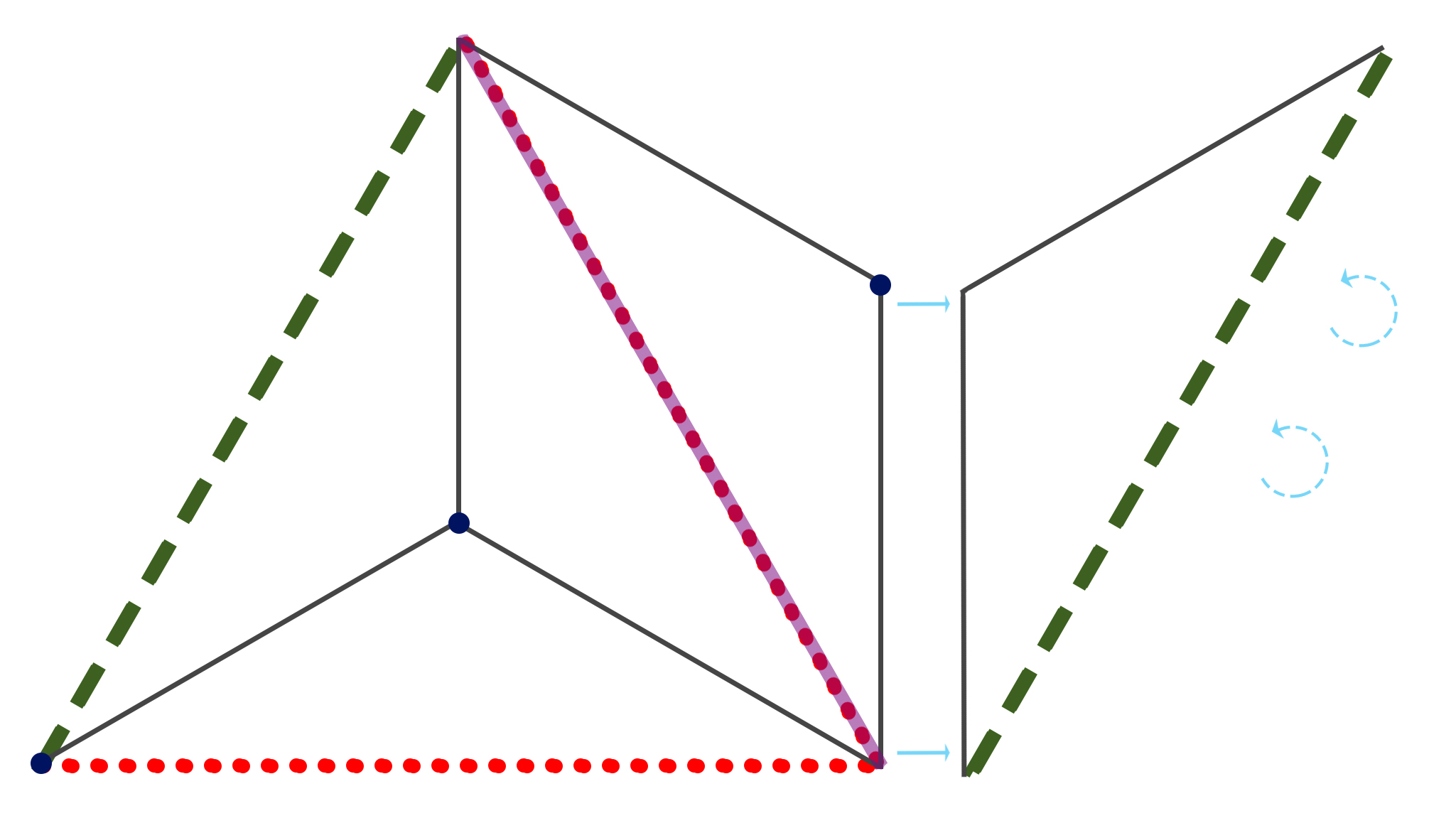}
	\includegraphics[scale=0.27]{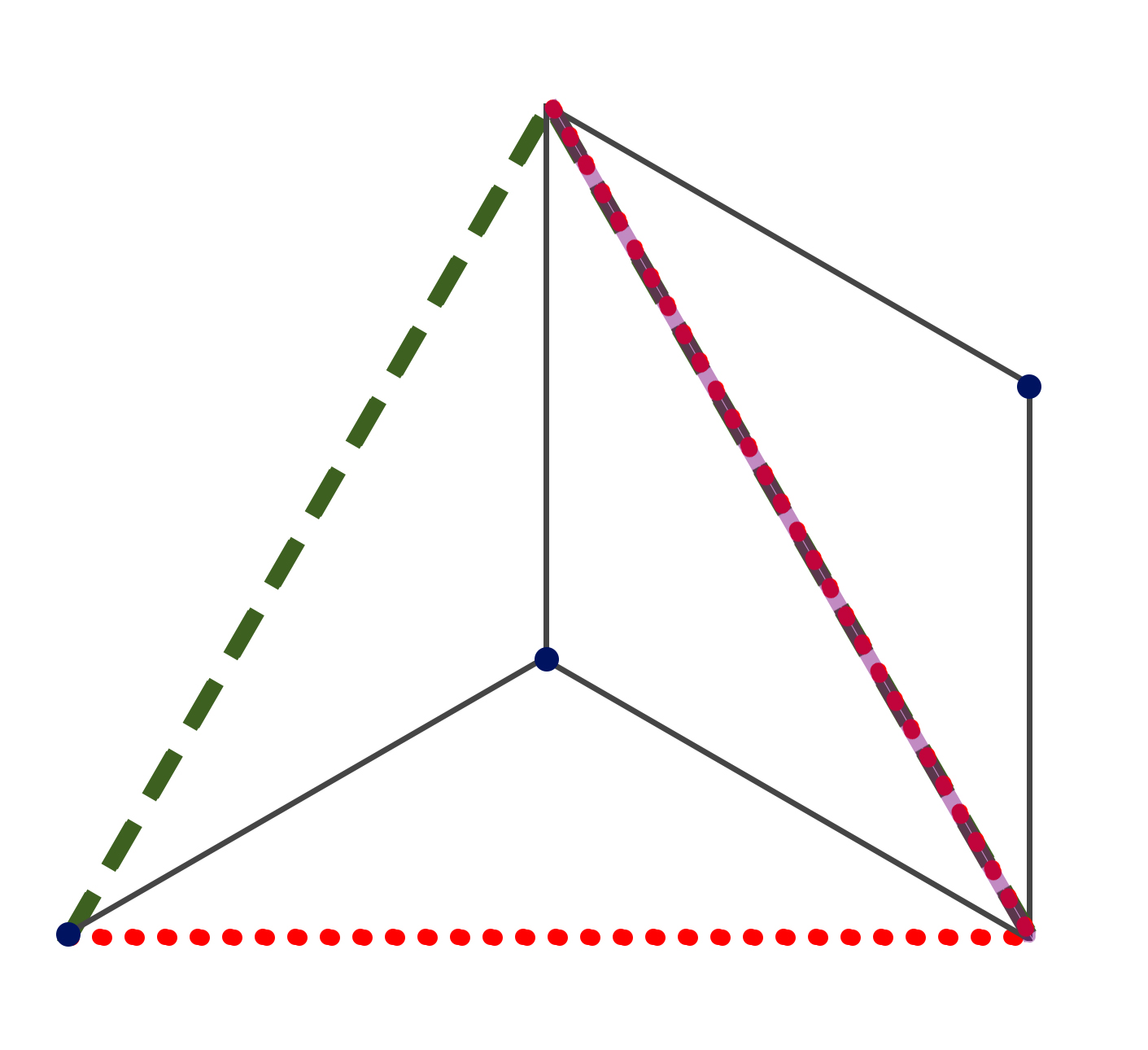}
	\includegraphics[scale=0.27]{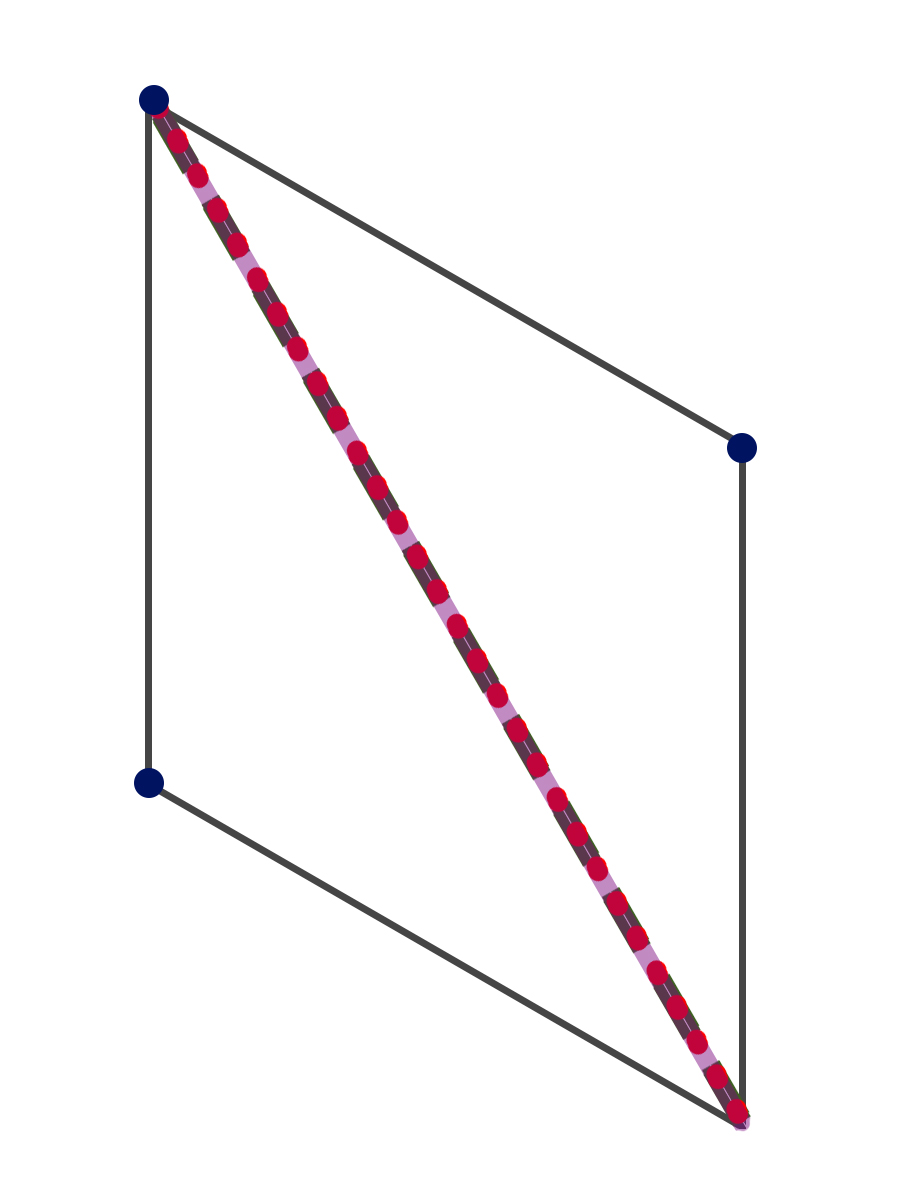}
	\begin{minipage}[t]{1.\linewidth}
    \centering
    \caption*{(c) Folding the orbifold.}
    \end{minipage}
    \\
    \includegraphics[width=0.4\textwidth]{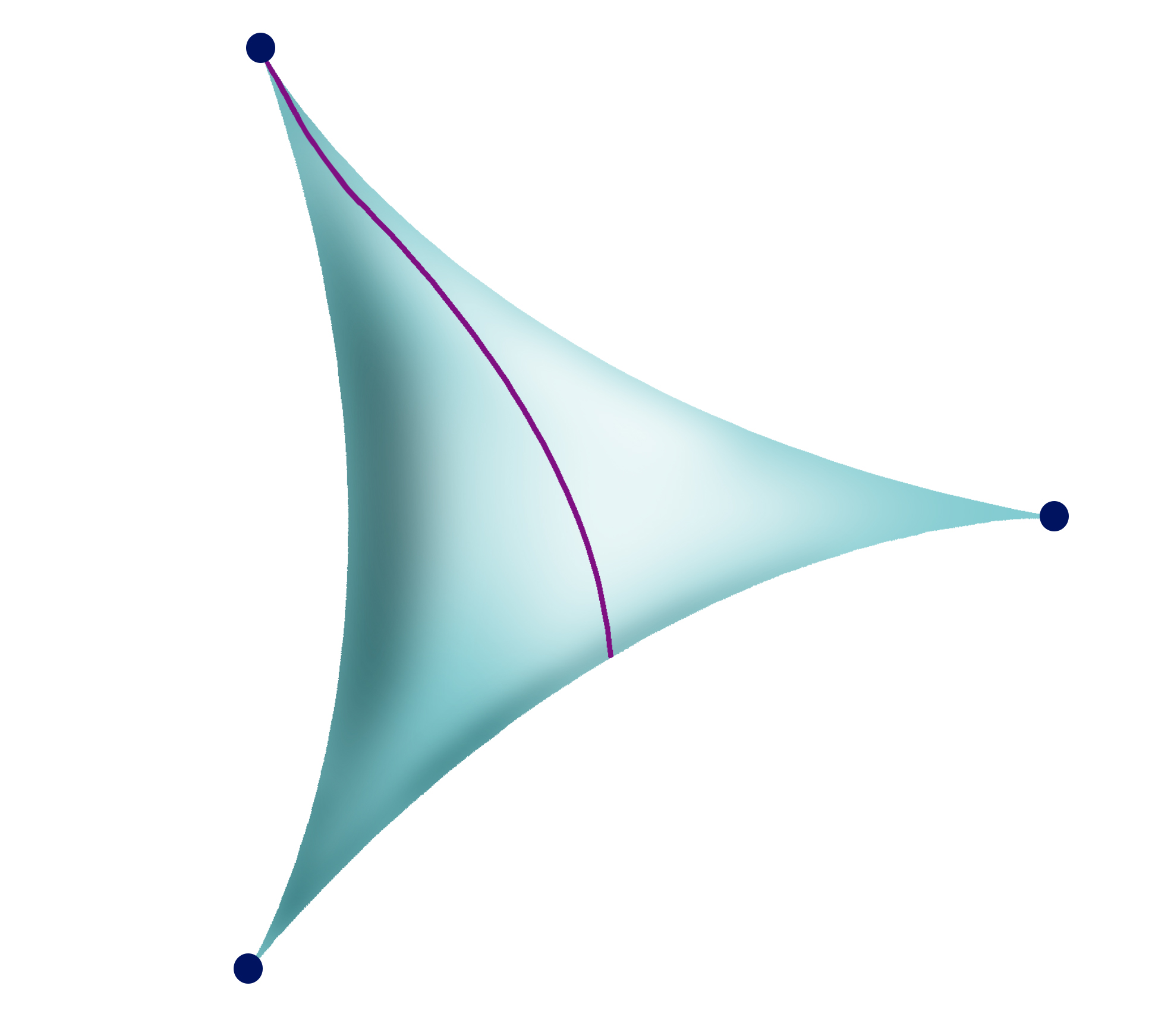}
	\begin{minipage}[t]{1.\linewidth}
    \centering
    \caption*{(d) Illustrative visualization of the orbifold. }
    \end{minipage}    
\caption{Visualization of a single $\mathbb{T}^2/\mathbb{Z}_3$ orbifold. The actual $\mathbb{T}^6/(\ZZ)$ orbifold employed in this work consists of the direct product of three copies of this orbifold.}
\label{fig:orbi}
\end{figure}

A visualization of such an orbifold is shown in \cref{fig:orbi}, which, for illustration, is focused on a single complex ED $z_i$. In (a), the unfolded compactified ED is shown where, to form the torus in (b), one identifies the green dotted lines with the red dotted lines, as required by \cref{eq:modtra}. Three fixed points from \cref{eq:fp} are represented by blue dots and the coordinates must be multiplied by the corresponding $R_i$. The same coloured lines and points correspond to each other in all the figures. The folding of a $\mathbb{Z}_3$ orbifold dictated by \cref{eq:modrot} is schematically represented in (c). In particular, this procedure is detailed as follows:
\begin{enumerate}
    \item First, one separates the whole space into two equilateral triangles.
    \item Each equilateral triangle is then divided into three equal isosceles triangles.
    \item Then, the three of them are rotated and identified together resulting in the rightmost of the four pictures in \cref{fig:orbi} (c). Note that the dotted lines from (a) and (b) must be identified with the purple one.
\end{enumerate}
Even though the projective space can not be properly visualized, (d) is merely illustrative, and the $\mathbb{T}^2/\mathbb{Z}_3$ would appear as shown here. The three conical singularities correspond to the three fixed points, and the purple line also corresponds to the one in (a) and (b) as soon as the identification with the dotted ones is performed.
%%%%%%%%%%%%%%%%%%%%%%%%%%%%%%%%

As detailed in \cref{app:orb}, each field receives a different phase under the orbifold action depending on its charges defined in \cref{eq:modrot}. The chiral superfields $\phi_i$ receive an extra phase contribution coming from the one multiplying each $z_i$. Each field receives two charges, one for each $\mathbb{Z}_3$, and $\mathcal{V}_{\textbf{248}}$ is split into components as shown in \cref{tab:pe3}.
%%%%%
\begin{table}[htb!]
	\centering
	\footnotesize
	\renewcommand{\arraystretch}{1.1}
	\begin{tabular}[t]{l|llll}
		\hline
		 & $V$ & $\phi_1$ & $\phi_2$ & $\phi_3$\\ 
			\hline
	$\mathcal{V}_{\textcolor{magenta}{(\textbf{8},\textbf{1},\textbf{1},0,0)}} $ & $1,1$ &  $\omega, 1$& $\omega, \omega^2$ & $\omega, \omega$
	\\
		$\mathcal{V}_{\textcolor{magenta}{(\textbf{1},\textbf{8},\textbf{1},0,0)}} $ & $1,1$ &  $\omega, 1$& $\omega, \omega^2$ & $\omega, \omega$
		\\
		$\mathcal{V}_{\textcolor{magenta}{(\textbf{1},\textbf{1},\textbf{8},0,0)}} $ & $1,1$ &  $\omega, 1$& $\omega, \omega^2$ & $\omega, \omega$
		\\
		$\mathcal{V}_{\textcolor{magenta}{(\textbf{1},\textbf{1},\textbf{1},0,0)}} $ & $1,1$ &  $\omega, 1$& $\omega, \omega^2$ & $\omega, \omega$
		\\
		$\mathcal{V}_{\textcolor{magenta}{(\textbf{1},\textbf{1},\textbf{1},0,0)}} $ & $1,1$ &  $\omega, 1$& $\omega, \omega^2$ & $\omega, \omega$
		\\
		$\mathcal{V}_{\textcolor{brown}{(\textbf{1},\textbf{1},\textbf{1},1,3)}} $ & $1,\omega$ &  $\omega, \omega$& $\omega, 1$ & $\omega, \omega^2$
		\\
		$\mathcal{V}_{\textcolor{brown}{(\textbf{1},\textbf{1},\textbf{1},-1,3)}} $ & $1,\omega^2$ &  $\omega, \omega^2$& $\omega, \omega$ & $\omega, 1$
		\\
		$\mathcal{V}_{\textcolor{brown}{(\textbf{1},\textbf{1},\textbf{1},1,-3)}} $ & $1,\omega$ &  $\omega, \omega$& $\omega, 1$ & $\omega, \omega^2$
		\\
		$\mathcal{V}_{\textcolor{brown}{(\textbf{1},\textbf{1},\textbf{1},-1,-3)}} $ & $1,\omega^2$ &  $\omega, \omega^2$& $\omega, \omega$ & $\omega, 1$
		\\
		$\mathcal{V}_{\textcolor{brown}{(\textbf{1},\textbf{1},\textbf{1},2,0)}} $ & $1,\omega^2$ &  $\omega, 1$& $\omega, \omega$ & $\omega, 1$
		\\
		$\mathcal{V}_{\textcolor{brown}{(\textbf{1},\textbf{1},\textbf{1},-2,0)}} $ & $1,\omega$ &  $\omega, \omega$& $\omega, 1$ & $\omega, \omega^2$
		\\			
		$\mathcal{V}_{(\bar{\textbf{3}},\textbf{3},\textbf{3},0,0)} $ & $1,\omega$ &  $\omega, \omega$& $\omega, 1$ & $\omega, \omega^2$
		\\
		$\mathcal{V}_{({\textbf{3}},\bar{\textbf{3}},\bar{\textbf{3}},0,0)} $ & $1,\omega^2$ &  $\omega, \omega^2$& $\omega, \omega$ & $\omega, 1$
		\\
		$\mathcal{V}_{\textcolor{ForestGreen}{(\textbf{1},\bar{\textbf{3}},\textbf{3},1,1)}} $ & $\omega^2,1$ &  $1, 1$& $1, \omega^2$ & $1, \omega$
		\\
	$\mathcal{V}_{\textcolor{ForestGreen}{(\textbf{1},\bar{\textbf{3}},\textbf{3},-1,1)}} $ & $\omega^2,\omega$ &  $1, \omega$& $1, 1$ & $1, \omega^2$
	\\
	$\mathcal{V}_{\textcolor{ForestGreen}{(\textbf{1},\bar{\textbf{3}},\textbf{3},0,-2)}} $ & $\omega^2,\omega^2$ &  $1, \omega^2$& $1, \omega$ & $1, 1$
	\\
		\hline
	\end{tabular}
	\hspace*{0.5cm}
	\begin{tabular}[t]{l|llll}
		\hline
		 & $V$ & $\phi_1$ & $\phi_2$ & $\phi_3$\\ 
		\hline
	$\mathcal{V}_{\textcolor{blue}{({\textbf{3}},\textbf{3},\textbf{1},1,1)}} $ & $\omega^2,\omega^2$ &  $1, \omega^2$& $1, \omega$ & $1, 1$
	\\
	$\mathcal{V}_{\textcolor{blue}{({\textbf{3}},\textbf{3},\textbf{1},-1,1)}} $ & $\omega^2,1$ &  $1, 1$& $1, \omega^2$ & $1, \omega$
	\\
	$\mathcal{V}_{\textcolor{blue}{({\textbf{3}},\textbf{3},\textbf{1},0,-2)}} $ & $\omega^2,\omega$ &  $1, \omega$& $1, 1$ & $1, \omega^2$
	\\
	$\mathcal{V}_{\textcolor{blue}{(\bar{\textbf{3}},\textbf{1},\bar{\textbf{3}},1,1)}} $ & $\omega^2,\omega$ &  $1, \omega$& $1, 1$ & $1, \omega^2$
	\\
	$\mathcal{V}_{\textcolor{blue}{(\bar{\textbf{3}},\textbf{1},\bar{\textbf{3}},-1,1)}} $ & $\omega^2,\omega^2$ &  $1, \omega^2$& $1, \omega$ & $1, 1$
	\\
	$\mathcal{V}_{\textcolor{blue}{(\bar{\textbf{3}},\textbf{1},\bar{\textbf{3}},0,-2)}} $ & $\omega^2,1$ &  $1, 1$& $1, \omega^2$ & $1, \omega$
	\\
	$\mathcal{V}_{\textcolor{orange}{(\textbf{1},\textbf{3},\bar{\textbf{3}},-1,-1)}} $ & $\omega,1$ &  $\omega^2, 1$& $\omega^2, \omega^2$ & $\omega^2, \omega$
	\\
	$\mathcal{V}_{\textcolor{orange}{(\textbf{1},\textbf{3},\bar{\textbf{3}},1,-1)}} $ & $\omega,\omega^2$ &  $\omega^2, \omega^2$& $\omega^2, \omega$ & $\omega^2, 1$
	\\
	$\mathcal{V}_{\textcolor{orange}{(\textbf{1},\textbf{3},\bar{\textbf{3}},0,2)}} $ & $\omega,\omega$ &  $\omega^2, \omega$& $\omega^2, \omega$ & $\omega^2, \omega^2$
	\\
		$\mathcal{V}_{\textcolor{red}{(\bar{\textbf{3}},\bar{\textbf{3}},\textbf{1},-1,-1)}} $ & $\omega,\omega$ &  $\omega^2, \omega$& $\omega^2, \omega$ & $\omega^2, \omega^2$
		\\
		$\mathcal{V}_{\textcolor{red}{(\bar{\textbf{3}},\bar{\textbf{3}},\textbf{1},1,-1)}} $ & $\omega,1$ &  $\omega^2, 1$& $\omega^2, \omega^2$ & $\omega^2, \omega$
		\\
		$\mathcal{V}_{\textcolor{red}{(\bar{\textbf{3}},\bar{\textbf{3}},\textbf{1},0,2)}} $ & $\omega,\omega^2$ &  $\omega^2, \omega^2$& $\omega^2, \omega$ & $\omega^2, 1$
		\\
		$\mathcal{V}_{\textcolor{red}{({\textbf{3}},\textbf{1},\textbf{3},-1,-1)}} $ & $\omega,\omega^2$ &  $\omega^2, \omega^2$& $\omega^2, \omega$ & $\omega^2, 1$
		\\
		$\mathcal{V}_{\textcolor{red}{({\textbf{3}},\textbf{1},\textbf{3},1,-1)}} $ & $\omega,\omega^2$ &  $\omega^2, \omega^2$& $\omega^2, \omega$ & $\omega^2, 1$
		\\
		$\mathcal{V}_{\textcolor{red}{({\textbf{3}},\textbf{1},\textbf{3},0,2)}} $ & $\omega,1$ &  $\omega^2, 1$& $\omega^2, \omega^2$ & $\omega^2, \omega$\\
		\hline
	\end{tabular}
	\caption{Charges of each $\mathcal{N}=1$ superfield being $ \SU{3}{C}\times \SU{3}{L}\times \SU{3}{R}\times \U{F3}\times \U{F8}$ multiplet under a $\ZZ$ orbifolding. Only the fields with both charges equal to unity have zero modes. The representations are color coded as \textcolor{magenta}{adjoint fields}, \textcolor{ForestGreen}{Higgs and leptons}, \textcolor{blue}{quarks}, \textcolor{orange}{mirror Higgs and mirror fermions}, \textcolor{red}{mirror quarks}, and {\bf SM exotics}.} 
	\label{tab:pe3}
\end{table}
%%%%%%%%%%%%%%
The zero (massless) modes containing the SM particle spectrum coming from the purely rotational boundary conditions (i.e. the ones with charge $(1,1)$), for each chiral multiplet, are then
\begin{equation}\begin{split}
V_\mu &: \textcolor{magenta}{(\textbf{8},\textbf{1},\textbf{1},0,0)}+\textcolor{magenta}{(\textbf{1},\textbf{8},\textbf{1},0,0)}+\textcolor{magenta}{(\textbf{1},\textbf{1},\textbf{8},0,0)}+2\times\textcolor{magenta}{(\textbf{1},\textbf{1},\textbf{1},0,0)} \,,
\\
\phi_1&:\textcolor{ForestGreen}{(\textbf{1},\bar{\textbf{3}},\textbf{3},1,1)}
+\textcolor{blue}{(\bar{\textbf{3}},\textbf{1},\bar{\textbf{3}},0,-2)}+\textcolor{blue}{({\textbf{3}},\textbf{3},\textbf{1},-1,1)} \,, \\
\phi_2&:\textcolor{ForestGreen}{(\textbf{1},\bar{\textbf{3}},\textbf{3},-1,1)}+\textcolor{blue}{(\bar{\textbf{3}},\textbf{1},\bar{\textbf{3}},1,1)}+\textcolor{blue}{({\textbf{3}},\textbf{3},\textbf{1},0,-2)} \,, \\
\phi_3&: \textcolor{ForestGreen}{(\textbf{1},\bar{\textbf{3}},\textbf{3},0,-2)}+\textcolor{blue}{(\bar{\textbf{3}},\textbf{1},\bar{\textbf{3}},-1,1)}+\textcolor{blue}{({\textbf{3}},\textbf{3},\textbf{1},1,1)} \,.
\label{eq:zmt}
\end{split}\end{equation}
%%%%%%%%%
So far, we have only considered the rotational boundary conditions while additional mass-splitting effects emerging from the Wilson-line effective VEVs must also be incorporated. It is very important to note that the spectrum in \cref{eq:zmt} does not correspond to the massless field content of the theory at any point. It is shown here because it helps visualize the way in which the symmetry is being broken by the rotational boundary conditions. The {\it complete} $\E{8}$ symmetry breaking down to the SM gauge symmetry group does indeed involve also the translation boundary conditions in the form of Wilson lines together with the rotational boundary conditions. Hence, the latter two breakings are superimposed and happen simultaneously.

For ease of notation, in the remainder of this article, the representations in \cref{eq:zmt} are named as
\begin{equation}\begin{split}
V_\mu &: \textcolor{magenta}{\Delta_C}+\textcolor{magenta}{\Delta_L}+\textcolor{magenta}{\Delta_R}+\textcolor{magenta}{Z_{F3}}+\textcolor{magenta}{Z_{F8}} \,, \\
\phi_1&:\textcolor{ForestGreen}{\textbf{L}_1}+\textcolor{blue}{\textbf{Q}_{R3}}+\textcolor{blue}{\textbf{Q}_{L2}} \,, \\
\phi_2&:\textcolor{ForestGreen}{\textbf{L}_2}+\textcolor{blue}{\textbf{Q}_{R1}}+\textcolor{blue}{\textbf{Q}_{L3}} \,, \\
\phi_3&:\textcolor{ForestGreen}{\textbf{L}_3}+\textcolor{blue}{\textbf{Q}_{R2}}+\textcolor{blue}{\textbf{Q}_{L1}} \,.
\label{eq:zmt2}
\end{split}\end{equation}
%%%%%%%%%
Expressing the chiral superfields in terms of their gauge indices, where upper indices correspond to an antitriplet representation of $\SU{3}{}$ while down indices stand for a triplet one, one can further split them in terms of their components as
\begin{equation}
\begin{array}{llll}
\textcolor{ForestGreen}{\textbf{L}^i_{\ jk}}:& \textcolor{ForestGreen}{\textbf{L}^1_{\ 1k}}=\textcolor{OliveGreen}{H_{dk}^0}\,, & \textcolor{ForestGreen}{\textbf{L}^1_{\ 2k}}=\textcolor{OliveGreen}{H_{uk}^-}\,, & \textcolor{ForestGreen}{\textbf{L}^1_{\ 3k}}=\textcolor{blue}{\nu_k} \,,  \\
& \textcolor{ForestGreen}{\textbf{L}^2_{\ 1k}}=\textcolor{OliveGreen}{H_{dk}^+}\,, & \textcolor{ForestGreen}{\textbf{L}^2_{\ 2k}}=\textcolor{OliveGreen}{H_{uk}^0}\,, & \textcolor{ForestGreen}{\textbf{L}^2_{\ 3k}}=\textcolor{blue}{e_k}\,,  \\
& \textcolor{ForestGreen}{\textbf{L}^3_{\ 1k}}=\textcolor{violet}{\nu^c_k}\,, & \textcolor{ForestGreen}{\textbf{L}^3_{\ 2k}}=\textcolor{blue}{e^c_k}\,, & \textcolor{ForestGreen}{\textbf{L}^3_{\ 3k}}=\textcolor{brown}{\varphi_k}\,,  \\
\textcolor{blue}{\textbf{Q}_{R\ j}^{\ \ i}}: &  \textcolor{blue}{\textbf{Q}_{R\ j}^{\ \ 1}}=\textcolor{blue}{d^c_j}\,, & \textcolor{blue}{\textbf{Q}_{R\ j}^{\ \ 2}}=\textcolor{blue}{u^c_j}\,, & \textcolor{blue}{\textbf{Q}_{R\ j}^{\ \ 3}}=D^c_j\,, \\
\textcolor{blue}{\textbf{Q}_{Li j}}:&  \textcolor{blue}{\textbf{Q}_{L1 j}}=\textcolor{blue}{d_j}\,, & \textcolor{blue}{\textbf{Q}_{L2 j}}=\textcolor{blue}{u_j}\,, & \textcolor{blue}{\textbf{Q}_{L3 j}}=D_j\,, \\
\label{eq:fieldnames}
\end{array}
\end{equation}
where the color indices are not explicitly shown and the family ones are not expanded. To complete the notation, the complex conjugated chiral superfield representations are named as $\textcolor{orange}{\overline{\textbf{L}}},\textcolor{red}{\overline{\textbf{Q}}_R},\textcolor{red}{\overline{\textbf{Q}}_L}$, while the remaining superfields transforming as $(\bar{\textbf{3}},\textbf{3},\textbf{3},\textbf{1})$ and $({\textbf{3}},\bar{\textbf{3}},\bar{\textbf{3}},\textbf{1})$ are denoted by $\textbf{X}$ and $\bar{\textbf{X}}$, respectively.
The representations are color-coded as follows: \textcolor{magenta}{adjoint fields}, \textcolor{OliveGreen}{Higgs}, \textcolor{blue}{SM fermions}, \textcolor{orange}{mirror Higgs}, \textcolor{red}{mirror fermions}, \textcolor{violet}{right handed neutrinos}, \textcolor{brown}{flavons}, and {\bf family-neutral exotics}.

As stated before in \cref{eq:vwl}, there are three Wilson lines $U_i^r \equiv U_i$, where both real degrees of freedom $r=1,2$ form a complex one. They correspond to a gauge transformation which must comply with the consistency conditions coming from the Poincar\'e algebra
\begin{equation}
[U_i,U_j]=0 \,,\ \ \ [U_i,U^{\rm F}_i]=0 \,,\ \ \  [U_i,U^{\rm C}_i]=0 \,,
\label{eq:wlcom}
\end{equation}
where $U^{\rm C,F}_i$ denote the gauge transformation that accompanies the orbifold action of $\ZCF$ on a chiral superfield
\begin{equation}
    \mathcal{Z}^{\rm C,F}[\phi_i(x,z)]=U^{\rm C,F}_i \phi_i(x,\mathcal{Z}^{\rm C,F}[z]) \,.
\end{equation}

There are two distinct ways to solve the last two constraints in \cref{eq:wlcom}. The first one is by requiring
\begin{equation}
 U_i^3=1 \,,
\end{equation}
which corresponds to the case of discrete Wilson lines resulting in additional boundary conditions on the fixed points. For instance, before one considers the effect of such discrete Wilson lines, one has the $U^{\rm C,F}_i$ gauge transformation as a boundary condition in \cref{eq:ZCF} at the zero brane, while at the $(1+e^{i\pi/3})/3$ brane they read as $U_i U^{\rm C,F}_i$ and, finally, at the $2(1+e^{i\pi/3})/3$ brane, one has $ U_i^2 U^{\rm C,F}_i$. Any other boundary condition would result in an incomplete SM field content as it would leave a whole column, row or family from \cref{eq:fieldnames} without zero modes. This means that, in the considered GUT, a phenomenologically viable low-scale spectrum is not compatible with the presence of discrete Wilson lines. 

The second possibility is to restrict the Wilson lines to representations where the orbifold action is trivial, i.e.~zero modes. One can choose a continuous Wilson line which generates an effective VEV in the three $\phi_{1,2,3}$ zero modes. However, due to invariance under translations defined in \cref{eq:modtra}, from where the effective VEVs result, they have to commute with each other, complying with the first constraint of \cref{eq:wlcom}. The restrictions on the effective VEVs come from the $\E{8}$ commutation relations in \cref{eq:e8comm} given in \cref{app:e8gen}. 

Note, the Wilson line is assumed to be aligned with $\textcolor{ForestGreen}{\textbf{L}^1_{\ 3k}}\sim \textcolor{violet}{\nu^c_k}$ and $\textcolor{ForestGreen}{\textbf{L}^3_{\ 3k}}\sim \textcolor{brown}{\varphi_k}$, where all corresponding generators commute with each other as seen in \cref{eq:e8comm}. These generate effective VEVs which are naturally scaled as
\begin{equation}
    \braket{\textcolor{brown}{\varphi_i}},\braket{\textcolor{violet}{{\nu}^c_i}}\sim 
    \frac{1}{R_i} \,.
\end{equation}
From these components, the $\braket{\textcolor{violet}{{\nu}^c_1}}$ can always be rotated away by an $\SU{3}{R}$ rotation while the $\braket{\textcolor{violet}{{\nu}^c_{2,3}}}$ can be made real by family symmetry ($\U{F3}$ and $\U{F8}$) transformations. There are no other possible rotations, therefore, the Wilson line is defined by three complex and two real parameters.

The Higgs states are assumed to obtain a VEV radiatively, hence, generating a much smaller scale and breaking the EW symmetry of the SM but preserving the $\U{e.m.}$ of electromagnetism, with the corresponding charge, 
\begin{equation}
 Q_{\rm EM}=\tfrac{1}{6}[\sqrt{3}T^{\rm L}_8+\sqrt{3}T^{\rm R}_8-3T^{\rm L}_3-3T^{\rm R}_3] \,,
\end{equation}
where the normalization ${\rm Tr}\ T_a T_b=2\delta_{ab}$ is adopted.

The chosen Wilson line configuration breaks the gauge trinification symmetry down to that of the SM, i.e.
\begin{equation}
   \SU{3}{C}\times \SU{3}{L}\times \SU{3}{R}\times \U{F3}\times \U{F8} \to \SU{3}{C}\times \SU{2}{L}\times \U{Y}\,.
\end{equation}
Such a breaking leaves the following massless fields in the low-energy SM-like EFT:
 \begin{equation}\begin{split}
V_\mu &:  \textcolor{magenta}{(\textbf{8},\textbf{1},0)}+\textcolor{magenta}{(\textbf{1},\textbf{3},0)}+\textcolor{magenta}{(\textbf{1},\textbf{1},0)} \,, \\
\phi_1&: \textcolor{blue}{(\textbf{1},\textbf{2},-3)}+3\times\textcolor{blue}{(\textbf{1},\textbf{1},6)}+ \textcolor{OliveGreen}{(\textbf{1},\textbf{2},3)}+ \textcolor{OliveGreen}{(\textbf{1},\textbf{2},-3)}+ \textcolor{brown}{(\textbf{1},\textbf{1},0)}+ \textcolor{violet}{(\textbf{1},\textbf{1},0)}  \\
& \ +\textcolor{blue}{(\bar{\textbf{3}},\textbf{1},-4)}+\textcolor{blue}{(\bar{\textbf{3}},\textbf{1},2)}+(\bar{\textbf{3}},\textbf{1},2) +\textcolor{blue}{(\textbf{3},\textbf{2},1)}+ (\textbf{3},\textbf{1},-2) \,, \\
\phi_2&: \textcolor{blue}{(\textbf{1},\textbf{2},-3)}+3\times\textcolor{blue}{(\textbf{1},\textbf{1},6)}+ \textcolor{OliveGreen}{(\textbf{1},\textbf{2},3)}+ \textcolor{OliveGreen}{(\textbf{1},\textbf{2},-3)}+ \textcolor{brown}{(\textbf{1},\textbf{1},0)}+ \textcolor{violet}{(\textbf{1},\textbf{1},0)}  \\
& \ +\textcolor{blue}{(\bar{\textbf{3}},\textbf{1},-4)}+\textcolor{blue}{(\bar{\textbf{3}},\textbf{1},2)}+(\bar{\textbf{3}},\textbf{1},2) +\textcolor{blue}{(\textbf{3},\textbf{2},1)}+ (\textbf{3},\textbf{1},-2) \,, \\
\phi_3&:\textcolor{blue}{(\textbf{1},\textbf{2},-3)}+3\times\textcolor{blue}{(\textbf{1},\textbf{1},6)}+ \textcolor{OliveGreen}{(\textbf{1},\textbf{2},3)}+ \textcolor{OliveGreen}{(\textbf{1},\textbf{2},-3)}+ \textcolor{brown}{(\textbf{1},\textbf{1},0)}+ \textcolor{violet}{(\textbf{1},\textbf{1},0)}  \\
& \ +\textcolor{blue}{(\bar{\textbf{3}},\textbf{1},-4)}+\textcolor{blue}{(\bar{\textbf{3}},\textbf{1},2)}+(\bar{\textbf{3}},\textbf{1},2) +\textcolor{blue}{(\textbf{3},\textbf{2},1)}+ (\textbf{3},\textbf{1},-2) \,,
\label{eq:zmff3}
\end{split}\end{equation}
which in what follows are respectively denoted as
\begin{equation}\begin{split}
V_\mu &:  \textcolor{magenta}{G_\mu}+\textcolor{magenta}{W_\mu}+\textcolor{magenta}{B_\mu} \,, \\
\phi_i&:\textcolor{blue}{L_i}+\textcolor{blue}{e^c_i}+\textcolor{OliveGreen}{h_{ui}}+\textcolor{OliveGreen}{h_{di}}+\textcolor{brown}{\varphi_i}+ \textcolor{violet}{\nu^c_i} \, \\
&+ \textcolor{blue}{u^c_{i-1}}+\textcolor{blue}{d^c_{i-1}}+D^c_{i-1}+\textcolor{blue}{Q_{i+1}}+D_{i+1} \,,
\end{split}\end{equation}
with $i=1,2,3$ and the index rules $0\to 3$ and $4\to 1$ are implied. These are the only massless superfields resulting from the compactification procedure described above, containing the SM fermions and gauge bosons, as well as three-right handed neutrinos $\textcolor{violet}{\nu^c_i}$, six Higgs doublets $\textcolor{OliveGreen}{h_{ui}},\textcolor{OliveGreen}{h_{di}}$, three flavons $\textcolor{brown}{\varphi_i}$, three vector-like color triplet pairs $D^c_i,D_i$, and the associated superpartners. Such a field content will be further split in the mass spectrum as a result of the Wilson line effective VEVs, leaving solely the SM fermions, color coded in blue, and gauge bosons, color coded in magenta, as the only massless states in the theory at low energies (i.e. below the Wilson-line breaking scale). Thus, the considered $\E{8}$ GUT fully reproduces the SM structure at low energies as it should.

%%%%%%%%%%%%%%%%%%%%%%%%%%%%%%%%%%%%%%%%%%%%%%%%%%
\subsection{Gauge anomaly cancellation}
\label{sec:anom}
%%%%%%%%%%%%%%%%%%%%%%%%%%%%%%%%%%%%%%%%%%%%%%%%%%

The original theory is based on the $\E{8}$ gauge group with matter contained in a single adjoint representation. The only 10d Weyl/Majorana fermion lies in this real representation. Thus, at the fundamental level, the model is free from gauge anomalies. As mentioned above, the 10d superfield decomposes into an infinite tower of KK modes of 4d superfields, and each $\textbf{248}$ representation is broken into nine different sets -- one for each of the possible orbifold charges given in \cref{tab:pe3}. One can then expand the chiral superfields as
  \begin{equation}
    \phi_i(x,z_i)=\sum_{\textbf{s}} \sum_{a,b = 0,1,2}\phi^{\textbf{s}ab}_{i}(x)f(z_j)_{\textbf{s}ab} \,,
    \label{eq:decom}
\end{equation}
with $\textbf{s}=(s_1,s_2,s_3,s_4,s_5,s_6)$ denoting the KK level coming from a schematic Fourier-like decomposition of the 6d states, while $a,b = 0,1,2$ arising from a splitting of each $\mathbb{Z}_3^{\rm C,F}$ orbifolding. Note that the decomposition into different $\textbf{s}$ modes does not break the $E_8$ symmetry, while the decomposition into different $a,b$ modes does actually break it. The zero modes correspond to the configurations with $s_i=a=b=0$. Note that the above series can be separated into an infinite set of decomposed $\textbf{248}$'s, where the anomalies cancel in each such set.

As was emphasised above, the $\E{8}$ gauge group is broken simultaneously by the rotation and translation boundary conditions below the compactification scale, which is defined by $\Lambda=1/R_l$, where $R_l$ is the largest radius of $R_{1,2,3}$. Specifically, they break $\E{8}\to \SU{3}{C}\times \SU{2}{L}\times \U{Y}$ yielding the massless field content shown in \cref{eq:zmff3} which comprises the SM fields as well as additional SM singlets, vector-like fermions and their superpartners. Therefore, the massless sector of the theory is also anomaly free \cite{Aranda:2020zms}.

One must remark that the Wilson lines behave like effective VEVs but they are not related to spontaneous symmetry breaking. They come from boundary conditions on translations, just as the orbifolding comes from the boundary conditions on rotations. As alluded to above, both rotation and translation boundary conditions simultaneously induce the full breaking of $\E{8}$ and thus they both contribute to the particles' masses that would be integrated out when studying the corresponding low-energy SM-like EFT. Therefore, even though the model is built in terms of the fields listed in \cref{eq:zmt2}, there is no point in the theory when this spectrum appears as entirely massless, and therefore no anomalies are present at any stage.

As this model is determined on an orbifold, which is not a continuous spacetime, one must pay special attention to the discontinuous points. This orbifold has 27 fixed ED points listed in \cref{eq:fp}. At these points, which are 4d branes, the boundary conditions force some of the 4d fields to vanish (i.e.~their ED profiles are zero in all the fixed points), so that they do not have the full $\E{8}$ symmetry nor the full $\E{8}$ representation. The gauge transformations in a fixed point have less freedom as the gauge parameters are only functions of the four open dimensions. To avoid having multiply valued functions, the ED degrees freedom for the gauge symmetry must be fixed. Their symmetry, which is defined in a 4d spacetime, can only be studied in a fixed specific gauge (which only fixes the ED degrees of freedom) where the Wilson line behaves like an effective VEV as they are disconnected in EDs. In this gauge, all the fields are periodic in the compact dimensions as described in \cref{eq:wilgauge}. In any other gauge, the fields are discontinuous multiply-valued functions, and their study requires techniques beyond the standard complex analysis. Most importantly, the gauge symmetry in these points is $\SU{3}{C}\times \SU{2}{L}\times \U{Y}$, with the massless field content specified in \cref{eq:zmff3}, with no extra gauge freedom. Therefore, there are no anomalies generated in the fixed points either \cite{Scrucca:2004jn}. As the considered GUT has no localized fields, there are no extra contributions to the anomaly, rendering the model anomaly free at all steps without the need for any additional fields.

%%%%%%%%%%%%%%%%%%%%%%%%%%%%%%%%%%%%%%%
\section{Effective Lagrangian at low energies}
\label{sec:trilag}
%%%%%%%%%%%%%%%%%%%%%%%%%%%%%%%%%%%%%%%

In order to study phenomenological implications of the considered GUT, the next step is to build the EFT Lagrangian from the fields given in \cref{eq:zmt} that are relevant below the compactification scale. They must respect the trification symmetry supplemented by the family symmetry, i.e.~$ \SU{3}{C}\times \SU{3}{L}\times \SU{3}{R}\times \U{F3}\times \U{F8}$, that can only be broken through the effective, Wilson line, VEVs in $\braket{\textcolor{ForestGreen}{\textbf{L}}}\sim \braket{\textcolor{brown}{\varphi_i}},\braket{\textcolor{violet}{\nu^c_i}}$. Furthermore, the higher-order operators are mediated by heavy KK modes. In \cref{app:lag}, the basic steps to obtain an effective Lagrangian from a general superpotential like the one that emerges in the considered GUT are described. The general superpotentials of the GUT under discussion are obtained in \cref{app:modlag}. From \cref{eq:genac}, the low-energy EFT Lagrangian is then defined in terms of the potentials $\mathcal{K},\,\mathcal{W}$ and $\mathcal{H}$.

At leading order, the gauge field strength superpotential $\mathcal{H}$ takes the standard form, i.e. $\mathcal{H}_A^{\ B} W^AW_B=W^A W_A$, the K\"ahler potential $\mathcal{K}$ is also the standard one, while the chiral superpotential $\mathcal{W}$ reads as follows:
\begin{equation}
\begin{split}
\mathcal{W} \sim & \, g\phi_1\phi_2\phi_3 
\\
= & \, g [\epsilon^{ijk}\textcolor{ForestGreen}{\textbf{L}^l_{\ m i}}\textcolor{blue}{\textbf{Q}_{R\ j}^{\ m}}\textcolor{blue}{\textbf{Q}_{Llk}}
+\epsilon_{abc}\epsilon^{mno}\textcolor{ForestGreen}{\textbf{L}^a_{\ m 1}}\textcolor{ForestGreen}{\textbf{L}^b_{\ n 2}}\textcolor{ForestGreen}{\textbf{L}^c_{\ o 3}}
 +\epsilon_{abc}\textcolor{blue}{\textbf{Q}_{R\ 1}^{\ a}}\textcolor{blue}{\textbf{Q}_{R\ 2}^{\ b}}\textcolor{blue}{\textbf{Q}_{R\ 3}^{\ c}} +\epsilon^{abc}\textcolor{blue}{\textbf{Q}_{La1}}\textcolor{blue}{\textbf{Q}_{Lb2}}\textcolor{blue}{\textbf{Q}_{Lc3}}]
 \\
= & \, g\varepsilon^{ijk} \Big( \textcolor{OliveGreen}{h_{ui}}\textcolor{blue}{u^c_j}\textcolor{blue}{Q_k}+ \textcolor{OliveGreen}{h_{di}}\textcolor{blue}{d^c_j}\textcolor{blue}{Q_k}+[\textcolor{brown}{\varphi_i}+\braket{\textcolor{brown}{\varphi_i}}]D^c_j D_k+[\textcolor{violet}{\nu^c_i}+\braket{\textcolor{violet}{\nu^c_i}}]\textcolor{blue}{d^c_j}D_k +\textcolor{blue}{e_i^c}\textcolor{blue}{u^c_j}D_k+\textcolor{blue}{L_i Q_j}D^c_k
\\
+ & \, \textcolor{OliveGreen}{H_{di}}\textcolor{blue}{L_j}\textcolor{blue}{e^c_k}+\textcolor{OliveGreen}{H_{ui}}\textcolor{blue}{L_j}[\textcolor{violet}{\nu^c_k}+\braket{\textcolor{violet}{\nu^c_k}}]+\textcolor{OliveGreen}{H_{di}}\textcolor{OliveGreen}{H_{uj}}[\textcolor{brown}{\varphi_k}+\braket{\textcolor{brown}{\varphi_k}}]
 +\textcolor{blue}{d^c_i}\textcolor{blue}{u^c_j}D^c_k+\textcolor{blue}{Q_i}\textcolor{blue}{Q_j}D_k\Big) \,,
\label{eq:quyu}
\end{split}
\end{equation}
where $g$ is the gauge coupling of the $\E{8}$ theory evolved to an appropriate energy-scale and the corresponding Clebsch-Gordan coefficients (CGC) are ignored here.

The fermion mass terms are generated from SUSY breaking effects in the K\"ahler potential as can be seen in \cref{app:lag}. This takes place already at leading order via the $\mathcal{D}$-term,
\begin{equation}
    \braket{\mathcal{D}_A}=\sum_{ij} \braket{\textcolor{violet}{\nu^c_i}}^\dagger t_A\braket{\textcolor{violet}{\nu^c_j}}+\braket{\textcolor{brown}{\varphi_i}}^\dagger t_A \braket{\textcolor{brown}{\varphi_j}}+\braket{\textcolor{brown}{\varphi_i}}^\dagger t_A\braket{\textcolor{violet}{\nu^c_j}}+\braket{\textcolor{violet}{\nu^c_i}}^\dagger t_A\braket{\textcolor{brown}{\varphi_j}}\neq 0 \,,
    \label{eq:dbreak}
\end{equation}
at the scale of the effective Wilson-line VEVs.

\begin{figure}[h]
\centering
\includegraphics[scale=0.2]{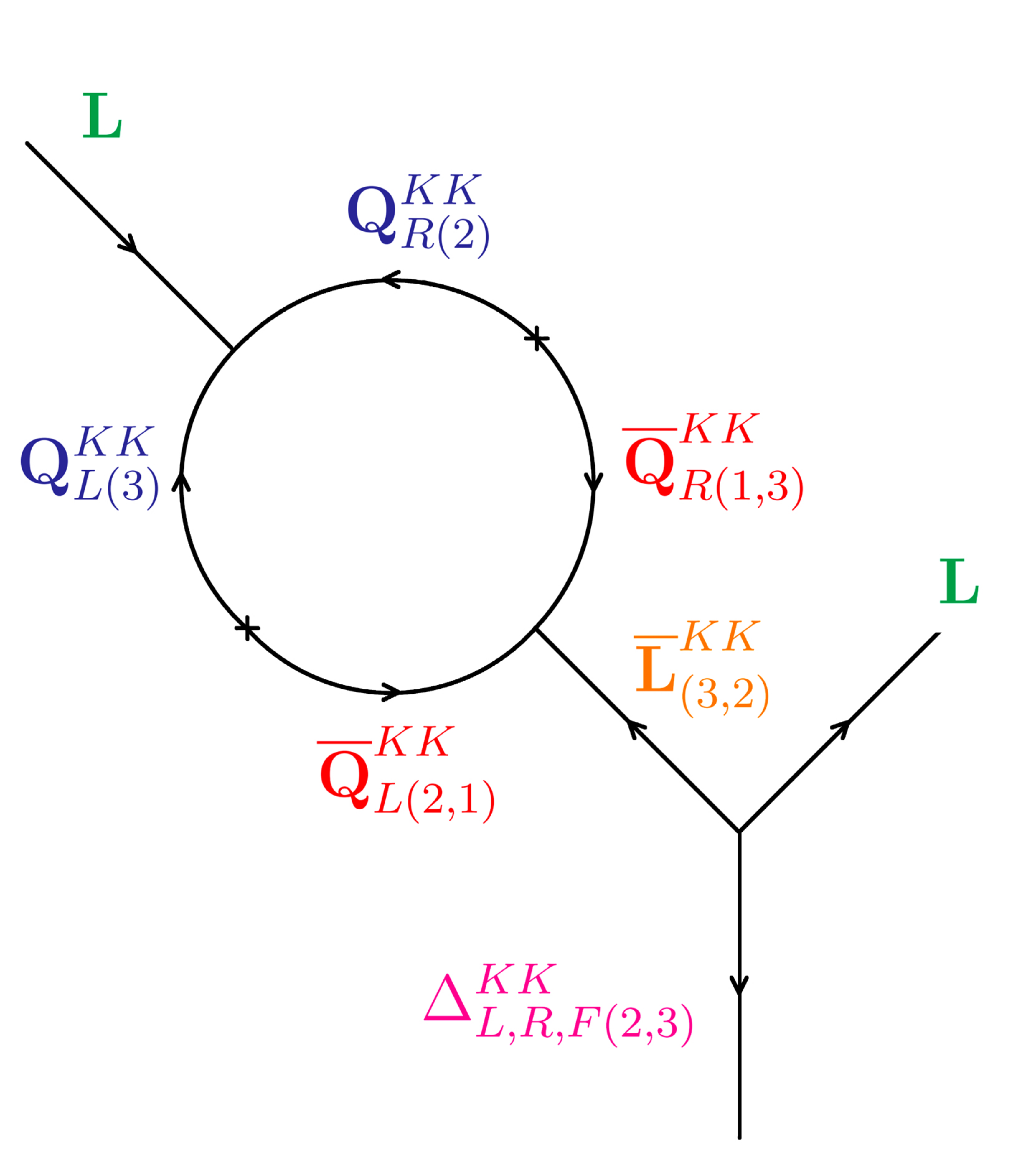}
\caption{The leading-order diagram for the effective adjoint superfields as  $\textcolor{Magenta}{\Delta_{\rm L,R,F}} = \textcolor{ForestGreen}{\textbf{L}^\dagger\textbf{L}}$ through an interaction with the corresponding KK mode. There are two possible KK modes here that come from different chiral fields $\phi_i$. }
\label{fig:del}
\end{figure}

There are no adjoint scalars as zero modes in this model but there are KK modes which can mediate the Wilson line effective VEVs to provide a richer structure of the couplings. They can enter as in diagram \cref{fig:del}. This way one can have effective adjoint VEVs. Note that even when in the considered model the gauge family symmetry is $\U{F3}\times \U{F8}$, one generate adjoints $\textcolor{Magenta}{\Delta_{\rm F}}$, as such fields exist inside $\E{8}$ and become KK modes.

%%%%%%%%%%%%%%%%%%%%%%%%%%%%%%%%%%%%%%%%%%%%%%%%%%%%%%%%%%%
%%%%%%%%%%%%%%%%%%%%%%%%%%%%%%%%%%%%%%%%%%%%%%%%%%%%%%%%%%%

In order to obtain the SM-like EFT Lagrangian, one does not need to consider the full Lagrangian of the GUT since not every term of the latter is relevant at low-energy scales. In fact, the part of the superpotential that is relevant at low energies is obtained by truncating the higher-order terms and yielding \cref{eq:quyu}, where only the second and third generation obtain their masses through the superpotential. However, their mass matrix is completely antisymmetric and can neither generate the observed flavour structure nor the fermion mass spectra that would fit the SM. The correct physical mass spectra and the flavour mixing structure of light fermions are then obtained after SUSY breaking effects take place. Upon compactification of the EDs, the Wilson lines behave as effective VEVs in $\textcolor{ForestGreen}{\textbf{L}}$. In turn, one sees from \cref{eq:susybreak} in \cref{app:lag} that such effective VEVs cannot generate $\mathcal{F}$-term SUSY breaking, but rather, they do induce $\mathcal{D}$-term SUSY breaking already at the leading order. One then finds that, using \cref{eq:fulla}, the fermion masses are obtained through the terms
\begin{equation}
\begin{split}
  -\frac{1}{2}\left\langle\left.\frac{\partial^2 \mathcal{W}}{\partial\phi_i\partial\phi_j}\right|_S\right\rangle\bar{\psi}_i\psi_{Lj}-i\sqrt{2}\left\langle\left.\frac{\partial^2 \mathcal{K}}{\partial\phi_i^\dagger\partial\phi_j}\right|_S S^\dagger_i t_A\right\rangle\bar{\psi}_j\lambda_{AL}
  +\frac{i}{2\sqrt{2}}\left\langle\left.\frac{\partial \mathcal{H}_{AB}}{\partial\phi_i}\right|_S\mathcal{D}_A\right\rangle \bar{\lambda}_B \psi_{Li} \,,
  \label{eq:ferma}
  \end{split}
\end{equation}
where the first one encodes the usual SUSY preserving fermion mass term coming from the superpotential, the second term is the usual SUSY preserving mass term for gauginos, whereas the last one contains the new $\mathcal{D}$-term SUSY breaking effects. In the remainder of this section, we use a notation expressed in terms of usual fields, rather than superfields, charged under the SM gauge symmetry. Following the standard notation adopted in the literature, the $\textcolor{blue}{Q},\textcolor{blue}{L},\textcolor{blue}{u^c},\textcolor{blue}{d^c},\textcolor{blue}{e^c},\textcolor{violet}{\nu^c},D,D^c$ are defined to be fermions, with their scalar partners having an extra tilde $\sim$ above their labels. The $\textcolor{OliveGreen}{H_{u,d}},\textcolor{brown}{\varphi}$ are defined to be scalars while their fermion partners are denoted by an extra tilde $\sim$ above their labels.

%%%%%%%%%%%%%%%%%%%%%%%%%%%%%%%%%%%%%%
\subsection{Chiral fermions}
\label{sec:chiral-ferm}
%%%%%%%%%%%%%%%%%%%%%%%%%%%%%%%%%%%%%%

\begin{figure}[h]
\centering
\includegraphics[scale=0.24]{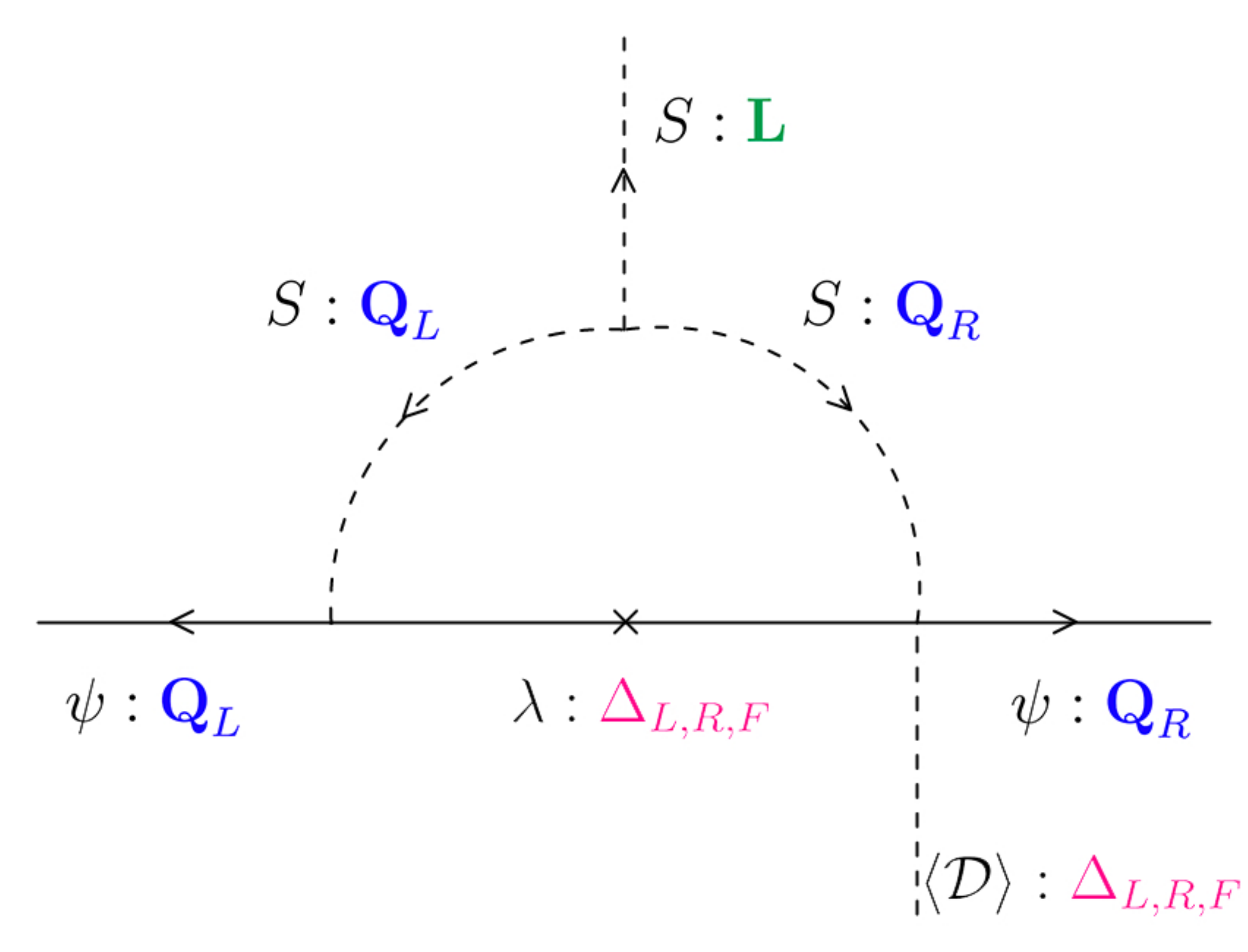}
\caption{The one-loop diagrams providing SUSY-breaking contributions to the light fermion masses built in terms of usual fields, not superfields . Here, $\psi$ and $S$ denote the fermion and scalar components of the corresponding superfields, $\lambda$ stands for the gaugino and $\braket{\mathcal{D}}$ labels a adjoint VEV built from the $\braket{\textcolor{ForestGreen}{\textbf{L}}}$. The three-point vertices come from the $\mathcal{D}\mathcal{D}$ and $\mathcal{F}^\dagger \mathcal{F}$ terms in \cref{eq:fdsol,eq:fulla}, with at least one scalar $S$ taking a Wilson line effective VEV.}
\label{fig:newmas}
\end{figure}

The last term in \cref{eq:ferma} generates corrections to the fermion mass spectra proportional to $\braket{\mathcal{D}}$ through the one-loop diagrams shown in \cref{fig:newmas}.
%%%%%%%%%%%%%%%%%%%%%%%%%%%%%%%%%%%%%%%%%%%%%%%%%%
%%%%%%%%%%%%%%%%%%%%%%%%%%%%%%%%%%%%%%%%%%%%%%%%%%

In order to simplify the corresponding expressions, one defines the shorthand notations
\begin{equation}
    \begin{split}
        \textcolor{Sepia}{(\textbf{w}^2)^{i}_{\ j} }\equiv&\braket{\textcolor{violet}{\tilde{\nu}^{c\dagger i}\tilde{\nu}^c_j}}+\braket{\textcolor{brown}{\varphi^{\dagger i}\varphi_j}} \,, \\
         \textcolor{Sepia}{(\textbf{w}^2_D)^{i}_{\ j} }\equiv&\delta^i_j\ (\braket{\textcolor{violet}{\tilde{\nu}^{c\dagger j}\tilde{\nu}^c_j}}+\braket{\textcolor{brown}{\varphi^{\dagger j}\varphi_j}}) \,, \quad\quad ({\rm no\ sum\ over}\ j)\, ,\\
\textcolor{Sepia}{\textbf{w}^2}\equiv&\textcolor{Sepia}{(\textbf{w}^2)^i_{\ i} } \,.
    \end{split}
\end{equation}
The mass terms for quarks and their vector-like counterparts then read as
    \begin{align}
        \mathcal{L}_{mq}&\sim \epsilon^{ijk}\left\{\textcolor{blue}{u^c_a}\textcolor{blue}{Q_b}\braket{\textcolor{OliveGreen}{H_{ui}}}\left[\delta^a_j\delta^b_k+\frac{1}{\Lambda^2}(\textcolor{Sepia}{(\textbf{w}^2)^{a}_{\ j}}\delta^b_k-\textcolor{Sepia}{(\textbf{w}^2)^{b}_{\ k}}\delta^a_j)\right]\right. \nonumber \\
       &
       + \textcolor{blue}{d^c_a}\textcolor{blue}{Q_b}\braket{\textcolor{OliveGreen}{H_{di}}}\left[\left(1+\frac{\braket{\textcolor{violet}{\tilde{\nu}^{c\dagger s}\tilde{\nu}^c_s}}}{\Lambda^2}\right)\delta^a_j\delta^b_k
       +\frac{1}{\Lambda^2}(\textcolor{Sepia}{(\textbf{w}^2)^{a}_{\ j}}\delta^b_k-\textcolor{Sepia}{(\textbf{w}^2)^{b}_{\ k}}\delta^a_j)\right] \nonumber \\
       &
       + D^c_a D_b\Big[\braket{\textcolor{brown}{\varphi_i}}\left(1+\frac{\textcolor{Sepia}{\textbf{w}^2}}{\Lambda^2}+\frac{\braket{\textcolor{brown}{\varphi^{\dagger s}}\textcolor{brown}{\varphi_s}}}{\Lambda^2}\right)\delta^a_j\delta^b_k+\braket{\textcolor{violet}{\tilde{\nu}^c_i}}\frac{\braket{\textcolor{violet}{\tilde{\nu}^{c\dagger s}}\textcolor{brown}{\varphi_s}}}{\Lambda^2}\delta^a_j\delta^b_k \nonumber \\
    &       
       +\frac{1}{\Lambda^2}\braket{\textcolor{brown}{\varphi_i}}(\textcolor{Sepia}{(\textbf{w}^2)^{a}_{\ j}}\delta^b_k-\textcolor{Sepia}{(\textbf{w}^2)^{b}_{\ k}}\delta^a_j)\Big] \label{eq:mq} \\
       & 
       + \textcolor{blue}{d^c_a} D_b\Big[\braket{\textcolor{violet}{\tilde{\nu}^c_i}}\left(1+\frac{\textcolor{Sepia}{\textbf{w}^2}}{\Lambda^2}+\frac{\braket{\textcolor{violet}{\tilde{\nu}^{c\dagger s}\tilde{\nu}^c_s}}}{\Lambda^2}\right)\delta^a_j\delta^b_k+\braket{\textcolor{brown}{\varphi_i}}\frac{\braket{\textcolor{brown}{\varphi^{\dagger s}}\textcolor{violet}{\tilde{\nu}^{c }_s}}}{\Lambda^2}\delta^a_j\delta^b_k  \nonumber \\
    &
       +\frac{1}{\Lambda^2}\braket{\textcolor{violet}{\tilde{\nu}^c_i}}(\textcolor{Sepia}{(\textbf{w}^2)^{a}_{\ j}}\delta^b_k-\textcolor{Sepia}{(\textbf{w}^2)^{b}_{\ k}}\delta^a_j)\Big]
       %\\
       %&
       \left.+ D^c_a\textcolor{blue}{Q_b}\braket{\textcolor{OliveGreen}{H_{di}}}\left[\frac{\braket{\textcolor{violet}{\tilde{\nu}^{c\dagger s}}\textcolor{brown}{\varphi_s}}}{\Lambda^2}\delta^a_j\delta^b_k\right]\right\}
       %\\
        %&
        + \mathrm{h.c.} + \mathcal{O}(1/\Lambda^4) \,, \nonumber 
    \end{align}
where dimensionless constants coming from loop factors, gauge couplings and CGC have been ignored, for simplicity.  The first line contains the up-type quark mass terms which are antisymmetric at the leading order and only contribute to two physical quark masses. The next-to-leading order corrections fill up the entire mass matrix generating a hierarchy between the lightest up quark and the heavier charm and top quarks. The second line contains the down-type quark masses which depend on up to three different Higgs VEVs and a new term that differs from the up quarks. This is enough to create a different mass structure and a viable Cabibbo–Kobayashi–Maskawa (CKM) mixing matrix similarly to the discussion in \cite{Morais:2020ypd}. The third line contains vector-like $\SU{3}{C}$-triplet masses generated at the compactification scale, which is desirable. The fourth line contains a large mixing between one of the extra vector-like states and the down-type quarks, meaning that the light down-type quark lies in a linear combination between them, further contributing to the CKM matrix. Finally, the fifth line contains an EW scale mixing between the vector-like and the down-type quarks. 

The leading contributions to the masses of all leptons, both chiral and vector-like, can then be expressed as
\begin{equation}
    \begin{split}
        \mathcal{L}_{ml}\sim&\epsilon^{ijk}\bigg\{\textcolor{violet}{\nu^c_a}\textcolor{blue}{L_b}\braket{\textcolor{OliveGreen}{H_{ui}}}\Big[\Big(1+\frac{1}{\Lambda^2}\textcolor{Sepia}{\textbf{w}^2}+\frac{1}{\Lambda^2}\braket{\textcolor{brown}{\varphi^{\dagger s}}\textcolor{brown}{\varphi_s}}\Big)\delta^a_j\delta^b_k+\frac{1}{\Lambda^2}\Big(\textcolor{Sepia}{(\textbf{w}^2)^{a}_{\ j}}\delta^b_k-\textcolor{Sepia}{(\textbf{w}^2)^{b}_{\ k}}\delta^a_j\Big)\Big]
        \\
        &
        +\textcolor{blue}{e^c_a}\textcolor{blue}{L_b}\braket{\textcolor{OliveGreen}{H_{di}}}\Big[\Big(1+\frac{1}{\Lambda^2}\textcolor{Sepia}{\textbf{w}^2}+\frac{1}{\Lambda^2}\braket{\textcolor{brown}{\varphi^{\dagger s}\varphi_s}}\Big)\delta^a_j\delta^b_k+\frac{1}{\Lambda^2}\Big(\textcolor{Sepia}{(\textbf{w}^2)^{a}_{\ j}}\delta^b_k-\textcolor{Sepia}{(\textbf{w}^2)^{b}_{\ k}}\delta^a_j\Big)\Big]
        \\
        &
        +\textcolor{OliveGreen}{\tilde{H}_{ua}}\textcolor{blue}{L_b}\Big[\Big(\braket{\textcolor{violet}{\tilde{\nu}^c_{i}}}\Big[1+\frac{1}{\Lambda^2}\braket{\textcolor{brown}{\varphi^{\dagger s}\varphi_s}}\Big]+\frac{1}{\Lambda^2}\braket{\textcolor{brown}{\phi_i}}\braket{\textcolor{brown}{\varphi^{\dagger s}}\textcolor{violet}{\tilde{\nu}^c_s}}\Big)\delta^a_j\delta^b_k+\frac{1}{\Lambda^2}\Big(\textcolor{Sepia}{(\textbf{w}^2)^{a}_{\ j}}\delta^b_k-\textcolor{Sepia}{(\textbf{w}^2)^{b}_{\ k}}\delta^a_j\Big)\Big]
        \\
        &
        +\textcolor{OliveGreen}{\tilde{H}_{da}}\textcolor{OliveGreen}{\tilde{H}_{ub}}\Big[\Big(\frac{1}{\Lambda^2}\braket{\textcolor{violet}{\tilde{\nu}^c_i}}\braket{\textcolor{violet}{\tilde{\nu}^{c\dagger s}}\textcolor{brown}{\varphi_s}}+\braket{\textcolor{brown}{\phi_i}}\Big[1+\frac{1}{\Lambda^2}\braket{\textcolor{violet}{\tilde{\nu}^{c\dagger s}\tilde{\nu}^c_s}}\Big]\Big)\delta^a_j\delta^b_k+\frac{1}{\Lambda^2}\Big(\textcolor{Sepia}{(\textbf{w}^2)^{a}_{\ j}}\delta^b_k-\textcolor{Sepia}{(\textbf{w}^2)^{b}_{\ k}}\delta^a_j\Big)\Big]
        \\
        &
        +\textcolor{brown}{\tilde{\varphi}_a}\textcolor{blue}{L_b}\braket{\textcolor{OliveGreen}{H_{ui}}}\braket{\textcolor{brown}{\varphi^{\dagger s}}\textcolor{violet}{\tilde{\nu}^{c }_s}}\frac{1}{\Lambda}\delta^a_j\delta^b_k
        +
        \textcolor{blue}{e^c_a}\textcolor{OliveGreen}{\tilde{H}_{db}}\braket{\textcolor{OliveGreen}{H_{di}}}\braket{\textcolor{violet}{\tilde{\nu}^{c\dagger s}}\textcolor{brown}{\varphi_s}}\frac{1}{\Lambda}\delta^a_j\delta^b_k
        \\ &
        +
        \textcolor{violet}{\nu^c_a}\textcolor{OliveGreen}{\tilde{H}_{da}}\braket{\textcolor{OliveGreen}{{H}_{ui}}}\braket{\textcolor{violet}{\tilde{\nu}^{c\dagger s}}\textcolor{brown}{\varphi_s}}\frac{1}{\Lambda}\delta^a_j\delta^b_k
        +
        \textcolor{violet}{\nu^c_a}\textcolor{OliveGreen}{\tilde{H}_{ua}}\braket{\textcolor{OliveGreen}{{H}_{di}}}\braket{\textcolor{violet}{\tilde{\nu}^{c\dagger s}}\textcolor{brown}{\varphi_s}}\frac{1}{\Lambda}\delta^a_j\delta^b_k
        \\
        &
        +\textcolor{brown}{\tilde{\varphi}_a}\textcolor{OliveGreen}{\tilde{H}_{ub}}\braket{\textcolor{OliveGreen}{{H}_{di}}}\Big[\Big(1+\frac{1}{\Lambda^2}\textcolor{Sepia}{\textbf{w}^2}+\frac{1}{\Lambda^2}\braket{\textcolor{brown}{\varphi^{\dagger s}\varphi_s}}\Big)\delta^a_j\delta^b_k+\frac{1}{\Lambda^2}\Big(\textcolor{Sepia}{(\textbf{w}^2)^{a}_{\ j}}\delta^b_k-\textcolor{Sepia}{(\textbf{w}^2)^{b}_{\ k}}\delta^a_j\Big)\Big]
          \\
        &
        +\textcolor{brown}{\tilde{\varphi}_a}\textcolor{OliveGreen}{\tilde{H}_{db}}\braket{\textcolor{OliveGreen}{{H}_{ui}}}\Big[\Big(1+\frac{1}{\Lambda^2}\textcolor{Sepia}{\textbf{w}^2}+\frac{1}{\Lambda^2}\braket{\textcolor{violet}{\tilde{\nu}^{c\dagger s}\tilde{\nu}^c_s}}\Big)\delta^a_j\delta^b_k+\frac{1}{\Lambda^2}\Big(\textcolor{Sepia}{(\textbf{w}^2)^{a}_{\ j}}\delta^b_k-\textcolor{Sepia}{(\textbf{w}^2)^{b}_{\ k}}\delta^a_j\Big)\Big]
        \bigg\}
        \\
        &
        + \mathrm{h.c.} + \mathcal{O}(1/\Lambda^6) \,,
    \end{split}
    \label{eq:ml}
\end{equation}
where dimensionless constants coming from loop factors, gauge couplings and CGC have been again ignored, for simplicity. The first line contains the Dirac mass terms for neutrinos, the second one provides the charged lepton mass terms whereas the third one includes R-parity violating Higgsino-lepton mixing, which can be rotated into the Higgsino (or vector-like lepton) masses. This means that the light SM-like charged leptons lie in a linear combination between them. The fourth line contains the Higgsino mass terms and the fifth one offers a mixing term between the neutrinos and the flavinos. Note that the latter will behave as three extra right-handed neutrinos. The remaining lines contain EW scale mixing between leptons and higgsinos.

As described in \cref{app:orb}, the rotational boundary conditions preserve a $\rm{U}(1)_\mathcal{R}$ related to simple SUSY. The Wilson line effective VEVs break the remaining SUSY, and hence also break $\rm{U}(1)_\mathcal{R}$ completely. In usual SUSY models a discrete symmetry $\mathbb{Z}_2^\mathcal{R}\subset \rm{U}(1)_\mathcal{R}$, denoted as R-parity, is typically preserved, which does not happen in the current model. The R-parity violating (RPV) terms $\sim \textcolor{OliveGreen}{\tilde{H}_{ua}}\textcolor{blue}{L_b}$ can be rotated away by an unitary transformation which then generates terms like $\sim y^{ijk}_{\rm RPV} \textcolor{blue}{Q_i L_j d^c_k}$, whose strongest constraint reads as $y^{111}_{\rm RPV}<0.001$ \cite{Dreiner:1997uz,Barbier:2004ez}. As this is proportional to the first-family neutrino Yukawa coupling, it is expected to be well below this constraint. There will be additional R-parity violating terms coming from $\sim \textcolor{blue}{\textbf{Q}_R}\braket{\textcolor{ForestGreen}{\Delta_R}}\textcolor{blue}{\textbf{Q}_R}\textcolor{blue}{\textbf{Q}_R}+\textcolor{blue}{\textbf{Q}_L}\braket{\textcolor{ForestGreen}{\Delta_L}}\textcolor{blue}{\textbf{Q}_L}\textcolor{blue}{\textbf{Q}_L}$, through $\textcolor{blue}{d}-D$ and $\textcolor{blue}{d^c}-D^c$ mixing. They will be smaller than the smallest quark Yukawa coupling which is enough to avoid the experimental constraints \cite{Dreiner:1997uz,Barbier:2004ez}. The terms that may generate proton decay will be studied in \cref{sec:prodec}.

%%%%%%%%%%%%%%%%%%%%%%%%%%%%%%%%%%%%%%%%%%%%%%%%%%%%%%%%%%%%%%%%
\subsection{Lepton-gaugino mixing: gaugino and Majorana masses}
\label{sec:gaugino-Majorana}
%%%%%%%%%%%%%%%%%%%%%%%%%%%%%%%%%%%%%%%%%%%%%%%%%%%%%%%%%%%%%%%%

The effective VEV $\braket{\textcolor{ForestGreen}{\textbf{L}}}$ generates a mixing between chiral fermions and gauginos through the second term in \cref{eq:ferma}. Provided that it is singlet under $\SU{3}{C}$, neither quarks nor gluons participate in such interaction terms as they are all of the form $\sim \braket{\textcolor{ForestGreen}{\textbf{L}}^\dagger}\textcolor{magenta}{\lambda_A}t_A \textcolor{ForestGreen}{\textbf{L}}$. The gauginos are then named as $\textcolor{magenta}{\lambda^{L,R,F}_{\tilde{a}}}$, with $\tilde{a}=1,...,8$ and the Gell-Mann matrices are denoted as $t_{\tilde{a}i}^{\ \ j}$. Thus, the leading terms that mix leptons and gauginos can be cast as
\begin{equation}
    \begin{split}
        \mathcal{L}_{\psi\lambda}\sim&\ \textcolor{magenta}{\lambda^{F}_{3,8}}\big[
        \textcolor{violet}{\nu^c_j}t_{(3,8)i}^{\ \ j}\braket{\textcolor{violet}{\tilde{\nu}^{c\dagger i}}}
        +\textcolor{brown}{\tilde{\varphi}_j}t_{(3,8)i}^{\ \ j}\braket{\textcolor{brown}{\varphi^{\dagger i}}}\big]\\
        &
        +\big[\textcolor{blue}{e^c_i}\braket{\textcolor{violet}{\tilde{\nu}^{c\dagger i}}}(\textcolor{magenta}{\lambda^{R}_{1}}-i\textcolor{magenta}{\lambda^{R}_{2}})+\textcolor{violet}{\nu^c_i}\braket{\textcolor{violet}{\tilde{\nu}^{c\dagger i}}}(\textcolor{magenta}{\lambda^{R}_{3}}+\textcolor{magenta}{\lambda^{R}_{8}}/\sqrt{3})-2\textcolor{brown}{\tilde{\varphi}_i}\braket{\textcolor{brown}{\varphi^{\dagger i}}}\textcolor{magenta}{\lambda^{R}_{8}}/\sqrt{3}\\
        &
        +\textcolor{brown}{\tilde{\varphi_i}}\braket{\textcolor{violet}{\tilde{\nu}^{c\dagger i}}}(\textcolor{magenta}{\lambda^{R}_{4}}-i\textcolor{magenta}{\lambda^{R}_{5}})+\textcolor{violet}{\nu^c_i}\braket{\textcolor{brown}{\varphi^{\dagger i}}}(\textcolor{magenta}{\lambda^{R}_{4}}+i\textcolor{magenta}{\lambda^{R}_{5}})+\textcolor{blue}{e^c_i}\braket{\textcolor{brown}{\varphi^{\dagger i}}}(\textcolor{magenta}{\lambda^{R}_{6}}+i\textcolor{magenta}{\lambda^{R}_{7}})
        \big]\\
        &
        +\big[\textcolor{OliveGreen}{\tilde{H}_{di}}\textcolor{magenta}{\lambda^L}\braket{\textcolor{violet}{\tilde{\nu}^{c\dagger i}}}+\textcolor{blue}{L_i}\textcolor{magenta}{\lambda^L}\braket{\textcolor{brown}{\varphi^{\dagger i}}}-2(\textcolor{violet}{\nu^c_i}\braket{\textcolor{violet}{\tilde{\nu}^{c\dagger i}}}+\textcolor{brown}{\tilde{\varphi}_i}\braket{\textcolor{brown}{\varphi^{\dagger i}}})\textcolor{magenta}{\lambda^L_8}/\sqrt{3}
        \big]
        \\
        & + {\rm h.c.} \,,
        \label{eq:lepgauginomix}
    \end{split}
\end{equation}
where $\textcolor{magenta}{\lambda^{L}}$ is the $\SU{2}{L}$ doublet defined as $\textcolor{magenta}{\lambda^{L}}=( \textcolor{magenta}{\lambda^{L}_{4}}-i\textcolor{magenta}{\lambda^{L}_{5}}, \ \textcolor{magenta}{\lambda^{L}_{6}}-i\textcolor{magenta}{\lambda^{L}_{7}})^\top$, and the dimensionless couplings coming from the extra CGC and gauge couplings have been omitted. This Lagrangian generates Majorana masses for gauginos, flavinos and right-handed neutrinos. They are allowed as $\braket{\textcolor{ForestGreen}{\textbf{L}}}$ breaks the $U(1)_\mathcal{R}$ symmetry as discussed in \cref{app:lag}.

From \cref{eq:lepgauginomix} one can obtain the gaugino masses that correspond to the masses of the heavy vector bosons associated with the broken gauge symmetry generators, $\textcolor{magenta}{\lambda_A} M^\dagger_A M_B \textcolor{magenta}{\lambda_B}=\textcolor{magenta}{\lambda_A\lambda_B}(\braket{\textcolor{ForestGreen}{\textbf{L}}^\dagger}t_A t_B\braket{\textcolor{ForestGreen}{\textbf{L}}})$, with the $A,B$ indices defined as $\lambda_A\equiv \lambda^{L,R,F}_{\tilde{a}}$. The vector boson and gaugino masses for a given broken generator are proportional to $\braket{\textcolor{brown}{\varphi^{\dagger i}\varphi_i}},\braket{\textcolor{violet}{\tilde{\nu}^{c\dagger i}}\textcolor{violet}{\tilde{\nu}^{c }_i}},\braket{\textcolor{brown}{{\varphi}^{\dagger  i}}\textcolor{violet}{\tilde{\nu}^{c }_i}},\braket{\textcolor{brown}{\varphi_i}}+\braket{\textcolor{violet}{\tilde{\nu}^{c }_i}}$, which are, in turn, given at the compactification scale. This is discussed in \cref{app:modlag} and detailed in \cref{eq:gauginomass}.

The leading SUSY breaking contributions to the non-diagonal gaugino mass terms from $M^\dagger_A M_B$ are of the order $\sim \braket{\mathcal{D}_A\mathcal{D}_B}$ and have the same mass structure as the SUSY preserving one, so they may be ignored. This comes from the fact that the same effective VEV that breaks the GUT symmetry is the one that breaks SUSY such that they are aligned. 

It can be seen from \cref{eq:lepgauginomix} that leptons and higgsinos mix with gauginos providing a contribution to their masses. More importantly, the mixing $\sim \braket{\textcolor{ForestGreen}{\textbf{L}}^\dagger}\textcolor{magenta}{\lambda_A}t_A \textcolor{ForestGreen}{\textbf{L}}$ of leptons with neutral gauginos (with Majorana mass $M_{AB}$) induces Majorana mass terms for leptons $\sim \textcolor{ForestGreen}{\textbf{L}}(t_A\braket{\textcolor{ForestGreen}{\textbf{L}}^\dagger})^T M_{AB}^{-1}\braket{\textcolor{ForestGreen}{\textbf{L}}^\dagger}t_B \textcolor{ForestGreen}{\textbf{L}}$ which can be decomposed as
\begin{equation}
\begin{split}
    \mathcal{L}_{mml}\sim &
    \frac{1}{\Lambda}\textcolor{violet}{\nu^c_i}\textcolor{violet}{\nu^c_j}\braket{\textcolor{violet}{\tilde{\nu}^{c\dagger i}\textcolor{violet}{\tilde{\nu}^{c\dagger j}}}}+\frac{1}{\Lambda}\textcolor{brown}{\tilde{\varphi}_i}\textcolor{brown}{\tilde{\varphi}_j}\braket{\textcolor{brown}{\tilde{\varphi}^{\dagger i}\textcolor{brown}{\tilde{\varphi}^{\dagger j}}}}
    +\frac{1}{\Lambda}\textcolor{violet}{\nu^c_i}\textcolor{brown}{\tilde{\varphi}_j}\braket{\textcolor{violet}{\tilde{\nu}^{c\dagger i}\textcolor{brown}{\tilde{\varphi}^{\dagger j}}}} \,,
    \end{split}
\end{equation}
where the extra CGC and gauge couplings have been omitted. Note that both right-handed neutrinos and flavinos get large Majorana masses providing a seesaw mechanism with three very light neutrinos and six heavy ones. However, as $\braket{\textcolor{violet}{\tilde{\nu}^c_1}}=0$, then $\textcolor{violet}{{\nu}^c_1}$ remains massless and corresponds to the goldstino. Therefore, this model offers five right-handed neutrinos with masses of the order of the compactification scale while the remaining one is an EW-scale sterile neutrino. The masses of the right-handed neutrinos are at the Wilson line scales which, as will be seen below, they are sufficiently large to generate small left-handed neutrino masses through the standard seesaw mechanism.

Finally, one can remark that the only fermions that obtain a mass through the terms $\sim \textcolor{ForestGreen}{\textbf{L}}\braket{\textcolor{ForestGreen}{\textbf{L}}^\dagger\textcolor{ForestGreen}{\textbf{L}}^\dagger} \textcolor{ForestGreen}{\textbf{L}}$ are the right-handed neutrinos and flavinos. There are no invariant mass terms $\sim \textcolor{ForestGreen}{\textbf{L}}\braket{\textcolor{ForestGreen}{\textbf{L}}^\dagger\textcolor{ForestGreen}{\textbf{L}}} \textcolor{ForestGreen}{\textbf{L}}$, therefore the leptons do not receive a large mass from the gaugino mixing.

%%%%%%%%%%%%%%%%%%%%%%%%%%%%%%%%%%%%%%
\subsection{Unbroken gauginos}
\label{sec:gauginos}
%%%%%%%%%%%%%%%%%%%%%%%%%%%%%%%%%%%%%%

In the previous section, it has been discussed how the gauginos, corresponding to broken gauge symmetry generators, get a mass. This mechanism leaves the gluino, winos and binos massless with eigenvectors,
\begin{equation}
    \textcolor{magenta}{\tilde{g}_{\tilde{a}}}=\textcolor{magenta}{\lambda^C_{\tilde{a}}} \,,\ \ \ \textcolor{magenta}{\tilde{W}_{1,2,3}}=\textcolor{magenta}{\lambda^L_{1,2,3}} \,, \ \ \ \textcolor{magenta}{\tilde{B}}=(\sqrt{3}\textcolor{magenta}{\lambda^L_8}-3\textcolor{magenta}{\lambda^R_3}+\sqrt{3}\textcolor{magenta}{\lambda^R_8})/\sqrt{15} \,.
\end{equation}
%%%%%%%%%%%%%%%%%%%%%%%%%%%%%%%%%%%%%%%%%%%%%%%%%%%%%%%%%%%%%%%%%%%
\begin{figure}[h]
\centering
\includegraphics[scale=0.3]{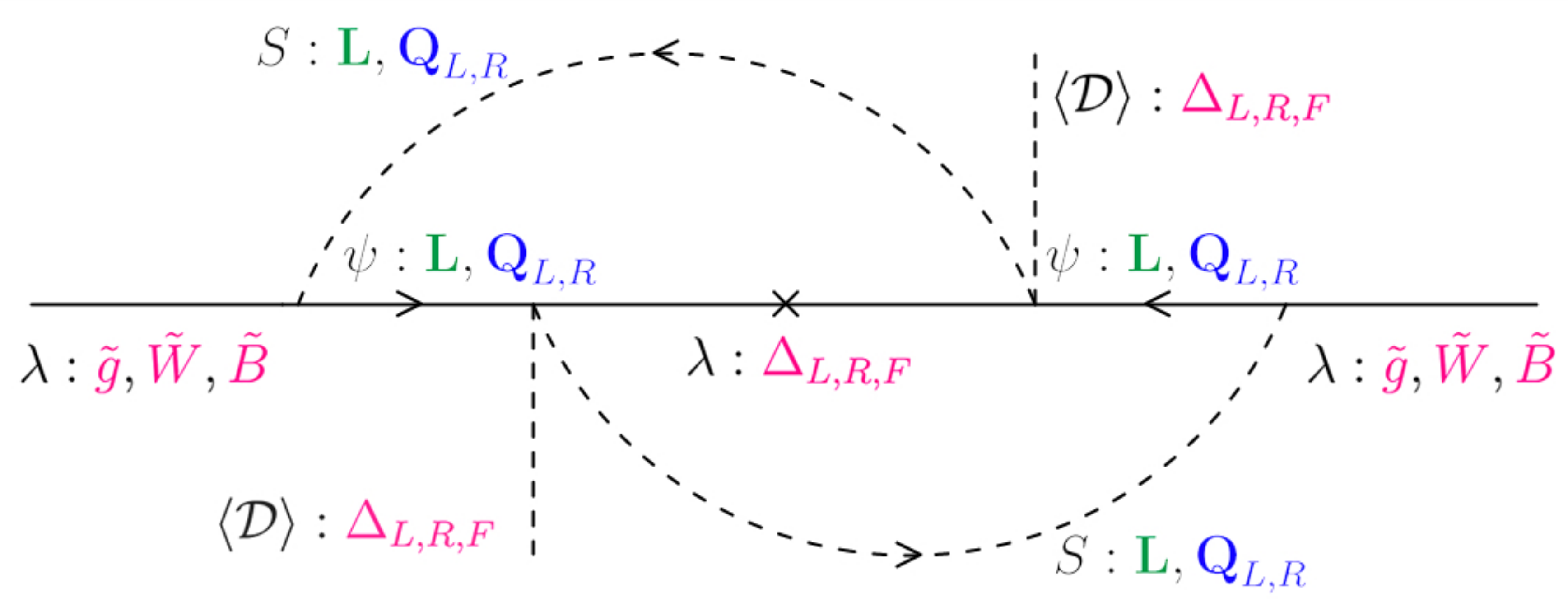}
\includegraphics[scale=0.3]{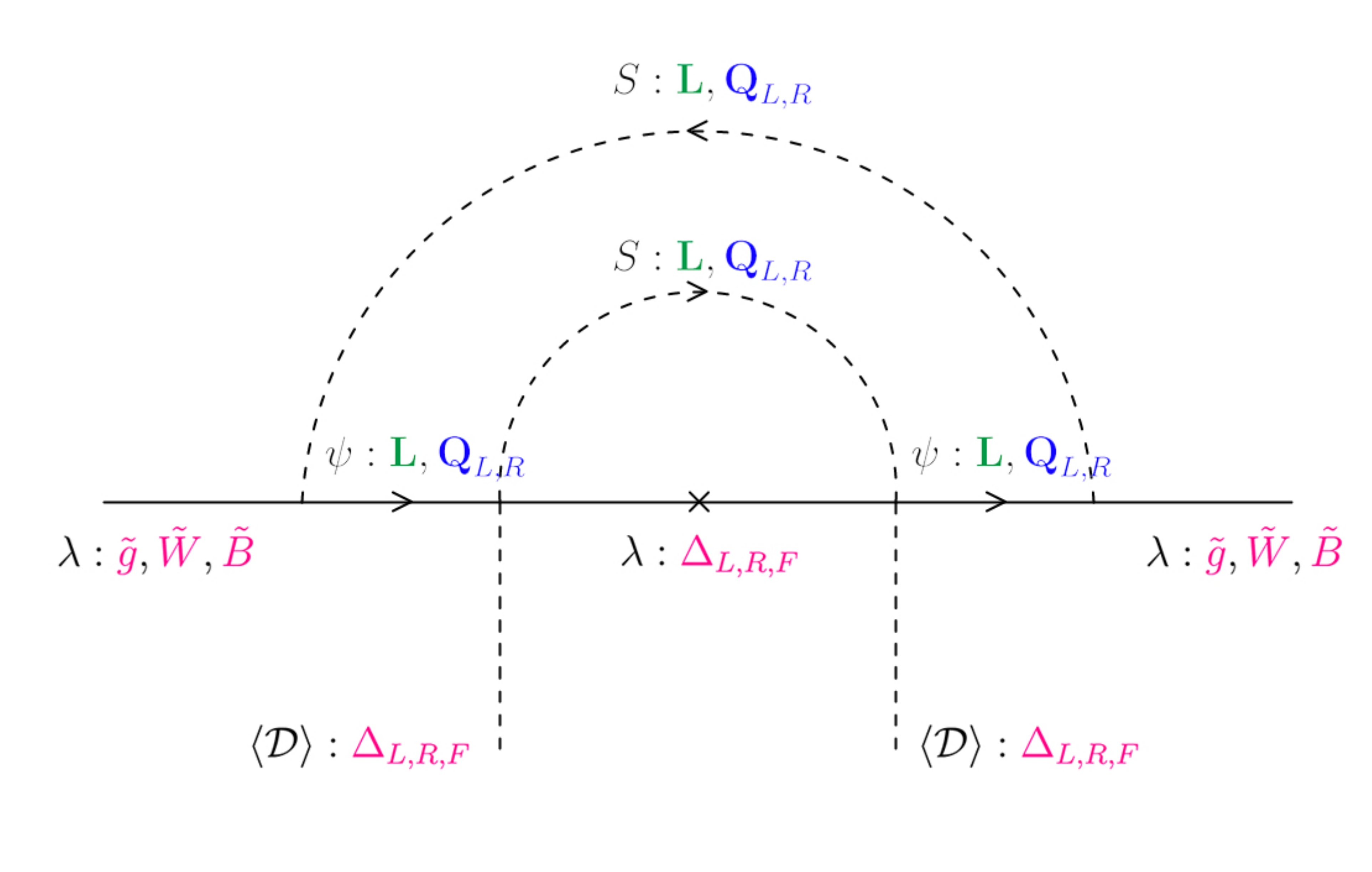}
\caption{The two-loop diagrams for SUSY-breaking gaugino masses. The $\braket{\mathcal{D}}$ does not have any SM charges so it must couple via a loop diagram.}
\label{fig:gauginomass}
\end{figure}
%%%%%%%%%%%%%%%%%%%%%%%%%%%%%%%%%%%%%%%%%%%%%%%%%%%%%%%%%%%%%%%%%%%

The unbroken gauginos, i.e. those corresponding to unbroken symmetry generators, on the other hand, cannot acquire a tree-level SUSY breaking mass. As $\braket{\mathcal{D}}$ preserves the SM, it does not communicate SUSY breaking directly. Instead, such a communication has to take place through the two loop diagrams shown in \cref{fig:gauginomass}, providing the mass terms for the remaining gauginos as
\begin{equation}
    \mathcal{L}_{mg}\sim\frac{\textcolor{Sepia}{\textbf{w}}^{5}}{\Lambda^4}\left(\textcolor{magenta}{\tilde{g}\tilde{g}}+\textcolor{magenta}{\tilde{W}\tilde{W}}+\textcolor{magenta}{\tilde{B}\tilde{B}}\right) \,,
\end{equation}
where, as usual, the effective couplings absorb the CGCs and loop factors, which are different for each gaugino as distinct fields run through the loops. This effectively defines the usual soft gaugino mass as
\begin{equation}
    m_{1/2}\sim g^4l^2\frac{\textcolor{Sepia}{\textbf{w}}^{5}}{\Lambda^4} \,,
    \label{eq:gluinomass}
\end{equation}
where $l^2$ is the two-loop suppression factor, and $g$ is the corresponding gauge coupling.

%%%%%%%%%%%%%%%%%%%%%%%%%%%%%%%%%%
\subsection{Scalar boson masses}
\label{sec:scalar-masses}
%%%%%%%%%%%%%%%%%%%%%%%%%%%%%%%%%%

The scalar mass terms come from the scalar potential which is obtained by considering \cref{eq:fulla} and substituting the truncated series from \cref{eq:fdsol} in
\begin{equation}
    \mathcal{L}_S= -\left.\frac{\partial^2 \mathcal{K}}{\partial\phi_j^\dagger\partial\phi_b}\right|_S
     \left.\frac{\partial \mathcal{W}}{\partial\phi_j}\right|_S\left.\frac{\partial \mathcal{W}}{\partial\phi_b^\dagger}\right|_S  
     -\frac{1}{2} \left[\left.\frac{\partial \mathcal{K}}{\partial\phi_i}\right|_S  t_A S_i+\left.\frac{\partial \mathcal{K}}{\partial\phi_i^\dagger}\right|_S  t_A S_i^\dagger\right]\left[\left.\frac{\partial \mathcal{K}}{\partial\phi_j}\right|_S  t_A S_j+\left.\frac{\partial \mathcal{K}}{\partial\phi_j^\dagger}\right|_S  t_A S_j^\dagger\right] \,.
     \label{eq:scamas}
\end{equation}
The effective VEVs $\braket{\textcolor{violet}{\tilde{\nu}^c}},\braket{\textcolor{brown}{\varphi}}$ enter the potential a priori, as they come from the Wilson lines, i.e.~the ED profiles of the fields, and are not determined by its minimization. The SUSY preserving mass terms, coming from the first term in \cref{eq:scamas}, read as
\begin{equation}
    \begin{split}
        \mathcal{L}_{mss}\sim&(\delta^a_i\delta^b_j-\delta^a_j\delta^b_i)\left\{
        \textcolor{blue}{\tilde{u}^c_a\tilde{u}^{c\dagger i}}\braket{\textcolor{OliveGreen}{H_{ub}H_{u}^{\dagger j}}}
        +(\textcolor{blue}{\tilde{Q}_a}\braket{\textcolor{OliveGreen}{H_{ub}}})(
        \textcolor{blue}{\tilde{Q}^{\dagger i}}\braket{\textcolor{OliveGreen}{H_{u}^{\dagger j}}})\right.\\
       &
        + \textcolor{blue}{\tilde{d}^c_a\tilde{d}^{c\dagger i}}(\braket{\textcolor{OliveGreen}{H_{db}H_{d}^{\dagger j}}}+\braket{\textcolor{violet}{\tilde{\nu}^c_b\tilde{\nu}^{c\dagger j}}})+(\textcolor{blue}{\tilde{Q}_a}\braket{\textcolor{OliveGreen}{H_{db}}})(\textcolor{blue}{\tilde{Q}^{\dagger i}}\braket{\textcolor{OliveGreen}{H_{d}^{\dagger j}}})
       \\
       &
        + \tilde{D}^c_a\tilde{D}^{c \dagger i}\braket{\textcolor{brown}{\varphi_b\varphi^{\dagger j}}} +\tilde{D}_a\tilde{D}^{\dagger i} (\braket{\textcolor{brown}{\varphi_b\varphi^{\dagger j}}}+\braket{\textcolor{violet}{\tilde{\nu}^c_b\tilde{\nu}^{c\dagger j}}})
        \\
        &
        +\textcolor{violet}{\tilde{\nu}^c_a\tilde{\nu}^{c\dagger i}}\braket{\textcolor{OliveGreen}{H_{ub}H^{\dagger j}_u}}+\textcolor{blue}{\tilde{e}^c_a \tilde{e}^{c\dagger i}}\braket{\textcolor{OliveGreen}{H_{db}H^{\dagger j}_d}}+\textcolor{brown}{\varphi_a\varphi^{\dagger i}}(\braket{\textcolor{OliveGreen}{H_{ub}H^{\dagger j}_u}}+\braket{\textcolor{OliveGreen}{H_{db}H^{\dagger j}_d}})
        \\
        &
        +( \textcolor{blue}{\tilde{L}_a}\braket{\textcolor{OliveGreen}{H_d^{\dagger j}}})( \textcolor{blue}{\tilde{L}^{\dagger b}}\braket{\textcolor{OliveGreen}{H_{d i}}})+( \textcolor{blue}{\tilde{L}_a}\braket{\textcolor{OliveGreen}{H_u^{\dagger j}}})( \textcolor{blue}{\tilde{L}^{\dagger b}}\braket{\textcolor{OliveGreen}{H_{u i}}})
        \\
        &
        +\textcolor{OliveGreen}{H_{ua}H_u^{\dagger i}}(\braket{\textcolor{brown}{\varphi_b\varphi^{\dagger j}}}+\braket{\textcolor{violet}{\tilde{\nu}^c_b\tilde{\nu}^{c\dagger j}}})+\textcolor{OliveGreen}{H_{da}H_d^{\dagger i}}\braket{\textcolor{brown}{\varphi_b\varphi^{\dagger j}}}
        \\
        & 
        \left.+
        (\textcolor{OliveGreen}{H_{ua}}\braket{\textcolor{OliveGreen}{H_{d b}}})(\textcolor{OliveGreen}{H_{u}^{\dagger i}}\braket{\textcolor{OliveGreen}{H_{d}^{\dagger j}}})+(\textcolor{OliveGreen}{H_{da}}\braket{\textcolor{OliveGreen}{H_{u b}}})(\textcolor{OliveGreen}{H_{d}^{\dagger i}}\braket{\textcolor{OliveGreen}{H_{u}^{\dagger j}}})
        \right\}\, .
    \end{split}
    \label{eq:mss}
\end{equation}

The second term in \cref{eq:scamas}, on the other hand, contains the SUSY breaking scalar masses through the $\braket{\mathcal{D}}$-term contribution which, in variance to the gaugino sector, can couple to every scalar at leading order,
\begin{equation}
\begin{split}
    \mathcal{L}_{msd}\sim&\ \Big(\textcolor{blue}{\tilde{Q}_a\tilde{Q}^{\dagger i}}
    +
    \textcolor{blue}{\tilde{u}^c_a\tilde{u}^{c\dagger i}}+\textcolor{blue}{\tilde{e}^{c}_a\tilde{e}^{c\dagger i}}\Big)\textcolor{Sepia}{(\textbf{w}^2_D)^{a}_{\ i}}
    +
    \Big(\textcolor{blue}{\tilde{d}^c_a\tilde{d}^{c\dagger i}} +\textcolor{blue}{\tilde{L}_{a}\tilde{L}^{\dagger i}}\Big)\Big[\textcolor{Sepia}{(\textbf{w}^2_D)^{a}_{\ i}}-\delta^a_i\braket{\textcolor{violet}{\tilde{\nu}^{c\dagger j}\tilde{\nu}^c_j}}\Big]
    \\
    &
    +\Big(\tilde{D}_a\tilde{D}^{\dagger i}+
    \textcolor{OliveGreen}{H_{ua}H_u^{\dagger i}}\Big)\Big[\textcolor{Sepia}{(\textbf{w}^2_D)^{a}_{\ i}}-\delta^a_i\textcolor{Sepia}{\textbf{w}^2}\Big]
    +
    \Big(\tilde{D}_a^c\tilde{D}^{c\dagger i} +\textcolor{OliveGreen}{H_{da}H_d^{\dagger i}}\Big)\Big[\textcolor{Sepia}{(\textbf{w}^2_D)^{a}_{\ i}}-\delta^a_i\braket{\textcolor{brown}{\varphi^{\dagger j}\varphi_j}}\Big]
    \\
    &
    +
    \textcolor{blue}{\tilde{d}^c_a}\tilde{D}^{c\dagger a}\braket{\textcolor{violet}{\tilde{\nu}^{c\dagger j}}\textcolor{brown}{\varphi_j}}
    +
    \tilde{D}^{c}_a\textcolor{blue}{\tilde{d}^{c\dagger a}}\braket{\textcolor{brown}{\varphi^{\dagger j}}\textcolor{violet}{\tilde{\nu}^{c}_j}}
    +
    \textcolor{blue}{\tilde{L}_a}\textcolor{OliveGreen}{H^{\dagger a}_d}\braket{\textcolor{violet}{\tilde{\nu}^{c\dagger j}}\textcolor{brown}{\varphi_j}}
    +
    \textcolor{OliveGreen}{H_{da}}\textcolor{blue}{\tilde{L}^{\dagger a}}\braket{\textcolor{brown}{\varphi^{\dagger j}}\textcolor{violet}{\tilde{\nu}^{c}_j}}
    \\ 
    &
    +
    \textcolor{violet}{\tilde{\nu}^{c}_a\tilde{\nu}^{c\dagger i}}\Big[\textcolor{Sepia}{(\textbf{w}^2_D)^{a}_{\ i}}+\delta^a_i\braket{\textcolor{violet}{\tilde{\nu}^{c\dagger j}\tilde{\nu}^c_j}}\Big]
    +
    \textcolor{brown}{\varphi_a\varphi^{\dagger i}}\Big[\textcolor{Sepia}{(\textbf{w}^2_D)^{a}_{\ i}}+\delta^a_i\braket{\textcolor{brown}{\varphi^{\dagger j}\varphi_j}}\Big]
    \\
    &
    +\textcolor{violet}{\tilde{\nu}^c_a}\textcolor{brown}{\varphi^{\dagger a}}\braket{\textcolor{violet}{\tilde{\nu}^{c\dagger j}}\textcolor{brown}{\varphi_j}}
    +\textcolor{brown}{\varphi_a}\textcolor{violet}{\tilde{\nu}^{c\dagger a}}\braket{\textcolor{brown}{\varphi^{\dagger j}}\textcolor{violet}{\tilde{\nu}^{c}_j}} \,,
\end{split}
\label{eq:scalar-masses}
\end{equation}
where the gauge couplings and CGCs have been omitted as usual. As we notice from this formula, all scalar mass terms receive $\braket{\mathcal{D}}$-term contributions at leading order. These effectively define the soft scalar masses to be $m_{0}^2\sim \textcolor{Sepia}{\textbf{w}^2}$. If there is a large hierarchy between the effective Wilson line VEVs, the stops and selectron $\textcolor{blue}{\tilde{u}^c},\textcolor{blue}{\tilde{Q}},\textcolor{blue}{\tilde{e}^c}$, can be significantly lighter, as they do not have a $\delta^a_i$ mass contribution proportional to the highest scale. Furthermore, the Higgs $\textcolor{OliveGreen}{H_{u1}}$ and $D_1$ fields do not acquire a contribution proportional to the highest scale, so they become lighter.

The existence of more than one neutral Higgs states may mediate Flavour Changing Neutral 
Currents (FCNCs) which are highly constrained as they have not been detected experimentally. Having one up-type and one down-type Higgs specific doublets does not generate FCNCs. In this model, the lightest neutral Higgs scalar lies mainly inside $\textcolor{OliveGreen}{H_{u1}}$ (although it is a linear combination of all of them). This is the only Higgs doublet that does not receive mass contributions from the highest scale and must have a mass of $125\ {\rm GeV}$. All the other Higgs doublets have a mass proportional to the highest scale and, as it will be described below, a large hierarchy between these scales will be assumed. As the highest scale will always be well above than $150 \ {\rm TeV}$, then the model avoids all FCNCs constraints, regardless of its couplings \cite{Branco:2011iw}.

%%%%%%%%%%%%%%%%%%%%%%%%%%%%%%%%
\section{Proton decay}
\label{sec:prodec}
%%%%%%%%%%%%%%%%%%%%%%%%%%%%%%%%

Like in many of the existing GUTs, in the considered model there will be high-energy processes that mix leptons with quarks, so there may be interactions that cause the proton to decay. It is straightforward to notice that the adjoint representations (for gauge and chiral superfields) $\textcolor{magenta}{\Delta_{\rm C,L,R,F}}$ preserve baryon and lepton number, therefore they do not mediate proton decay.

The first contribution to proton decay comes from the mirror representations. From the K\"ahler potential in \cref{eq:kah} one can extract the possible $QQQL$ terms mediated by KK modes
\begin{equation}
    \begin{split}
        \mathcal{K}'_p\sim&\tfrac{1}{\Lambda^3}\epsilon_{ijk}\ \textcolor{blue}{\textbf{Q}_{R\ (o-1)}^{\ i}} \textcolor{ForestGreen}{(\tilde{\Delta}_R)^j_{\ n}} \textcolor{blue}{\textbf{Q}_{R\ o}^{\ n}} \textcolor{blue}{\textbf{Q}_{R\ (o+1)}^{\ k}}
        +\tfrac{1}{\Lambda^3}\epsilon_{ijk}\ \textcolor{blue}{\textbf{Q}_{R\ (o-1)}^{\ i}}  \textcolor{blue}{\textbf{Q}_{R\ o}^{\ j}} \textcolor{ForestGreen}{(\tilde{\Delta}_F)^a_{\ (o+1)}}\textcolor{blue}{\textbf{Q}_{R\ a}^{\ k}}
        \\
        &
        +\tfrac{1}{\Lambda^3}\epsilon^{ijk}\ \textcolor{blue}{\textbf{Q}_{Li(o-1)}}  \textcolor{ForestGreen}{(\tilde{\Delta}_L)^a_{\ j}} \textcolor{blue}{\textbf{Q}_{Lao}} \textcolor{blue}{\textbf{Q}_{Lk(o+1)}}
        +\tfrac{1}{\Lambda^3}\epsilon^{ijk}\ \textcolor{blue}{\textbf{Q}_{Li(o-1)}}   \textcolor{blue}{\textbf{Q}_{Ljo}} \textcolor{ForestGreen}{(\tilde{\Delta}_F)^b_{\ (o+1)}}\textcolor{blue}{\textbf{Q}_{Lkb}}\, ,
        \label{eq:kpd1}
    \end{split}
\end{equation}
where we follow the index rules $o=0\to 3,\ o=4\to 1$ when summing over $o=1,2,3$. Since only one lepton is required, it follows from the structure of $\textcolor{ForestGreen}{\Delta}=\textcolor{ForestGreen}{\textbf{L}^\dagger \textbf{L}}$ that one of the $\textcolor{ForestGreen}{\textbf{L}}$ must acquire a VEV. If only one VEV is taken by $\textcolor{ForestGreen}{(\tilde{\Delta}_F)}$, only right-handed neutrinos or flavons can be introduced. If one takes a VEV and a lepton from $\textcolor{ForestGreen}{(\tilde{\Delta}_R)}$, all the proton decay related terms vanish due to antisymmetry. If one takes a VEV and a lepton from $\textcolor{ForestGreen}{(\tilde{\Delta}_L)}$, one of the terms that break the baryon number symmetry reads
\begin{equation}
    \begin{split}
        \mathcal{K}'_p\supset 
        &
        \tfrac{1}{\Lambda^5}\epsilon^{i3k}\ \textcolor{blue}{\textbf{Q}_{Li(o-1)}}  \textcolor{ForestGreen}{\braket{\textbf{L}^{\dagger 3n}_3}\textbf{L}^a_{\ 3n}} \textcolor{blue}{\textbf{Q}_{Lao}} \textcolor{blue}{\textbf{Q}_{Lk(o+1)}} \,.
        \label{eq:kpdd}
    \end{split}
\end{equation}
However, this and other similar terms do not contribute to proton decay since as they necessarily involve the third quark family (or vanish due to antisymmetry).
One may change the quark family by introducing an effective flavour adjoint (or, at low energies through CKM mixing, which will generate subleading corrections to the one studied)
\begin{equation}
    \begin{split}
        \mathcal{K}'_p\supset 
        &
        \tfrac{1}{\Lambda^5}\epsilon^{i3k}\epsilon^{opq}\ \textcolor{blue}{\textbf{Q}_{Lio}}  \textcolor{ForestGreen}{\braket{\textbf{L}^{\dagger 3n}_3}\textbf{L}^a_{\ 3n}} \textcolor{blue}{\textbf{Q}_{Lap}} \textcolor{ForestGreen}{(\tilde{\Delta}_F)^b_{\ q}}\textcolor{blue}{\textbf{Q}_{Lkb}} \\ 
        \supset &\tfrac{1}{\Lambda^5} \textcolor{blue}{u_1 d_2 d_1}\textcolor{violet}{\nu_n}\braket{\textcolor{ForestGreen}{(\tilde{\Delta}_F)^1_{\ 3}}}\braket{\textcolor{brown}{\varphi^{\dagger n}}} \,.
        \label{eq:kpdd}
    \end{split}
\end{equation}
Due to antisymmetry over generation indices, there can only be flavour-changing proton decays $p\to \bar{\nu}_{e,\mu}+ K^+$  \cite{Murayama:1994tc,deBoer:1994dg}. These processes are less constrained than the proton decay into pions \cite{Zyla:2020zbs}:
\begin{equation}
    \tau_{p\to K+L}=6.6\times 10^{-3}\ \tau_{p\to \pi+L} \,,
    \label{eq:prodecpil}
\end{equation}
with the most stringent bound on the proton lifetime \cite{Takhistov:2016eqm}
\begin{equation}
    \tau_{p\to \pi+L}<1.7\times 10^{34}\ {\rm yrs} \,.
\end{equation}

The proton decay process in \cref{eq:kpdd} involves two adjoints, where each one introduces a factor $\sim lg^4 \textcolor{Sepia}{\textbf{w}^2}/\Lambda^2$ as it follows from \cref{fig:del}. Therefore, the corresponding decay rate reads
\begin{equation}
    \Gamma_{p\to K+L}^{mf}\sim \frac{g^{18}l^4}{\Lambda^{10}}\braket{\textcolor{violet}{\tilde{\nu}^{c\dagger 1}\tilde{\nu}^c_3}\textcolor{brown}{\varphi^{\dagger}}}^2 m_p^5 \,,
    \label{eq:prodecmf}
\end{equation}
where $m_p$ is the proton mass, $l$ is the loop suppression factor and $g$ is the universal gauge coupling at the compactification scale $\Lambda$ where the mirror fermions are integrated out. As this is a decay into light neutrinos with negligible masses, there is no preference into which family such a decay occurs. So there is no explicit index in $\braket{\textcolor{brown}{\varphi^{\dagger}}}$, as it is meant to be the sum of the three possible decays.

The $\textbf{X},\overline{\textbf{X}}$ fields can potentially also mediate proton decay. These are the ones that usually mediate proton decay in SU(5)-based SUSY theories. The terms involving these massive gauge fields are
\begin{equation}
    \mathcal{K}_X\sim \textcolor{ForestGreen}{\textbf{L}} e^{-2\textbf{X}_V}\textcolor{blue}{\textbf{Q}_R^\dagger}+\textcolor{ForestGreen}{\textbf{L}} e^{-2\overline{\textbf{X}}_V}\textcolor{blue}{\textbf{Q}_L^\dagger}+\textcolor{blue}{\textbf{Q}_L}e^{-2\overline{\textbf{X}}_V}\textcolor{blue}{\textbf{Q}_R^\dagger}+ \mathrm{h.c.} \,,
\end{equation}
which could, in principle, trigger proton decay. However, these terms are not available for the physical quarks and leptons, i.e.~the zero modes. Each of the terms must be invariant under the orbifold charges, thus there can not be a term with two zero modes and one KK mode. Remarkably, the orbifold charges in \cref{tab:pe3} protect the broken gauge field terms from generating proton decay \cite{Kobakhidze:2001yk}.

One can also write the similar terms involving chiral superfields in the same representation
\begin{equation}
    \mathcal{W}_X\sim \textcolor{ForestGreen}{\textbf{L}} \textbf{X} \textcolor{red}{\overline{\textbf{Q}}_R}+\textcolor{ForestGreen}{\textbf{L}} \overline{\textbf{X}} \textcolor{red}{\overline{\textbf{Q}}_L}
    +\textcolor{orange}{\overline{\textbf{L}}} \overline{\textbf{X}} \textcolor{blue}{\textbf{Q}_R}+\textcolor{orange}{\overline{\textbf{L}}} \textbf{X} \textcolor{blue}{\textbf{Q}_L}
    +\textcolor{blue}{\textbf{Q}_L}\overline{\textbf{X}} \textcolor{red}{\overline{\textbf{Q}}_R}+\textcolor{blue}{\textbf{Q}_R}\textbf{X} \textcolor{red}{\overline{\textbf{Q}}_L}+\mathrm{h.c.} \,,
    \label{eq:pospd}
\end{equation}
which do not generate proton decay themselves but can, through a loop, provide the following terms
\begin{equation}
    \mathcal{K}_{pX}\sim \frac{g^6l^2}{\Lambda^2}\Big( \textcolor{ForestGreen}{\textbf{L}} \textcolor{blue}{\textbf{Q}_R^\dagger}\textcolor{ForestGreen}{\textbf{L}} \textcolor{blue}{\textbf{Q}_L^\dagger}+\textcolor{ForestGreen}{\textbf{L}} \textcolor{blue}{\textbf{Q}_R^\dagger}\textcolor{blue}{\textbf{Q}_L}
    \textcolor{blue}{\textbf{Q}_R^\dagger}\Big) + \mathrm{h.c.} \,.
\end{equation}
Here, the second term is a typical dimension-six SUSY proton decay term. By expanding over its indices
\begin{equation}
    \begin{split}
         \mathcal{K}_{pX}\sim& \frac{g^{6}l^2}{\Lambda^2}\textcolor{ForestGreen}{\textbf{L}^a_{\ b i}} \textcolor{blue}{\textbf{Q}_{Lak}}\textcolor{blue}{\textbf{Q}_{Re}^{\dagger\ \ i}}\braket{\textcolor{ForestGreen}{\tilde{\Delta}^m_{\ h}}}\textcolor{blue}{\textbf{Q}_{Rm}^{\dagger\ \ k}}\epsilon^{beh}+{\rm h.c.}\\
         \sim &\frac{g^{6}l^3}{\Lambda^4}\textcolor{blue}{L_i} \textcolor{blue}{Q_{k}}\textcolor{blue}{u^{c i\dagger}}\textcolor{blue}{d^{c k\dagger}}\braket{\textcolor{violet}{\tilde{\nu}^{c\dagger s}\tilde{\nu}^c_s}}+\mathrm{h.c.} \,,
         \label{eq:prodd}
    \end{split}
\end{equation}
one recognises that this process generates proton decay, with the rate
\begin{equation}
    \Gamma_{p\to K+L}^{X}\sim \frac{g^{12}l^4}{\Lambda^{8}}\braket{\textcolor{violet}{\tilde{\nu}^{c\dagger s}\tilde{\nu}^c_s}}^2 m_p^5 \,,
    \label{eq:prodecx}
\end{equation}
which is more important than the one mediated by only mirror fermions.

The vector-like triplets inside $\textcolor{blue}{\textbf{Q}_L},\textcolor{blue}{\textbf{Q}_R}$ may also mediate proton decay. Their couplings to the SM fermions is detailed in  \cref{eq:quyu}. For proton decay to happen, it needs to couple solely to quarks, i.e.~there must be a term $DQQ, D^c u^c d^c$, which is already present at leading order. These couplings thus give rise to the main proton decay channel \cite{Antusch:2014poa},
\begin{equation}
    \mathcal{W}_{pD}\sim g\Big[\textcolor{blue}{\textbf{Q}_{L12}}\textcolor{blue}{\textbf{Q}_{L21}}\textcolor{blue}{\textbf{Q}_{L33}}
    +
    \braket{\textcolor{ForestGreen}{\textbf{L}^3_{\ 31}}}\textcolor{blue}{\textbf{Q}_{R\ 2}^{\ 3}}\textcolor{blue}{\textbf{Q}_{L33}}
    +
    \textcolor{ForestGreen}{\textbf{L}^3_{\ 13}}\textcolor{blue}{\textbf{Q}_{R\ 2}^{\ 3}}\textcolor{blue}{\textbf{Q}_{L11}}\Big] \,,
\end{equation}
where $D,D^c$ fields get a large mass from the effective Wilson line VEV and hence can be integrated out, such that
\begin{equation}
\begin{split}
    \mathcal{W}_{pD}\sim &g\textcolor{blue}{\textbf{Q}_{L21}}\textcolor{blue}{\textbf{Q}_{L12}}\textcolor{blue}{\textbf{Q}_{L11}}\braket{\textcolor{ForestGreen}{\textbf{L}^{-1\ \ 31}_3}}\textcolor{ForestGreen}{\textbf{L}^1_{\ 33}}\\
    =&g\textcolor{blue}{u_1 d_2 d_1 \nu_3}/\braket{\textcolor{brown}{\varphi_1}} \,.
    \label{eq:prodecop}
    \end{split}
\end{equation}
This is the standard proton decay channel generated by extra vector-like triplets, with the decay rate
\begin{equation}
    \Gamma_{p\to K+L}^{D}\sim \frac{g^{2}}{\braket{\textcolor{brown}{\varphi_1}}^4} m_p^5 \,,
    \label{eq:dd}
\end{equation}
which is therefore the dominate decay channel. As the VEV $\braket{\textcolor{brown}{\varphi_1}}\sim 1/R_1$ defines the proton decay scale, it is assumed to be the largest. The other, smaller scales $\sim 1/R_{1,2}$ can also contribute through adding a family adjoint as shown in \cref{fig:del}. However, the presence of an extra factor makes such contributions subleading as long as $g^4 l \braket{\textcolor{brown}{\varphi_1}}/\braket{\textcolor{brown}{\varphi_{2,3}}} \ll 1$. So the extra contributions can be neglected if one considers a small enough gauge coupling, even with a large hierarchy between the effective VEVs.

It is important to note that there is a KK tower for each of the mediators which enhances the decay. The KK masses are much larger (they lack the Wilson line factor and the small gauge coupling in their masses) than those of the considered $D,\,D^c$ mediators.. In further numerical analysis below, it will be seen that the gauge coupling is sufficiently small so that these contributions can be indeed neglected. Therefore, for our purposes it is sufficient to study only the leading contribution to proton decay in \cref{eq:dd}.

The proton decay constraints set a limit on the GUT/compactification scale $\Lambda$ from \cref{eq:dd} by using $\braket{\textcolor{brown}{\varphi_i}}\sim 1/R_i$ and a phenomenologically viable condition $R_1\ll R_2 \ll R_3$. Therefore, we set the compactification scale to be the largest one $\Lambda\equiv 1/R_1$. In this case,
\begin{equation}
    \Lambda > 7.7\sqrt{g}\times 10^{15}\ {\rm GeV} \,.
    \label{eq:conspd}
\end{equation}
This constraint provides a way to bring the compactification scale down somewhat as long as the gauge coupling at that scale becomes sufficiently small as will be discussed below in \cref{sec:low-uni}. While $\Lambda$ still needs to be quite high compared to the EW scale, the scales defined by the other two radii can be much smaller as long as they fulfill the condition
\begin{equation}
    \Lambda < \frac{1}{l R_{2,3} g^4} \,.
    \label{eq:condL}
\end{equation}
This constrain comes from the fact that the other $D,D^c$ (with masses $\sim 1/R_{2,3}$) may also mediate proton decay by introducing an effective flavour adjoint as in \cref{fig:del}, which introduces a loop factor and four powers of the gauge coupling at the unification scale. Proton decay processes other than the dominant one considered above, such as $n$-$\bar{n}$ oscillations and the flavour-changing transitions, would be mediated by the same fields but appear at a higher order of up to $\mathcal{O}(1/\Lambda^{11})$, such that the corresponding experimental constraints can be easily satisfied \cite{Bandyopadhyay:2015fka}.

%%%%%%%%%%%%%%%%%%%%%%%%%%%%%%%%%%%%%%%%%%%%%%
 \section{Gauge couplings evolution}
 \label{sec:gcu}
%%%%%%%%%%%%%%%%%%%%%%%%%%%%%%%%%%%%%%%%%%%%%%

The gauge couplings of the SM that characterize the interaction strengths of the strong and EW forces evolve with the energy scale at a rate that depends on the field content. The presence of extra spacetime dimensions, and thus KK modes in the mass spectrum, has an impact on the running of the gauge couplings. In particular, such KK modes become important above a certain cut-off scale $\mu_\mathrm{KK}$ where they start playing a dominant role in the RG flow. Typically, the usual logarithmic running is modified acquiring a power-law structure which depends on the number of extra compact dimensions as discussed in Refs.~\cite{Dienes:1998vg,Dienes:1998vh}. 

In general terms, one-loop beta functions of a certain parameter $\mathcal{P}$ should be modified according to
 \begin{equation}
    \beta^{(1)}_\mathcal{P} \to \beta^{(1)}_{\mathcal{P}} + \left[S(\mu,\delta) -1\right] \tilde{\beta}^{(1)}_{\mathcal{P}} \,,
     \label{eq:beta}
 \end{equation}
where the usual logarithmic running is represented by $\beta^{(1)}_{\mathcal{P}}$ while the contribution of the KK modes is encoded in $\tilde{\beta}^{(1)}_{\mathcal{P}}$. The power-law running is governed by the $\left[S(\mu,\delta) -1\right]$ term with \cite{Dienes:1998vg,Dienes:1998vh,Liu:2012mea}
 \begin{equation}
     S(\mu,\delta) = X_\delta \left(\frac{\mu}{\mu_\mathrm{KK}}\right)^\delta \qquad \text{for} \qquad \mu \geq \mu_\mathrm{KK}\,,
     \label{eq:power}
 \end{equation}
and where $\delta$ represents the number of extra spacetime dimensions, and $\mu_\mathrm{KK}$ -- the scale at which the KK modes enter the particle spectrum. The $X_\delta$ factor is given by
 \begin{equation}
    X_\delta = \frac{2 \pi^{\delta/2}}{\delta \Gamma(\delta/2)} \,,
\end{equation}
with $\Gamma(x)$ being the Euler gamma function. 
 
A study of the evolution of the gauge couplings has been performed here, in order to show how the known interactions of the SM behave at high energy scales. At one-loop level it is convenient to recast the gauge couplings in terms of the inverse structure constants whose evolution in 4d is simply given by
 \begin{equation}
\alpha_\mathrm{A}^{-1}\left(\mu{}{}\right) = \alpha_0^{-1} - \frac{b_\mathrm{A}}{2 \pi} \log \frac{\mu}{\mu_0} \,.
\label{eq:RGE-log}
\end{equation}
The label $\mathrm{A}$ identifies a given gauge group with the gauge coupling $g_\mathrm{A}$ such that $\alpha_\mathrm{A} = g_\mathrm{A}^2/(4 \pi)$, while $\alpha_0^{-1}$ denotes the value of the inverse structure constant at the initial energy-scale $\mu_0$. Once the mass threshold scale $\mu_\mathrm{KK}$ (where the KK modes become relevant) is reached the structure constants of a theory with $\delta$ EDs evolve according to the power-law
\begin{equation}
    \alpha_\mathrm{A}^{-1}\left(\mu{}{}\right) = \alpha_0^{-1} - \frac{b_\mathrm{A}}{2 \pi} \log\frac{\mu}{\mu_0}
    +\frac{\tilde{b}_\mathrm{A}}{2 \pi} \log \frac{\mu}{\mu_\mathrm{KK}} - \frac{\tilde{b}_\mathrm{A} X_\delta}{2 \pi \delta} \left[ \left( \frac{\mu}{\mu_\mathrm{KK}} \right)^\delta - 1 \right]\,.
  \label{eq:RGE-pow}
\end{equation}
The value of the $b_\mathrm{A}$ and $\tilde{b}_\mathrm{A}$ coefficients will determine how fast a given gauge coupling evolves between any two scales. These coefficients depend on the field content decomposed into 4d Lorentz representations as well as their corresponding gauge representations. For non-Abelian gauge groups they are both given by
\begin{equation}
    b_\mathrm{A},\tilde{b}_\mathrm{A} = -\frac{11}{3} C_2\left(G\right) + \frac{4}{3} \kappa_\mathrm{F} T\left(F\right) + \frac{1}{3} \kappa_\mathrm{S} T\left(S\right)\,,
    \label{eq:SUrun}
\end{equation}
where $\kappa_\mathrm{F} = \tfrac{1}{2}$ for Weyl fermions, $\kappa_\mathrm{S} = 1\; (\tfrac12)$ for complex (real) scalars, $C_2\left(G\right)$ is a group Casimir in the adjoint representation while $T\left(F\right)$ and $T\left(S\right)$ are the Dynkin indices for fermions and scalars, respectively. Similarly, for the case of $\U{}$ symmetries the beta-function coefficients read as
\begin{equation}
    b^\prime_\mathrm{A},\tilde{b}^\prime_\mathrm{A} = -\frac{11}{3} \sum_v \left(\frac{Q_v}{2}\right)^2 +\frac{4}{3} \kappa_\mathrm{F} \sum_f \left(\frac{Q_f}{2}\right)^2 + \frac{1}{3} \kappa_\mathrm{S} \sum_s \left(\frac{Q_s}{2}\right)^2 \,,
    \label{eq:U1run}
\end{equation}
with $Q_v$, $Q_f$ and $Q_s$ being the Abelian charges of the vector, fermion and scalar degrees of freedom of the theory. 

The hierarchy between the radii of the EDs is arbitrary, and in this case it will be assumed
\begin{eqnarray} 
    R_1\ll R_2<R_3  ,
\end{eqnarray}
that would match the family hierarchy structure. 
As there are different radii from the extra dimensions, the result of the sum of the infinite series is different for each radius and the actual gauge coupling running is described by \cite{Dienes:1998vg,Dienes:1998vh}
\begin{equation}
       \alpha_i^{-1}(\Lambda) ~=~ \alpha_i^{-1}(\mu_0) ~-~
            {b_i-\tilde b_i\over 2\pi}\,\ln{\Lambda\over \mu_0} 
          ~-~ {\tilde b_i\over 4\pi}\,
             \int_{r\Lambda^{-2}}^{r\mu_0^{-2}} {dt\over t} \,
      \vartheta_3\left( {it\over \pi R^2_1} \right) 
     ^2 \vartheta_3\left( {it\over \pi R^2_2} \right) 
     ^2 \vartheta_3\left( {it\over \pi R^2_3} \right) 
     ^2~,
\label{eq:KKresult}
\end{equation}
where $\vartheta_3$ denotes a Jacobi theta function. In the regime $\Lambda R_i\gg 1$, this function can be approximated as
\begin{equation}
\vartheta_3\left( {it\over \pi R^2_i} \right) \approx R_i\sqrt{\frac{\pi}{t}}.
\end{equation}
Using this approximation \cref{eq:RGE-pow} can be recovered with the scale
\begin{equation}
    \mu_\mathrm{KK}=(R_1 R_2 R_3)^{-1/3},
    \label{eq:mukk}
\end{equation}
which defines the scale where the KK modes start acting as a power law running. These stop acting at the scale $1/R_1$, as above that scale, the theory is 10 dimensional.

Therefore there are 6 different physical scales given as follows:
\begin{itemize}
     \item $\Lambda_1=1/R_1$ is defined by the smallest radius $R_1$ related to the unification scale. Above it, the theory is 10 dimensional.
     \item $\Lambda_2=g/R_1$ is defined as the largest Wilson line scale where some of the fields obtain a mass.
     \item $\Lambda_3=(R_1 R_2 R_3)^{-1/3}$ is defined by the scale where all KK effects wear out. Below this scale there is no power law running
     \item $\Lambda_4=g/R_2$ defines the second Wilson line scale. 
     \item $\Lambda_5=g/R_3$ defines the third Wilson line scale.
     \item $\Lambda_6$ defines the EW scale where the Higgs mass is considered to run negative thus triggering the EWSB radiatively (see below).
 \end{itemize}
 
It is important to emphasise that the symmetry breaking happens in a single step as $\E{8}\to \SU{3}{C}\times\SU{2}{W}\times \U{Y}$, through the 10d-to-4d orbifold compactification, with different fields having different masses defined by the different $\Lambda_{1...6}$ scales. While it may seem that at certain energy scales there is an intermediate unbroken anomalous gauge symmetry, in fact, all anomalous symmetry generators are explicitly broken by the underlying space-time topology (i.e.~orbifolding) and the Wilson lines at {\it any} energy scale. Therefore, there are only two anomaly-free gauge symmetries that should be considered, one emerges in the UV and the other one -- in the IR, with different field content. The first one is the $\E{8}$ symmetry restored on the flat 10d background but is explicitly broken by orbifolding and hence can only be considered approximate in the UV regime. The second one is the SM symmetry $\SU{3}{C}\times\SU{2}{W}\times \U{Y}$ expected to the broken radiatively via the conventional Higgs mechanism. The detailed effective field content at each energy scale is described in \cref{app:ensca}.

%%%%%%%%%%%%%%%%%%%%%%%%%%%%%%%%%%%%%%%%%%%%%%%%%%
\subsection{Origin of low-scale Unification}
\label{sec:low-uni}
%%%%%%%%%%%%%%%%%%%%%%%%%%%%%%%%%%%%%%%%%%%%%%%%%%

This section presents a study of the unification of gauge couplings in the model. The analysis follows a purely agnostic approach without assuming any prior unification scale. The aim is to derive it from the theoretical constraints and then confront it with the proton decay bound and other phenomenological limits. The values of all the coefficients used in this section are shown in \cref{app:ensca}.

The SM gauge couplings are expressed as $g_1 \equiv g_\mathrm{Y}$, $g_2 \equiv g_\mathrm{L}$ and $g_3 \equiv g_\mathrm{C}$ at the EW scale in terms of the approximate $\E{8}$ gauge coupling $g_8$ and the intermediate symmetry breaking scales as
\begin{equation}
    \begin{aligned}
    \alpha_{i}^{-1}(m_{Z})  = & \alpha_{8}^{-1} +\frac{b_{i}^{\rm V} - \tilde{b}_{i}^{\rm V}}{2\pi}\log\left(\frac{\Lambda_{1}}{\Lambda_{2}}\right)+\frac{\tilde{b}_{i}^{\rm V}\pi^{2}}{72}\left[10^{6\log\left(\frac{\Lambda_{1}}{\Lambda_{2}}\right)}-1\right]
    \\
    &+\frac{b_{i}^{\rm IV} - \tilde{b}_{i}^{\rm IV}}{2\pi}\log\left(\frac{\Lambda_{2}}{\Lambda_{3}}\right)+\frac{\tilde{b}_{i}^{\rm IV}\pi^{2}}{72}\left[10^{6\log\left(\frac{\Lambda_{2}}{\Lambda_{3}}\right)}-1\right]
    \\
    &+\frac{b_{i}^{\rm III}}{2\pi}\log\left(\frac{\Lambda_{3}}{\Lambda_{4}}\right)
    +\frac{b_{i}^{\rm II}}{2\pi}\log\left(\frac{\Lambda_{4}}{\Lambda_{5}}\right)
    +\frac{b_{i}^{\rm I}}{2\pi}\log\left(\frac{\Lambda_{5}}{m_{Z}}\right)
    \,,
    \end{aligned}
    \label{eq:gRGEs}
\end{equation}
with $m_Z = \Lambda_6$ being the SM $Z$-boson mass scale where one defines the input values of the inverse structure constants and $i = 1,2,3$. Notice that the coefficients $\tilde{b}_i^\mathrm{V}$ have all the same sign which means that as the renormalization scale increases, the three inverse structure constants asymptotically converge to $\alpha^{-1}_8$. The concept of asymptotic unification was recently discussed in \cite{Cacciapaglia:2020qky} for an $\SU{5}{}$ theory in five space-time dimensions. \cref{eq:gRGEs} provides a second example of this same concept but for the case of a full $\mathrm{E}_8$ unification in ten dimensions.

\begin{figure}[]
    \centering
    \hspace{-1.2cm}
    \includegraphics[width=0.48\textwidth]{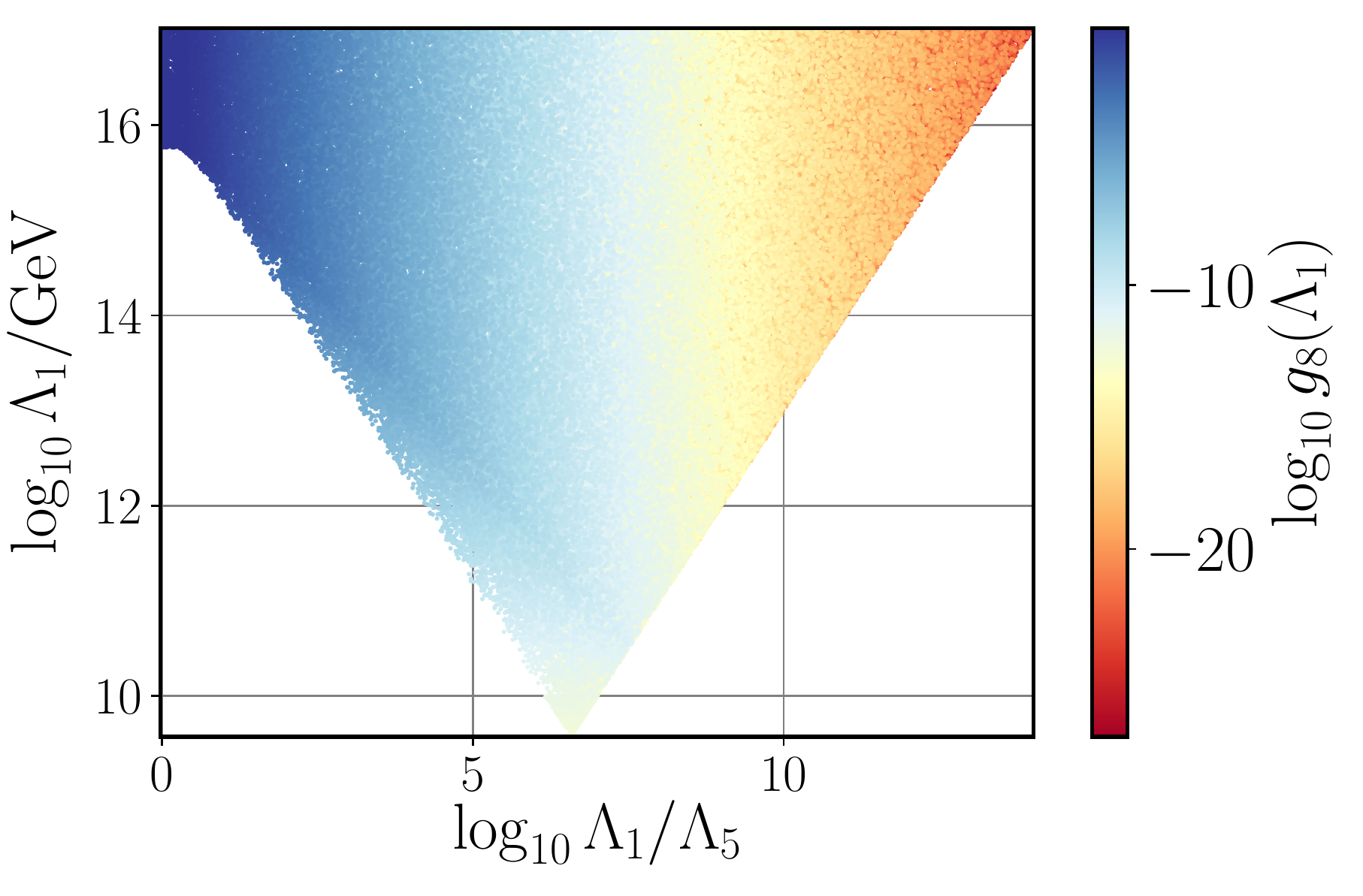}
    \includegraphics[width=0.48\textwidth]{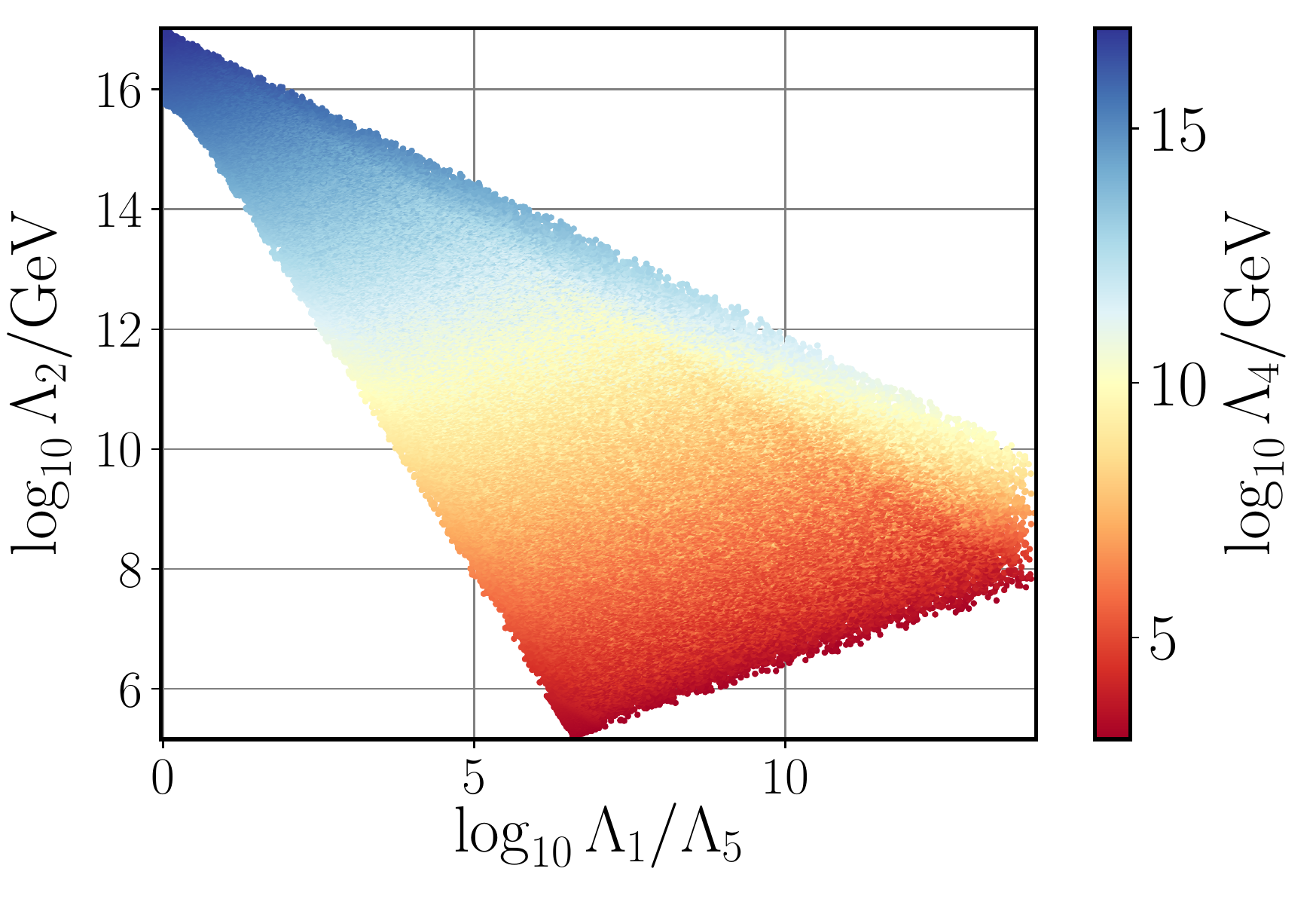}
    \\
    \includegraphics[width=0.48\textwidth]{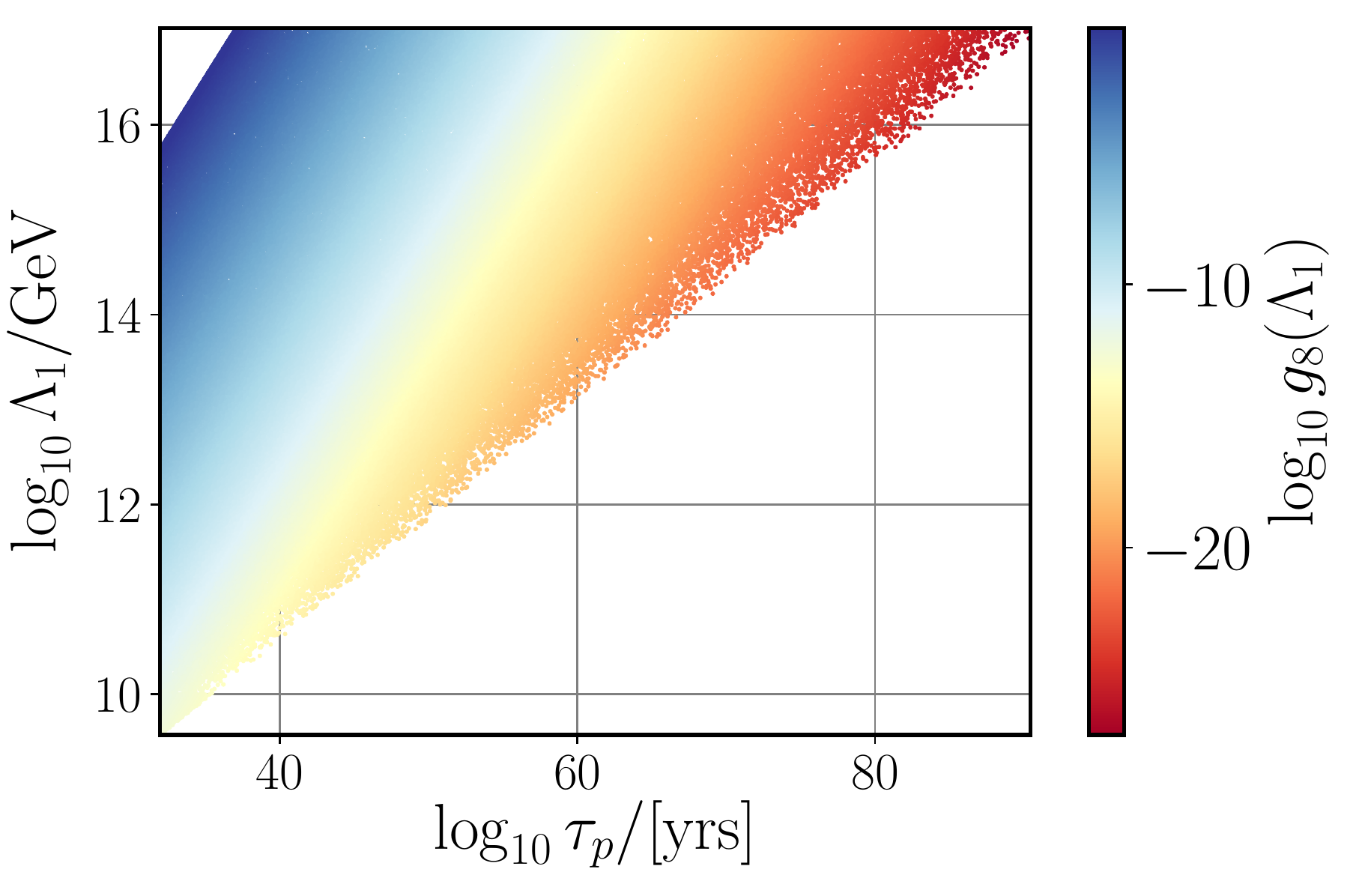}
	%\
	\caption{Scenarios consistent with asymptotic unification. The top-left panel depicts the relation between the Unification, $\Lambda_1$, and the lowest Wilson-line, $\Lambda_5$, scales. The colour gradation represents the value of the asymptotic $\mathrm{E}_8$ gauge coupling. The top-right panel showcases the relation between the RG scales involved in the gauge coupling's running. On the bottom panel we show the relation between the proton lifetime in years, the $\E{8}$ scale and the asymptotic $g_8$ coupling.}
	\label{fig:Scatter-E8-g8-g6}
\end{figure}

Using~\cref{eq:gRGEs} one can numerically solve it with respect to the unification scale as well as the values of the asymptotic unified coupling $g_8(\Lambda_1)$. Considering the scale $\mu_\mathrm{KK}$ defined in \cref{eq:mukk} above which the KK modes contribute to the running, a numerical routine written in \texttt{Python} engineered to find the roots of non-linear systems was developed and processed at the \texttt{blafis}\footnote{Technical details can be found at the Gr@v's website \cite{blafis}.} and \texttt{ARGUS} computer clusters as part of the overall computing infrastructure at the University of Aveiro. Considering a scenario with compressed $\Lambda_2 = \Lambda_3$, one has set the unification scale to randomly vary in the range $10^{6} < \Lambda_1/\mathrm{GeV} < 10^{17}$ while the remaining scales can take any value in this interval provided that $\Lambda_1 > \Lambda_2 = \Lambda_3 > \Lambda_4 > \Lambda_5 > \Lambda_6 = m_Z$. The results obtained that are consistent with proton decay life-time through \cref{eq:conspd} and that obey \cref{eq:condL} are shown in \cref{fig:Scatter-E8-g8-g6}.
%%%%%%%
%%%%%%%

It is rather relevant to highlight that, from the top-left panel in \cref{fig:Scatter-E8-g8-g6}, consistent scenarios in the model that obey all theoretical constraints allow a subset of solutions where the unification scale can be as low as, approximately, 7-orders of magnitude lighter than that of conventional GUT scenarios. This happens when there is an hierarchy of approximately six orders of magnitude between $\Lambda_1$ and $\Lambda_5$, featuring a tiny gauge coupling smaller than $10^{-10}$. In fact, independently of the scales involved in the running, the considered model always features asymptotic safety. It is also particularly significant to mention that no preferred GUT scale in the calculation was imposed. Therefore, all points obtained unveil the space of viable solutions that the model can offer.
In essence, the unification of families, matter and forces under $\E{8}$ can be achieved at scales as low as $1000~\mathrm{PeV}$ and the model can be classified as a \textit{low to intermediate-scale Grand Unified Theory} potentially falsifiable (at least partially) at the next generations of particle colliders. Furthermore, it has been verified that all such points pass one of the most stringent constraints known for GUT models, the proton lifetime. This can be seen in the bottom panel of \cref{fig:Scatter-E8-g8-g6} where the prediction, using \cref{eq:dd}, is that the proton must decay to kaons with a lifetime ranging from $\tau_{p\to K+L} \sim \mathrm{O}(10^{32})~\mathrm{yrs}$ all the way up to $\tau_{p\to K+L} \sim \mathcal{O}(10^{90})~\mathrm{yrs}$. The sizes of all scales involved in the running are represented in the top-right panel, where one can notice that the edge of the parameter space that allows the lowest unification scales requires relatively compressed $\Lambda_2$, $\Lambda_4$ and $\Lambda_5$ but a maximal hierarchy of about six orders of magnitude between $\Lambda_1$ and $\Lambda_5$.

Once again, no prior constraint regarding proton decay has been imposed in the analysis and it is rather tantalizing to note that the lowest unification scale value found, denoted as scenario (a) in \cref{tab:vals}, corresponds to a proton lifetime marginally above the current experimental lower bound $\tau_{p\to K+L}>1.1\times 10^{32}\ {\rm yrs}$, thus falsifiable in the near future.

The proton decay calculations above are performed at the messenger mass scale, which is $\Lambda_1$ in our scheme. The composite dimension-6 operator responsible for proton decay would have to run down to a low-energy scale where the experimental measurements are done. Since such a running includes that of Yukawa couplings \cite{Celis:2017hod}, the corresponding analysis lies well beyond the goals of this work. However, the RG evolution of the proton-decay Wilson coefficients is proportional to the product of the gauge couplings and the Wilson coefficient itself~\cite{Nagata:2013sba}, which is extremely small to start with. The precise calculation of this running may restrict the allowed unification scale but would not rule out the considered model.

The power-law behaviour of the RG equations (RGEs) is indeed responsible for the possibility of consistently achieving low-scale unification solutions. Similar arguments were previously noted as e.g.~in (\cite{Dienes:1998vg,Dienes:1998vh}). However, the model presented in this work is, to the best of the authors' knowledge, the first low to intermediate-scale $\E{8}$-GUT realization that incorporates the SM gauge interactions and the family replication observed in nature, both within the same framework, and that is consistent with proton decay limits.

\begin{figure}[]
    \subfloat[]{{\hspace{-1cm}\includegraphics[width=0.48\textwidth]{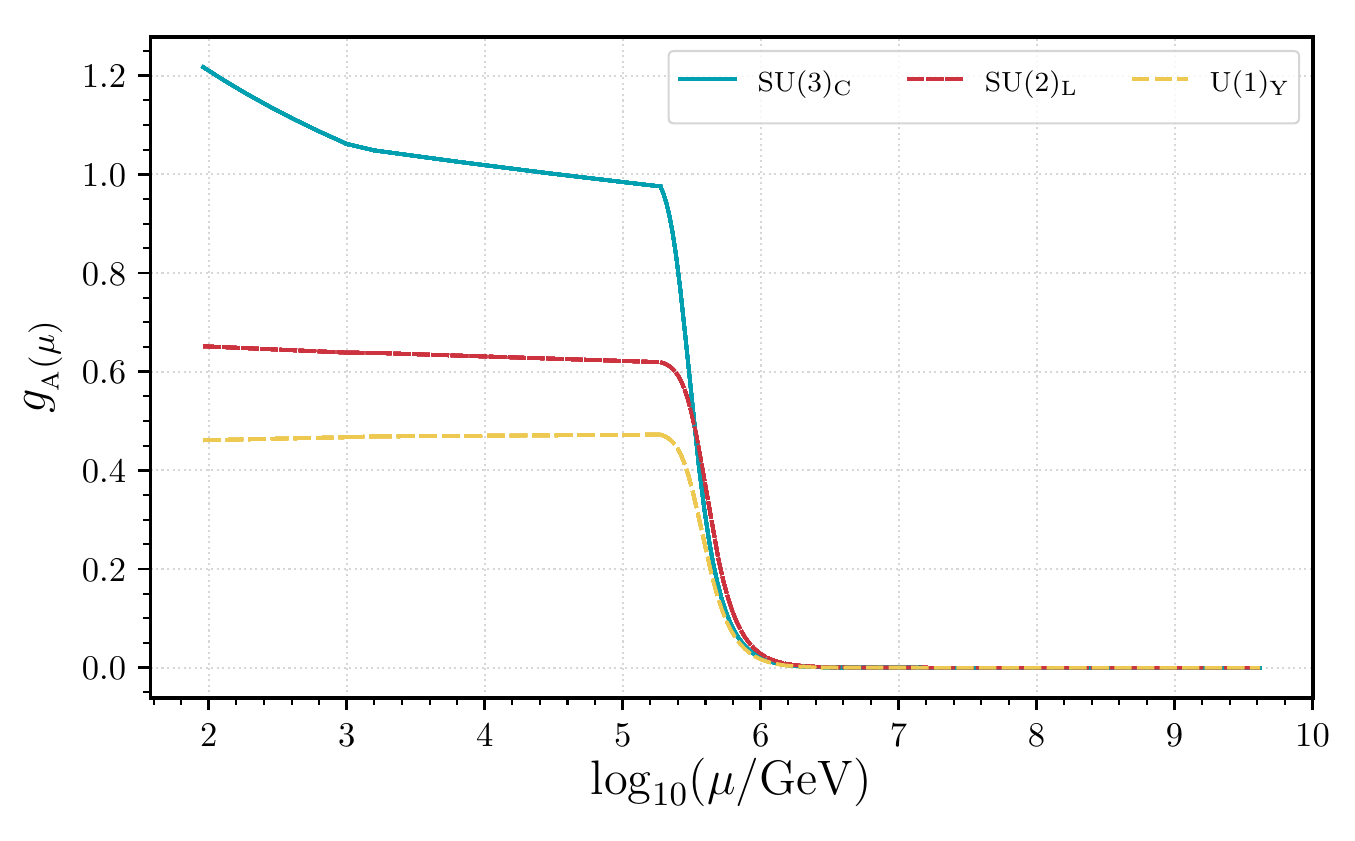} }} 
	\subfloat[]{{\includegraphics[width=0.48\textwidth]{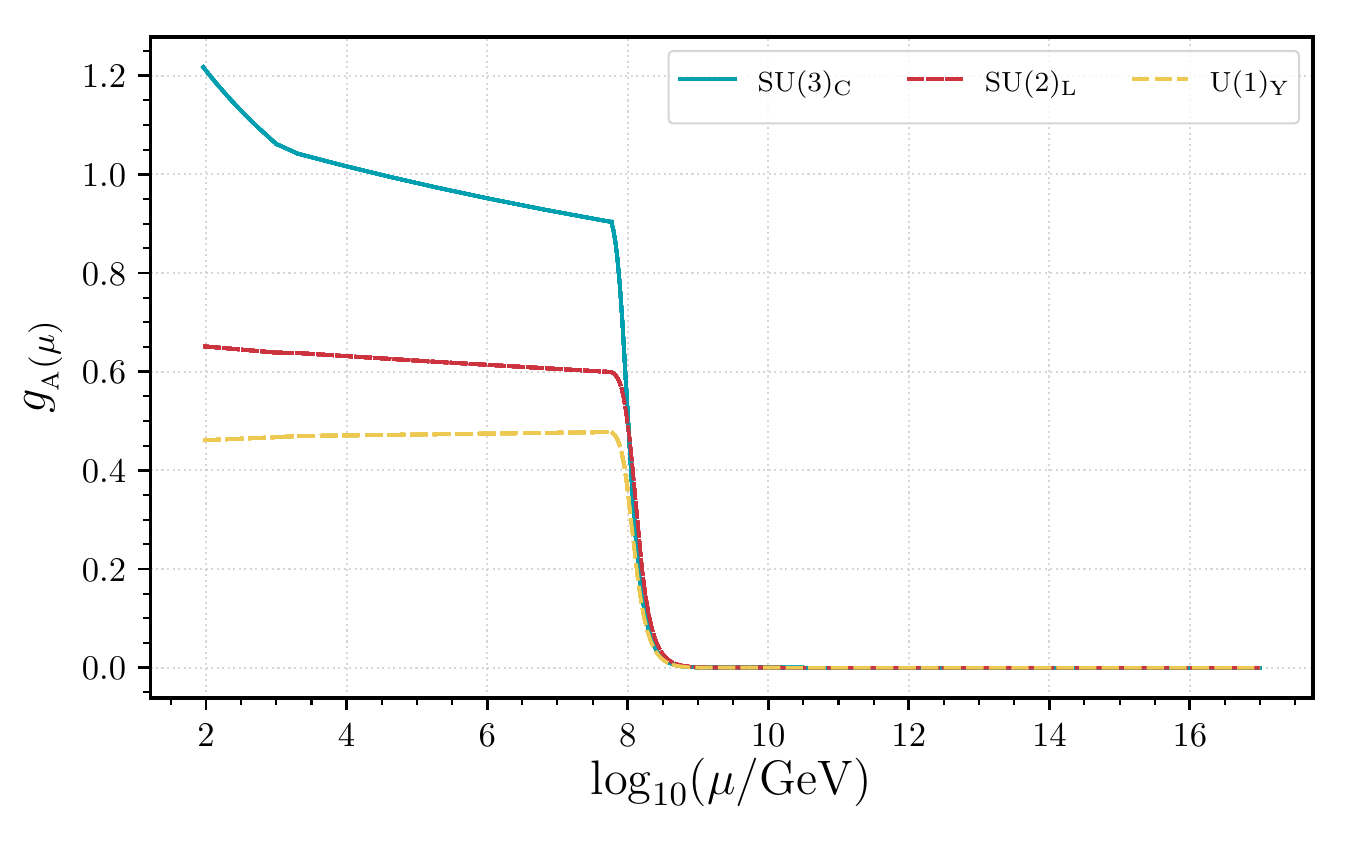} }}
	\\
	\subfloat[]{{\hspace{-1cm}\includegraphics[width=0.48\textwidth]{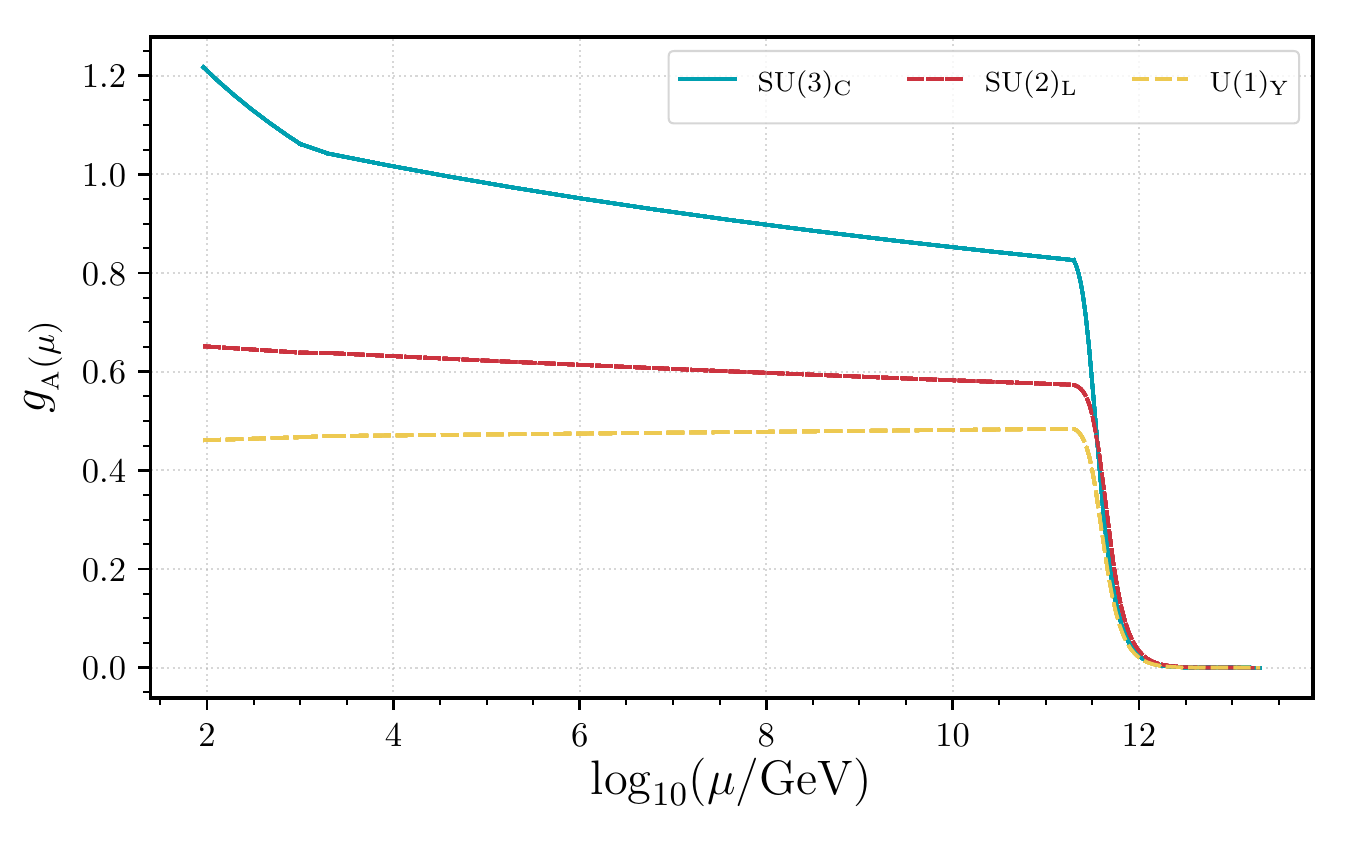} }} 
	\subfloat[]{{\includegraphics[width=0.48\textwidth]{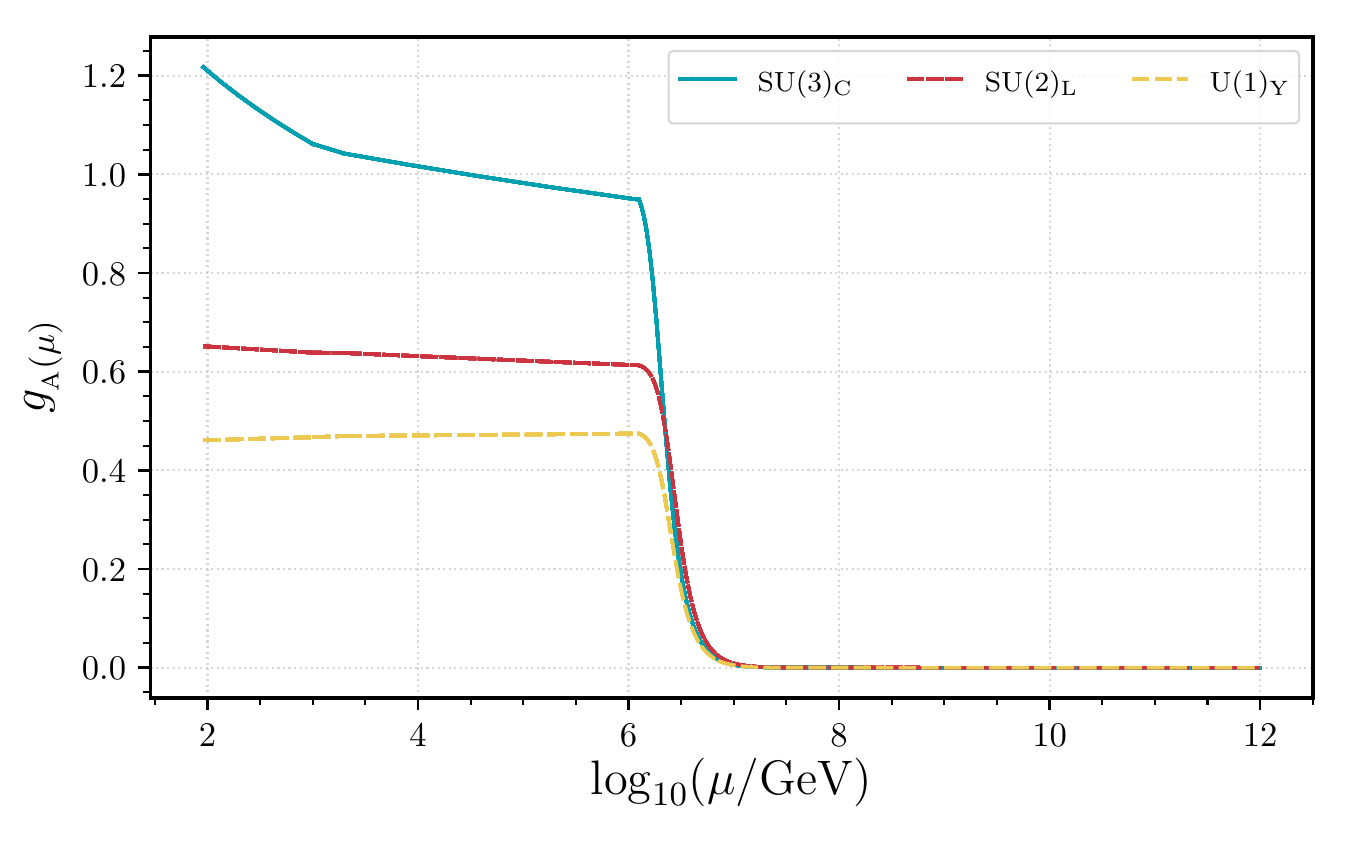} }}
	%\
	\caption{RG evolution of the gauge couplings for four selected scenarios. The $\U{Y}$ line denotes the running of the GUT-normalized $\sqrt{\frac{5}{3}} g_\mathrm{Y}$ gauge coupling.}
	\label{fig:RGEs}
\end{figure}

For a more concrete visualization \cref{fig:RGEs} shows four selected benchmark scenarios of the evolution of the gauge couplings from the unification scale down to the EW one. The scales involved in the running, the values of the $g_8$ gauge coupling and the proton lifetime are specified in \cref{tab:vals} (note that the scales are meant to show benchmark scenarios with different orders of magnitude for the energy scales, where the exact values are not relevant for the precision of this work). All four panels clearly demonstrate that the model can asymptotically unify all fundamental interactions, including family, in $\E{8}$ and that the unification scale can be realized well below the conventional GUT scale. It is clear that the effect of KK modes is crucial to realize such an Unification picture resulting in a ultra-weakly coupled theory at high-energy scale.
%%%%%%
%%%%%
\begin{table}[htb]
	\begin{center}
	%\resizebox{\columnwidth}{!}{%
		\begin{tabular}{c|cccccc}
			\toprule                     
			Scenario & $\log_{10}\tfrac{\Lambda_1}{\mathrm{GeV}}$ & $\log_{10}\tfrac{\Lambda_{2,3}}{\mathrm{GeV}}$ & $\log_{10}\tfrac{\Lambda_4}{\mathrm{GeV}}$ & $\log_{10}\tfrac{\Lambda_5}{\mathrm{GeV}}$ & 
			$g_8(\Lambda_1)$ &
			$\tau_p/[\mathrm{yrs}]$\\
			\midrule
		(a) & $9.62$ & $5.27$ & $3.20$ & $3.0$ & $3.0 \times 10^{-13}$ & $1.14 \times 10^{32}$\\
		(b) & $17.0$ & $7.77$ & $3.30$ & $3.0$ & $6.0 \times 10^{-28}$ & $7.88 \times 10^{90}$ \\
		(c) & $13.3$ & $11.3$ & $3.30$ & $3.0$ & $3.0 \times 10^{-6}$ & $4.97 \times 10^{32}$ \\
		(d) & $12.0$ & $6.1$ & $3.30$ & $3.0$ & $6.0 \times 10^{-18}$ & $7.88 \times 10^{50}$ \\
			\bottomrule
		\end{tabular}%}
		\caption{Threshold scales and values of the $\E{8}$ and $\E{6}$ gauge couplings at their respective breaking scales for the four plots in \cref{fig:RGEs}. The last column provides an estimate for the proton lifetime in its kaon decay channel.}
		\label{tab:vals}  
	\end{center}
\end{table}
%%%%%%%%%%
An appealing feature of the proposed unification paradigm is that the hierarchy problem can be remarkably alleviated. If GUT scale corrections to the Higgs boson mass are negligible as the values of $g_8(\Lambda_1)$ in \cref{tab:vals} suggest, scenarios such as (a) are also not expected to provide uncontrollably large corrections.

While the $\E{8}$ breaking scale, $\Lambda_1$, meets the proton decay limits, other energy-scales involved in the running may also have their own experimental constraints. Particularly relevant are the Wilson line scales whose largest contribution sets the size of heavy vector bosons. Current experimental limits bound $W_\mathrm{R}$ gauge bosons, contained in $E_8$, to be heavier than \cite{Aaboud:2017yvp}
\begin{equation}
    m_{W_R}>3.7\ {\rm TeV}\,,
\end{equation}
while the strongest bound on extra $Z'$ bosons is \cite{Zyla:2020zbs}
\begin{equation}
    m_{Z'}> 4.5\ {\rm TeV}\,.
\end{equation}
Since all points generated in the numerical scan result in $\Lambda_{2,3} > 100~\mathrm{TeV}$, the model is also unconstrained by direct searches for heavy gauge bosons at the LHC. In fact, it turns out that for the lower scale unification scenarios (see scenario (a) in \cref{tab:vals}) such particles must manifest themselves, at future colliders, at scales beyond approximately $100~\mathrm{TeV}$.

The masses of all KK modes in the model are integer multiples of the $\Lambda_2$ and $\Lambda_4$ and $\Lambda_5$ scales. The strongest phenomenological constraints on the size of EDs result from Flavour Changing Neutral Currents (FCNC), essentially caused by KK gluons as well as EW $S,T,U$ precision observables, which can be mostly affected by KK $W_\mathrm{R}$ bosons. Current searches pose a lower bound on the size of KK excitations that reads as \cite{Cheung:2001mq,Barbieri:2004qk}~\footnote{Note that the benchmark scenarios in \cref{tab:vals} show only the precision of orders of magnitude but are assumed to meet the previous constraints.}
\begin{equation}
\frac{1}{R_{1,2,3}}>6\ {\rm TeV} \,.
\label{eq:compac}
\end{equation}  

Finally, gaugino masses are generated at two-loop level and thus are expected to be lighter than the Wilson-line scales. The strongest restriction may come from direct searches for gluino masses which, according to Fig.~90.2 (right-panel) of \cite{Zyla:2020zbs} must be bounded as
 \begin{equation}
     m_{\tilde{g}}>2\ {\rm TeV} \,.
     \label{eq:gluino}
 \end{equation}
Exclusion limits for charginos and neutralinos should also be seen in a context where the only light scalar is the Higgs boson. Therefore, the most stringent constraints on their mass can be read off from Fig.~90.5 (right-panel) of \cite{Zyla:2020zbs}. For instance, with a neutralino mass up to approximately $250~\mathrm{GeV}$ the limit on the chargino mass reads
 \begin{equation}
     m_{\tilde{\chi}^\pm}>650\ {\rm GeV} \,.
     \label{eq:chargino}
 \end{equation}
The latter becomes unconstrained when the neutralino mass is heavier than $300~\mathrm{GeV}$. Finally, neutralino masses are bounded from below as
\begin{equation}
    m_{\tilde{\chi}^0}>116\ {\rm GeV} \,.
     \label{eq:nutralino}
\end{equation}
For a better indication of the excluded regions one can see the red contour in Fig.~90.5 (right-panel) of \cite{Zyla:2020zbs} . However, it is relevant to mention that these limits are merely indicative since current searches assume that neutralinos are stable particles, which does not apply in this case. In fact, the model proposed in this article does not possess R-parity and both the gaugino and Higgsino (or vector-like lepton) components of the charginos and neutralinos do decay into lighter states, including SM-like leptons. The gauginos obtain a mass as in \cref{eq:gluinomass}, which is at the $\Lambda_2$ scale and which should be sufficient to easily avoid all the existing constraints.

%%%%%%%%%%%%%%%%%%%%%%%%%%%%%%
\section{Outlook into the UV}
\label{sec:string-diff}
%%%%%%%%%%%%%%%%%%%%%%%%%%%%%%

The fact that the considered GUT is built into 10d spacetime with SUSY and has $\E{8}$ as a gauge group may bring superstring theory (ST) into consideration. However, some of the concepts discussed in this work would not be compatible with the conventional ST approach but still comprise a phenomenologically viable QFT framework. Let us describe key differences between the current model and a ST based one.

In the standard ST approach, the spacetime dimension is fixed, since only in 10d the conformal anomaly of the worldsheet is canceled. In the model presented in this article, 10d is chosen since $\mathcal{N}=1$ SUSY can be decomposed into $\mathcal{N}=4$ SUSY in 4d allowing a full (gauge-matter) unification of the SM. Even though this also happens in 7d, 8d and 9d theories \cite{Arkani-Hamed:2001vvu}, only 10d allows to obtain the complete SM matter sectors in the required gauge representations and with the observed family structure. Note that it is also possible to consider more than ten dimensions at the cost of loosing minimality.

In ST, an incorporation of quantum gravity implying the existence of graviton lies at the core of its foundation. For instance, in the heterotic string theory case, the gauge group $\E{8}\times \E{8}'$ plays a key role in the cancellation of anomalies in supergravity\footnote{$\SO{32}{}$ also solves this problem but is not considered to be a preferred option.}. For the GUT proposed in this article, a single $\E{8}$ is chosen since it unifies all the representations of the SM into a single one, while gravity lies beyond the scope of the current framework. Specifically, in variance to the conventional ST-based approach, here we refuse to unify gravity with other fundamental interactions. Our approach would be well justified in the case of a low orbifold compactification scale relative to the quantum gravity (Planck) scale such that any gravitational effects can be safely neglected at or below $\E{8}$ breaking scale.

Notice also that as gravity lies in the basis of any ST-based framework, one usually requires the gravitational anomaly cancellation between the gravitino, dilatino and gauginos which determines the gauge group of the fundamental theory. As there is no known consistent QFT of gravity at present, as well as there is no consensus in the literature of whether gravity is fundamental or emergent, the latter condition may be regarded as optional, and it is interesting to study QFT-based scenarios that do not rely on unification of gravity with other fundamental interactions. This is the main difference of our proposal from ST-based models.

As discussed above, the considered model is free from gauge anomalies which would not be the case in a ST based framework where inclusion of fundamental gravity is a must. First, the decomposition in \cref{eq:decom} would be incomplete in ST due to the presence of winding modes around each of the circles of each of the torus. Thus, since they do not add up to a complete $\textbf{248}$ irrep, this typically leads to the emergence of gauge anomalies. ST also contains twisted states which are inherently localized at the fixed points and equally lie in incomplete $\E{8}$ representations, also contributing to the gauge anomaly problem. Furthermore, there can be strings localized in a brane or connecting two of them. These would only ``feel'' a fraction of the Wilson line, or none at all, while the ST must be kept anomaly free for all values of the continuous Wilson line \cite{Ibanez:1987dw}. In a QFT-based scenario with only bulk states (like the one presented in the current work), the continuous Wilson lines do alleviate the gauge anomaly problem \cite{Scrucca:2004jn}. So, the gauge anomaly cancellation is a particularly important feature of our model as it is a QFT with no localized fields, which would not be possible in ST. 

The model presented in this work has a Lorentz anomaly since there is a single 10d chiral -- left-handed, by definition -- fermion in the $\textbf{248}$ $\E{8}$ representation\footnote{Note that 10d chirality is not the same as 4d chirality, as discussed in \cref{app:edpoinc}.}. Since it is a global symmetry, that is not viewed as a problem as the 10d Lorentz symmetry is broken by the orbifold compactification anyway. However, again, this would be considered a problem if one includes gravity. In this case, gravity would be added by upgrading the global Poincar\'e symmetry to a local one, which is equivalent to full coordinate invariance, i.e.~General Relativity. The simplest, but not unique, way to cancel the anomaly realised in ST is to add an extra $\E{8}'$ $\textbf{248}$ 10d right chiral supermultiplet or 248 $\E{8}$ singlets. If SUSY becomes local then there is also a spin-$3/2$ 10d chiral gravitino. The minimal way to cancel the anomaly is to add 247 singlet 10d left chiral superfields, although it is a more symmetric approach to add a single 10d right chiral superfield $\E{8}$ singlet, usually called dilaton, and an $\E{8}$ $\textbf{248}$ 10d left chiral supermultiplet \cite{Green:1984sg}. In ST an adjoint chiral multiplet can not be added as such, so one must add a second gauge symmetry $\E{8}'$ in order to fix this anomaly problem. From the viewpoint of a low-energy EFT and associated phenomenology, an inclusion of such an extra $\E{8}'$ in ST potentially leads to a large amount of extra massless fields and interactions that are not observed in nature. So, our model represents the simplest way to address this issue by simply {\it not} including gravity at the $\E{8}$ breaking scale, hence, eliminating the requirement of an additional $\E{8}'$.

As we have shown above, in our framework $\E{8}$ symmetry is, in fact, never realised exactly since the SM couplings never really unify, but disappear asymptotically leading to an approximately global $\E{8}$ in the UV regime. This result actually means that at any meaningful scales $\E{8}$ theory is only realised {\it approximately}; it is never an exact symmetry, in fact. Also, in our approach spacetime geometry is fixed, i.e. static and not dynamical, and is determined by the static global geometry of tori, in variance to quantum gravity approaches resorting to a dynamical spacetime. This is fully justified as our compactification scale appears to be way below the Planck scale in all scenarios we have studied here such that the quantum-gravitational effects are safely ignored at any step. So, together with the above arguments, this properly justifies our departure from the ST paradigm based upon the exact $\E{8}$ symmetry and a dynamical spacetime.

Finally, it is worth mentioning that if one sticks to a QFT approach as realised in this work, gravity can in principle be an emergent phenomenon instead of treating it as a fundamental one as in the ST-based framework \cite{DeAnda:2019uzq,Linnemann:2017hdo,Barcelo:2001ah}. In this case, the 10d super-Poincar\'e symmetry stays global and its anomaly would not be a problem, while it is broken by the orbifold compactification anyway. If an emergent local symmetry arises after compactification one would end up with an effective 4d theory of gravity at low energies. As the field content in this model will neither generate any 4d super-Poincar\'e nor mixed anomalies, no extra field content would be required in that case. 
However, none of the proposed QFT models of gravity are deemed to be fully consistent and each setup requires different constraints. The current approach is thus assumed free of gravity, extra work would be needed to include it and will be done elsewhere.

%%%%%%%%%%%%%%%%%%%%%%%%%%%%%%%%%
\section{Conclusion}
\label{sec:conclusion}
%%%%%%%%%%%%%%%%%%%%%%%%%%%%%%%%%
 
A model with a single 10d gauge $\E{8}$ superfield, effectively unifying all matter and gauge sectors of the SM, has been presented. The EDs are orbifolded as $\mathbb{T}^6/(\ZZ)$ whose extra dimensional rotational and translational boundary conditions break $\E{8}\to \SU{3}{C}\times \SU{2}{L}\times \U{Y}$ after compactification. Furthermore, SUSY is broken in the process generating specific soft masses while gauge anomalies are absent at all levels without the need for extra fields.
 
The model contains a single arbitrary complex parameter, the gauge coupling, while the orbifold further introduces thirteen additional arbitrary complex parameters. These parameters completely define the model's freedom and it is shown that they are sufficient to generate a variety of phenomenologically viable low-scale scenarios of New Physics generally reproducing the basic properties of the SM. The model contains a viable flavour structure for all SM-like fermion masses. Furthermore, the mass matrices are hierarchical. There are five right handed neutrinos which obtain a large Majorana mass so that a simple see-saw mechanism is in place.
 
Asymptotic unification of the gauge couplings is consistently achieved. In particular, the model admits scenarios where the unification scale can be as low as $10^9~{\rm GeV}$ while complying with proton decay constraints. The asymptotic value of the gauge couplings is ultra-small which results in a remarkable alleviation of the Higgs boson mass hierarchy problem.
 
The most appealing New Physics candidates emergent from the discussed theory are vector-like fermions (gauginos and higgsinos) as well as one generation of squarks and sleptons which can manifest themselves not too far from the TeV scale or at the reach of next generation of colliders. Most notably, even with a rather limited freedom but a noticeable predictive power, the model will allow for concrete New Physics predictions once the full SM-like sector is comprehensively derived from the details of the extra-dimensional space-time geometry. 

\acknowledgments
The authors thank M\'onica F. Ram\'irez for her support in generating the orbifold diagrams.  
This work has been supported in part by the MICINN (Spain) projects FIS2011-23000, FPA2011-27853-01,
FIS2014-52837-P, FPA2014-53375-C2-1-P, Consolider-Ingenio MULTIDARK CSD2009-00064, SNI-CONACYT (M\'exico), by the Swedish Research Council grant, contract number 2016-05996, by the European Research Council (ERC) under the European Union's Horizon 2020 research and innovation programme (grant agreement No 668679), and by the FCT (Portugal) projects PTDC/FIS-PAR/31000/2017, CERN/FIS-PAR/0027/2019 and CERN/FIS-PAR/0002/2019. This work is also supported by the Center for Research and Development in Mathematics and Applications (CIDMA) through the Portuguese Foundation for Science and Technology (FCT), references UIDB/04106/2020 and, UIDP/04106/2020, and by national funds (OE), through FCT, I.P., in the scope of the framework contract foreseen in the numbers 4, 5 and 6 of the article 23, of the Decree-Law 57/2016, of August 29, changed by Law 57/2017, of July 19.

\appendix

%%%%%%%%%%%%%%%%%%%%%%%%%%%%%%%%%%%%%%%%%%%%%%%%%%% 
\section{ED field representations}
\label{app:edpoinc}
%%%%%%%%%%%%%%%%%%%%%%%%%%%%%%%%%%%%%%%%%%%%%%%%%%% 
 
Compactifying the EDs implies reduction of the Poincar\'e symmetry. 
The full global symmetry of the model is $\mathcal{N}=1$ super-Poincar\'e in 10d which has generators
\begin{equation}
\begin{array}{ll}
    [P^S,P^T]=0,\ \ \ & [M^{ST},P^U]=i(P^S\eta^{TU}-P^T\eta^{SU}) \,,
    \\
    \multicolumn{2}{l}{[M^{ST},M^{UV}]=i(M^{SV}\eta^{TU}+M^{TU}\eta^{SV}-M^{SU}\eta^{TV}-M^{TV}\eta^{SU}),}
    \\
    \\
    \{Q_\alpha,\overline{Q}_{\dot{\beta}}\}=2\sigma^S_{\alpha\dot{\beta}}P_S\,,
    &
    [Q_\alpha,M^{ST}]=(\sigma^{ST})^\beta_{\ \alpha} Q_\beta\,,
    \\
    \{Q_\alpha,Q_\beta\}=\{\overline{Q}_{\dot{\alpha}},\overline{Q}_{\dot{\beta}}\}=0\,,\ \ \ 
    &
    [P_S,Q_\alpha]=[P_S,\overline{Q}_{\dot{\beta}}]=0\,,
\end{array}
\end{equation}
where $S=0,1,...,3,5,...10$ and $\alpha,\dot{\beta}=1,...8$.

Specially relevant is the reduction of the Lorentz symmetry as every field lies in an irreducible representation of the Lorentz group before and after compactification. In $N$ dimensions, the Lorentz group is $\SO{1,N-1}{}$. After compactifying $m$ spatial dimensions, it is reduced into $\SO{1,N-m-1}{}$. The scalar fields are singlets of $\SO{1,N-1}{}$ and remain singlets of $\SO{1,N-m-1}{}$ after compactification. The vector fields have an $(\textbf{N})$ representation under $\SO{1,N-1}{}$, as it is a vector with $N$ components. It decomposes into an $(\textbf{N-m})$ vector and $m\times (\textbf{1})$ scalars under $\SO{1,N-m-1}{}$ after compactification. It is important to remark that all the studied vector fields are massless gauge vector fields, which have two degrees of freedom less than a general vector due to their gauge freedom and masslessness.

%%%%%%%%%%%%%%%%%%%%%%%%%%%%%%%%%%%%
\subsection{Fermions in EDs}
\label{sec:fermions_EDs}
%%%%%%%%%%%%%%%%%%%%%%%%%%%%%%%%%%%%
 
Assuming a Minkowski metric with signature $\eta^{MN}=diag(-1,1,...,1)$, the kinetic term for fermions reads
\begin{equation}
     \mathcal{L}_\Psi=i\bar{\Psi}\Gamma^M\partial_M\Psi\,,
\end{equation}
where the Dirac matrices satisfy
\begin{equation}
     \{\Gamma^M,\Gamma^N\}=2\eta^{MN}\,.
\end{equation}

Starting from $n$ even-dimensional Dirac matrices, one can define the $(n+1)th$ matrix multiplying every matrix
\begin{equation}
     \Gamma^{n+1}=\Gamma^0\cdot \Gamma^1\cdot\cdot\cdot \Gamma^n \,.
\end{equation}
Starting from $n+1$ odd-dimensional Dirac matrices, the product $\Gamma^0\cdot \Gamma^1\cdot\cdot\cdot \Gamma^{n+1}=1,$
is trivial, so it can not be used to build the $n+2$ dimensional Dirac matrices. To do that one can define them in terms of the $n$ dimensional Dirac matrices
 \begin{equation}
     \Gamma^M_{(n+2)d}=i\left(\begin{array}{cc}0 & \Gamma^M_{nd} \\ \Gamma^M_{nd} & 0
     \end{array}\right)\,,\ \ \ \Gamma^{n+1}_{(n+2)d}=i\left(\begin{array}{cc}0 & \Gamma^{n+1}_{nd} \\ \Gamma^{n+1}_{nd} & 0
     \end{array}\right)\,,\ \ \ \Gamma^{n+2}_{(n+2)d}=i\left(\begin{array}{cc}0 & \mathbb{I}_{nd} \\ -\mathbb{I}_{nd} & 0
     \end{array}\right)\,,
 \end{equation}
which are $2^{n/2}\times 2^{n/2}$ matrices for even dimension $n$ and $2^{(n-1)/2}\times 2^{(n-1)/2}$ for odd dimensions $n$. The fermion would have $2^{n/2}$ complex components for even dimension or $2^{(n-1)/2}$ complex components for odd dimension. However, this can be a reducible representation.

In even $n$ dimensions, the $2^{n/2}$ component fermion can be separated by eigenvalues of the matrix $\Gamma^{n+1}_{nd}$ which defines the chirality. In odd $n$ dimensions there is no chirality. A fermion satisfying $\Psi=\pm\Gamma^{n+1}_{nd}\Psi$ is called a Weyl or a chiral fermion which exists for every even dimension. Note that chirality is different by a different matrix in every even dimension. One can also impose a real condition $\Psi=\Psi^c$, called Majorana condition, which can be satisfied non-trivially when $n=1,2,3,4,8,9,10$ (up to 10d).  Only in $n=2,10$ can both conditions be fulfilled simultaneously \cite{Sohnius:1985qm}. Note that for $\mathcal{N}=1$ SYM in $n$ dimensions, one has a $n$ dimensional vector and some representation of a $n$ dimensional fermion. For SUSY to be consistent they must have the same degrees of freedom. This happens exactly in $n=4,6,10$. In other spacetime dimensions, one has to also add scalar fields to match the degrees of freedom.

%%%%%%%%%%%%%%%%%%%%%%%%%%%%%%%%%%%%%%%%%%%%%%%%%%%%%%%%%%%%%
\subsection{$\mathcal{N}=1$ SUSY in six dimensions}
\label{sec:susy6d}
%%%%%%%%%%%%%%%%%%%%%%%%%%%%%%%%%%%%%%%%%%%%%%%%%%%%%%%%%%%%%

The $\mathcal{N}=1$ SYM theory in 6d is built from a gauge field $A_M$ and a chiral fermion $\Lambda$ in the adjoint representation with the Lagrangian
\begin{equation}
     \mathcal{L}_{6d}=-\frac{1}{4}F_{MN}F^{MN}-i\bar{\Lambda}\Gamma^MD_M\Lambda \,,
\end{equation}
where the Dirac matrices are are $8\times 8$ components, the Dirac fermion has eight complex components, but the chiral fermion $\Lambda=\tfrac{1}{2}(1-\Gamma_7)\Lambda$ has four complex components\footnote{Note that the 6d chirality is not the same as 4d chirality. A 6d chiral fermion decomposes into a left and a right 4d chiral fermion pair}. The equations of motion of the 6d chiral fermion reduce the degrees of freedom to four. The 6d vector has six real components, whose gauge freedom and masslessness reduce them to four degrees of freedom. Together they form a 6d gauge vector supermultiplet.

By decomposing the 6d vector $A_M$ into a 4d one $A_\mu$ and two scalar components that form a complex scalar $A=A_5+iA_6$, then writing the chiral fermion as $\Lambda=(\chi,0)^T$, and decomposing the 6d chiral superfield into two 4d Majorana fermions $\chi=(\lambda-i\psi)$, the Lagrangian can be expanded as
\begin{equation}\begin{split}
    \mathcal{L}_{6d}=&-\frac{1}{4}F_{\mu\nu}F^{\mu\nu}+(D_\mu A)^\dagger D^\mu A-i\bar{\lambda}\gamma^\mu D_\mu \lambda-i\bar{\psi}\gamma^\mu D_\mu \psi
    \\
    & -\sqrt{2}[\lambda,\psi] -\sqrt{2}[\bar{\lambda},\bar{\psi}]+\frac{1}{2}[A^\dagger,A]^2 \,,
\end{split}\end{equation}
this becomes a 4d $\mathcal{N}=2$ gauge supermultiplet built from a 4d $\mathcal{N}=1$ vector gauge supermultiplet and an adjoint chiral supermultiplet.

To build a matter 6d supermultiplet one would have to have a chiral fermion $\Psi=\tfrac{1}{2}(1+\Gamma_7)\Psi$ whose four degrees of freedom are matched by two complex scalars with the Lagrangian
\begin{equation}
    \mathcal{L}_{6dm}=(D^\mu A)^\dagger D_\mu A+(D^\mu B)^\dagger D_\mu B+\frac{i}{4}\bar{\Psi}\gamma^\mu D_\mu\Psi \,,
\end{equation}
where, by decomposing the 6d chiral fermion $\Psi$ into a 4d chiral left right fermion pair, one obtains an $\mathcal{N}=2$ hypermultiplet built from vector-like 4d chiral supermultiplet pair.
 
Note that there is no renormalizable 6d superpotential, just as there are no renormalizable $\mathcal{N}=2$ matter interactions. Therefore, an $\mathcal{N}=1$ SUSY 6d theory can be decomposed into an $\mathcal{N}=2$ SUSY 4d theory \cite{Sohnius:1985qm}.
 
%%%%%%%%%%%%%%%%%%%%%%%%%%%%%%%%%%%%%%%%%%%%%%%%%%%%%%%%%%%%%
\subsection{$\mathcal{N}=1$ SUSY in ten dimensions}
\label{sec:susy10d}
%%%%%%%%%%%%%%%%%%%%%%%%%%%%%%%%%%%%%%%%%%%%%%%%%%%%%%%%%%%%%
  
The $\mathcal{N}=1$ SYM theory in 10d is built from a gauge field $A_M$ and a fermion $\Lambda$ in the adjoint representation with the Lagrangian
 \begin{equation}
     \mathcal{L}_{10d}=-\frac{1}{4}F_{MN}F^{MN}-i\bar{\Lambda}\Gamma^MD_M\Lambda \,,
 \end{equation}
where the Dirac matrices have $32\times 32$ components, the Dirac fermion has 32 complex components. In 10d, a Weyl condition $\Lambda=\tfrac{1}{2}(1-\Gamma_{11})\Lambda$ can be imposed which reduces the fermion to sixteen components. In 10d (as well as in 2d) one can simultaneously impose the Majorana contition $\Lambda=C\bar{\Lambda}^T$, where $C$ is the charge conjugation matrix, and further reduce the fermion into eight complex components. The equations of motion reduce the degrees of freedom down to eight. The vector has ten real components which are reduced by gauge freedom and masslessness to eight. Therefore, the fermion/scalar degrees of freedom are matched.

The 10d Weyl/Majorana fermion can be decomposed into four 4d Weyl (or Majorana as they are massless) fermions $\lambda_{\bar{a}}$, with $\bar{a}=1,2,3,4$, where they transform as a $\textbf{4}$ of $SU(4)$, which is isomorphic to $SO(6)$, the ED rotation symmetry. The 10d vector can be decomposed into a 4d vector and six scalars $X_{\tilde{i}}=A_{4+\tilde{i}}$, where $\tilde{i}=1,...,6$, and they transform as a $\textbf{6}$ of $SO(6)$. They have the following Lagrangian
  \begin{equation}\begin{split}
      \mathcal{L}_{10d}=&-\frac{1}{4}F_{\mu\nu}F^{\mu\nu}+(D_\mu X^{\tilde{i}})^\dagger D^\mu X_{\tilde{i}}-i\bar{\lambda}^{\bar{a}}\gamma^\mu D_\mu \lambda_{\bar{a}}\\
      &
      +f^{\bar{a}\bar{b}}_{\tilde{i}}\lambda_{\bar{a}}[X^{\tilde{i}},\lambda_{\bar{b}}]+f_{\bar{a}\bar{b}\tilde{i}}\bar{\lambda}^{\bar{a}}[X^{\tilde{i}},\bar{\lambda}^{\bar{b}}]+\frac{1}{2}[X^{\tilde{i}},X^{\tilde{j}}]^2 \,,
  \end{split}\end{equation}
which is the Lagrangian of $\mathcal{N}=4$ SUSY in 4d \cite{Sohnius:1985qm}.
 
%%%%%%%%%%%%%%%%%%%%%%%%%%%%%%%%%%%%%%%%%%%
\section{Orbifolding and compactification}
\label{app:orb}
%%%%%%%%%%%%%%%%%%%%%%%%%%%%%%%%%%%%%%%%%%%

As seen in the previous appendix, the ED fermionic representations always decompose into 4d Dirac fermions. To generate the chirality one assumes an incomplete ED Poincar\'e group. The full ED Poincar\'e symmetry $\OO{6}\ltimes \T{6}$ can be orbifolded as
\begin{equation}
    (\OO{6}/\F{})\ltimes (\T{6}/\Gamma) \,,
\end{equation}
with $\F{}$ and $\Gamma$ being the discrete orbifold and lattice groups, associated to spacetime rotations and translations respectively. The subgroup $\SO{6}{}$ of the ED rotation symmetry $\OO{6}$ becomes the ${\SU{4}{}}_\mathcal{R}$ symmetry of the extended SUSY. One can define the orbifold upon modding it out by any discrete subgroup of $\OO{6}\simeq \SO{6}{}\times \mathbb{Z}_2\simeq \SU{4}{}\times\mathbb{Z}_2$. In particular, if one wants to preserve $\mathcal{N}=1$ SUSY, i.e.~leaving one ${\U{}}_\mathcal{R} \subset {\SU{4}{}}_\mathcal{R} $ unbroken, then the orbifold must be expressed in terms of a discrete subgroup of ${\SU{3}{}}_\mathcal{R}$ \cite{Dixon:1985jw,Dixon:1986jc}. While such a discrete subgroup can be, in general, non-Abelian, the involved commutation relations overly constrain the orbifold boundary conditions diminishing the possibility for obtaining a realistic model below the compactification scale \cite{DeAnda:2019cbq}. The most general Abelian discrete subgroup of $\SU{3}{}$ is $\mathbb{Z}_N \times \mathbb{Z}_M$ \cite{Fischer:2012qj} which enables to employ the latter as a discrete orbifold group in the considered approach, i.e.~$\F{}\simeq \mathbb{Z}_N\times \mathbb{Z}_M$.

It is worth noticing that a general $F=\mathbb{Z}_N$ orbifolding procedure is defined by identifying the ED coordinates
%%%%%
\begin{equation}
(x,z_1,z_2,z_3)\sim (x,e^{2i \pi n_1/N} z_1,e^{2i \pi n_2/N}z_2,e^{2i \pi n_3/N}z_3) \,,
\label{ni}
\end{equation}
%%%%%
with arbitrary integers $n_i$, satisfying $n_1+n_2+n_3=0\ {\rm mod}\ 2N$, so that $\mathcal{N}=1$ SUSY is preserved \cite{GrootNibbelink:2017luf}. The boundary conditions may also involve non-trivial gauge transformations. In fact, the considered orbifold discrete group $\F{}$ does not commute with the generators of the $\E{8}$ gauge group. A discrete $\mathbb{Z}_N \subset \E{8}$ transformation can be defined, in general, as a specific $\mathbb{Z}_N \subset {\U{}}_a \subset \E{8}$ subgroup where ${\U{}}_a$ can be any ${\U{}}$ subgroup of $\E{8}$. The $\mathbb{Z}_N$ implies a gauge transformation on a field $f$ denoted by a phase $e^{2i \pi q^f_a/N}$, 
where $q^f_a$ is the ${\U{}}_a$ charge the corresponding field $f$.

The boundary conditions act differently on vector $V$ and chiral $\phi$ components 
of the unique 10d gauge superfield $\mathcal{V}(x,z_1,z_2,z_3)$ as follows
\cite{Aranda:2020noz,deAnda:2020prd}
\begin{equation}\begin{split}
V(x,z_1,z_2,z_3)&=e^{2i \pi q^f_a/N} V(x,e^{2i\pi n_1/N} z_1,e^{2i\pi n_2/N} z_2,e^{2i\pi n_3/N} z_3),\\
\phi^i(x,z_1,z_2,z_3) &=e^{2i\pi n_i/N}e^{2i \pi q^f_a/N} \phi^i(x,e^{2i\pi n_1/N} z_1,e^{2i\pi n_2/N} z_2,e^{2i\pi n_3/N} z_3) \,,
\end{split}
\label{phi^i}
\end{equation}
so that each supermultiplet receives a phase associated to its charge and extended SUSY decomposition.
Here, the $V$ fields transform according to adjoint representation of the unbroken gauge group, 
i.e. the fields with identity boundary conditions. The representation of the light chiral 
superfield $\phi_i$ is the fundamental one of the unbroken group and contains the fields with 
charge $q^f_a=-n_i \mod N$. One then chooses a specific set of $n_i$ integers in order 
to single out the massless fields in consistency with the SM gauge and matter field content. 

To do an orbifolding $\mathbb{Z}_N\times\mathbb{Z}_M$ one applies two independent $\mathbb{Z}_N$ and $\mathbb{Z}_M$ orbifoldings just as described, each one associated to different ${\U{}}$ subgroups of $\E{8}$.  
Geometrically, the translation group is also modded out (compactified) to define 
the three tori as follows
\begin{equation}
    z_i\sim z_i+\tau^r_i \,,
\end{equation}
where $\tau^r_i$ is the period in compact EDs, with $r=1,2$ and $i=1,2,3$. The lattice group which 
compactifies the EDs is defined as $\Gamma=\{\tau^r_i\}\simeq \mathbb{Z}^6$. In this case, the 10d gauge 
superfield $\mathcal{V}(x,z_i)$ must also comply with periodicity boundary conditions up to a gauge 
transformation $U$ that is arbitrarily chosen for each $\tau^r_i$, i.e.
\begin{equation}
\mathcal{V}(x,z_i)=U^r_i \mathcal{V}(x,z_i+\tau^r_i) \,.
\label{eq:vwl-app}
\end{equation}
Therefore, similarly to the orbifold group $\F{}$, as a transformation of the discrete lattice group $\Gamma$ also 
implies an $\E{8}$ gauge transformation, it will not commute with all of the $\E{8}$ gauge group generators.

One can choose three arbitrary transformations $U^r_{i=1,2,3}$ separately -- one for each complex 
translation generator. A non-trivial $U^r_i$ gauge transformation is called a Wilson line and 
is explicitly defined as follows 
\begin{equation}
\mathcal{V}(x,z_i)=e^{i\alpha^{ar}_i z_i T_a}\tilde{\mathcal{V}}(x,z_i) \ \ \ {\rm with}\ \ \ \tilde{\mathcal{V}}(x,z_i)=\tilde{\mathcal{V}}(x,z_i+\tau^r_i) \,,
\label{eq:wilgauge}
\end{equation}
where $T_a$ are the generators of the algebra associated to $\E{8}$ and the $\alpha^{ar}_i$ are the corresponding coefficients that define the specific gauge transformation. The $\tilde{\mathcal{V}}$ are periodic functions of the EDs.
The Wilson lines can be reabsorbed through an inverse gauge transformation $\mathcal{V}\to U^{-1}\mathcal{V}$ generating an effective VEV in the ED components of the gauge vector in $\mathcal{V}$, which correspond to the 4d scalars in $\phi_i$. After the phase is reabsorbed, the components of $\mathcal{V}(x,z_i)$ become 
\begin{equation}
\begin{split}
V(x,z_i)&=\tilde{V}(x,z_i)\,, \\
\phi_i(x,z_i)&=\tilde{\phi}_i(x,z_i)+\sum_r \alpha^{ar}_{i} \tau^r_i  T_a \,.
\end{split}
\end{equation}
Therefore, the Wilson line can be seen as an effective VEV in the chiral superfields with the corresponding representation \cite{Candelas:1985en}. One must note that it is not a usual VEV found in the QFT approach, as it does not come from the minimization of a given potential. Instead, it comes from the ED profiles of the scalar fields and it triggers a splitting in the mass spectrum just as an actual VEV does.

To maintain the super-Poincar\'e structure in the resulting 4d theory, the mutual action of the discrete groups realises as $\F{} \Gamma=\Gamma$, i.e. the action of the orbifold rotations $\F{}$ must be equivalent to a lattice transformation $\Gamma$, then $\F{}$ must be an automorphism of $\Gamma$. As both, the discrete orbifold group $\F{}$ and the lattice group $\Gamma$, are assumed to be non-commuting with the $\E{8}$ gauge group, this restricts the possible alignment of the Wilson lines. This is represented in \cref{eq:wlcom}, which  implies that the continuous Wilson lines must ``live'' in the representations where the orbifold rotations are trivial. This means that the effective VEV must lie in the chiral supermultiplets with zero modes. For consistency with the translation group Abelian structure they must all commute with each other.

Note that one can identify specific Wilson lines that do not induce an effective VEV. These are called discrete Wilson lines. An orbifold $\mathbb{Z}_N$ is based on modding out a space rotation operator $R$ that complies with $R^N=I$. It will also follow $(R^q)^{N/q}=I$ when $q$ and $N/q$ are integers. If an orbifold rotation $R$ is accompanied by a gauge transformation $P$, it will satisfy $(P^q)^{N/q}=I$, then a discrete Wilson line (defined by a gauge transformation $U$) would comply with $(UP^q)^{N/q}=I$. It is therefore possible to treat them as extra rotation boundary conditions imposed on the non-origin fixed points. In the model presented in this paper no discrete Wilson lines are assumed.

As this paper is restricted to Abelian orbifolds, the orbifold action is only accompanied by Abelian gauge transformations, therefore they are generated by the Cartan subalgebra. The discrete Wilson lines must also commute with the orbifold rotations, therefore they must also be generated by the Cartan subalgebra. None of these orbifold transformations can reduce the rank. Continuous Wilson lines need not commute, therefore they are not generated by the Cartan subalgebra (or they are generated by a rotated basis). Continuous Wilson lines can reduce the rank \cite{Forste:2005rs} and they are the ones used in this model.

%%%%%%%%%%%%%%%%%%%%%%%%%%%%%%%%%%%%%%%%%%
\section{Full Lagrangian decomposition}
\label{app:lag}
%%%%%%%%%%%%%%%%%%%%%%%%%%%%%%%%%%%%%%%%%%

The original Lagrangian in 10d reads
\begin{equation}
\begin{split}
S=
&-\frac{1}{2}tr\int d^4x d^3z d^3\bar{z}d^2\boldsymbol\theta\left[\mathfrak{W}^\alpha_A \mathfrak{W}_{A\alpha}\right]+{\rm h.c.} \,,
\end{split}
\end{equation}
where $\mathfrak{W}^\alpha_A=\mathfrak{W}^\alpha_A(\mathcal{V})$ ($A$ runs through the gauge indices) is the standard 10d gauge superfield strength
\begin{equation}
    \mathfrak{W}^\alpha=\frac{i}{4}(\mathfrak{D}^T_\beta \epsilon^{\beta\gamma}\mathfrak{D}_\gamma)\mathfrak{D}_\alpha V(x,z_i,\boldsymbol\theta),\ \ \ {\rm where} \ \ \ \mathfrak{D}_\alpha=-\frac{\partial}{\partial\bar{\boldsymbol\theta}}-(\Gamma^M\boldsymbol\theta)_\alpha\frac{\partial}{\partial x^M} \,,
\end{equation}
where $M$ runs through $10$ dimensions, $\alpha$ is the spinor index and $\boldsymbol\theta$ is a 10d Majorana/Weyl fermion.
After compactification, the Lagrangian becomes \cite{Marcus:1983wb}
\begin{equation}
\begin{split}
S=&tr\int d^4x d^3z d^3\bar{z}d^4\theta\left[\phi^{i\dagger}e^{-2gV}\phi_i+i \phi^{i\dagger}(\partial_i e^{2gV})-i(\bar{\partial}^i e^{-2gV})\phi_i+\frac{1}{2} e^{2gV}\bar{\partial}^i\partial_i e^{-2gV}\right]\\
&-tr\int d^4x d^3z d^3\bar{z}d^2\theta\left[\frac{1}{2}W^\alpha_A W_{A\alpha}-\frac{ig}{6}\epsilon^{ijk}\phi_i[\phi_j,\phi_k]+\frac{g}{2}\epsilon^{ijk}\phi_i\partial_j\phi_k\right]+{\rm h.c.} \,,
\label{eq:4ds}
\end{split}
\end{equation}
where each $\phi_i,V$ are 4d chiral and gauge superfield, respectively but are functions on the 10d, while $g$ is the unique compactified gauge coupling. Now, $\theta$ is the standard 4d Majorana fermion coordinate. 

This action is invariant under global $\mathcal{R}$ symmetries coming from the compactification of $10d \mathcal{N}=1$ SUSY into a $4d \mathcal{N}=4$ SUSY.
The full $\mathcal{R}$ symmetry is $SU(4)_\mathcal{R}$ which is isomorphic to $SO(6)$ from the extra dimensional rotations. Therefore the $\mathcal{R}$ symmetry comes from the extra dimensional rotations.

The \cref{eq:4ds} only show explicitly the symmetry $SU(3)_\mathcal{R}\times U(1)_\mathcal{R}$ by being written in terms of superfields ($SU(4)_\mathcal{R}$ can be made explicit if the action were to be written in terms of scalar, fermion and vector fields).

The $SU(3)_\mathcal{R}$ symmetry corresponds to an arbitrary $\xi^i_{\ j}\in SU(3)$ transformation 
\begin{equation}
    \phi_j\to\xi^i_{\ j}\phi_i,\ \ \ z_j\to \xi^i_{\ j} z_i,
\end{equation}
where the $\phi$ and $z$ behave as triplets and must be rotated simultaneously.
The remaining symmetry is related to the $\mathcal{N}=1$ SUSY coordinate $\theta$ rotation $e^{i\zeta}\in U(1)_\mathcal{R}$together with
\begin{equation}
    \theta\to e^{i\zeta}\theta,\ \ \ \phi_i\to e^{2i\zeta/3}\phi_i,\ \ \ z_i\to e^{-2i\zeta/3}z_i.
\end{equation}
It can be seen that \cref{eq:4ds} is invariant under these transformations.

It is then required that the $\phi_i$ be equally $\mathcal{R}$ charged and the superpotential have an $\mathcal{R}$ charge $2$. Therefore the $\phi_i$ must have $\mathcal{R}$ charge $2/3$. Any VEV in the charged $\mathcal{R}$ scalar in $\phi_i$, breaks the $U(1)_\mathcal{R}$.  

Each field decomposed into an Fourier-like series of SO(6) modes
\begin{equation}
    \phi_i(x,z_i)=\sum_{\textbf{s}} \sum_{a,b }\phi^{sab}_{i}(x)f(z_j)_{sab} \,,
\end{equation}
where $\textbf{s}=(s_1,s_2,s_3,s_4,s_5,s_6)$ is a contracted index for the Fourier-like decomposition of 6d SO(6) modes . The orbifolding breaks the original representation into separate ones (of the unbroken gauge group), here denoted by the two indices $a,b$, one for each $\mathbb{Z}_N$ orbifolding.  There is a similar decomposition for the $V$. The $f$ are which are also eigenstates of the $\mathbb{Z}_3\times \mathbb{Z}_3$ 
\begin{equation}
    \partial_i f(z_j)_{\textbf{s}ab}=M_{i\textbf{s}ab}f_{\textbf{s}ab}(z_j) \,,
\end{equation}
with no sum over the indices intended, and $M_i\sim 1/R_i$, in terms of the radii $R_i$ defining each torus. These radii are arbitrary and independent. One can assume an arbitrary hierarchy between them. 

Each of the original four superfields $\phi_i, V$, each within a given $\textbf{248}$, is split into an infinite series where each $\textbf{248}$ split into nine different states. The relevant zero modes corresponding to $f_{000}$ has a zero eigenvalue mass and are the only ones surviving at low energies. Integrating out the massive superfields one would obtain the effective action in the general form
\begin{equation}
\begin{split}
S=\int d^4x d^4\theta\left[\frac{1}{2}\mathcal{K}(\phi,\phi^\dagger e^{-2gV})\right]+\int d^4x  d^2\theta \left[2\mathcal{W}(\phi)-\frac{1}{2}\mathcal{H}_{AB}(\phi)W^\alpha_A W_{\alpha B}\right]+{\rm h.c.} \,.
\label{eq:genac}
\end{split}
\end{equation}

Using the coordinate
\begin{equation}
    \hat{x}_\mu=x_\mu+\frac{1}{2}\bar{\theta}\gamma_5\gamma_\mu\theta \,,
\end{equation}
one can expand the left chiral superfields as 
\begin{equation}\begin{split}
    \phi(\tilde{x})=&S(\tilde{x})-\sqrt{2}\bar{\theta}\psi_L(\hat{x})+\bar{\theta}\theta_L \mathcal{F}(\hat{x}) \,,
    \end{split}
\end{equation}
where $S$ is a scalar field, $\psi_L$ is a left handed Weyl fermion and $\mathcal{F}$ is an auxiliary scalar field. We use $\eta^{\mu\nu}=diag(-1,1,1,1)$. The subindex $L$ for fermions means $\psi_L=P_L \psi=(1-\gamma_5)\psi/2$.
The gauge superfield and superfield strength (in the Wess-Zumino gauge and suppressing gauge indices)
\begin{equation}\begin{split}
    e^{-2gV(x)}=&1-ig(\bar{\theta}\gamma_5\gamma^\mu\theta) t_{A} A_{A\mu} +2ig(\bar{\theta}\gamma_5\theta)\bar{\theta} t_{A} \lambda_{A}(x)+\frac{g}{2}(\bar{\theta}\gamma_5\theta)^2 t_A\mathcal{D}_A\\
    &-\frac{g^2}{2}(\bar{\theta}\gamma_5\gamma^\mu\theta)(\bar{\theta}\gamma_5\gamma^\nu\theta) t_A A_{A\mu}(x) t_B A_{B\nu}(x),\\
    W_A(\hat{x})=&\lambda_{AL}(\hat{x})+\frac{1}{2}\gamma^\mu\gamma^\nu \theta_L F_{A\mu\nu} +\bar{\theta}\theta_L\gamma^\mu D_\mu \lambda_{AR}-i\theta_L\mathcal{D}_A(\hat{x}) \,,
    \end{split}
\end{equation}
where the spinor index $\alpha$ has been suppressed and $A$ denote the adjoint gauge indices. The $t$ are the gauge algebra generators,  $A_\mu$ is the gauge vector, $D_\mu=\partial_\mu +igt\cdot A_\mu$ is the covariant derivative, $F_{\mu\nu}=D_\mu A_\nu-D_\nu A_mu$ is the gauge field strength tensor, $\lambda$ is the gaugino Majorana fermion and $\mathcal{D}$ is the scalar gauge auxiliary field.
For convenience, one can define the evaluated function
\begin{equation}
    \left.\mathcal{K}\right|_S= \left.\mathcal{K}\right|_{\phi_i\to S_i,V\to 0},
\end{equation}
of superfields into only the scalar components.

With this superfield decomposition, one can write the general action $S=\int d^4 x \mathcal{L}$ from  \cref{eq:genac} as  \cite{Weinberg:2000cr}
\begin{equation}
    \begin{split}
        \mathcal{L}=&-\frac{1}{2}\left.\frac{\partial^2 \mathcal{K}}{\partial\phi_i^\dagger\partial\phi_j}\right|_S\left[(D_\mu S_i)^\dagger D^\mu S_j+\bar{\psi}_i\gamma^\mu  D_\mu  \psi_{Lj}\right]
        \\
        &-\frac{1}{8}\mathcal{H}_{AB}|_S\left[F_{A\mu\nu}F_B^{\mu\nu}-\frac{i\vartheta}{2}\epsilon_{\alpha\beta\gamma\delta}F_A^{\alpha\beta}F_B^{\gamma\delta}+4\bar{\lambda}_A\gamma^\mu D_\mu  \lambda_{BR}\right]
        \\
        &-\frac{1}{2}\left.\frac{\partial^2 \mathcal{W}}{\partial\phi_i\partial\phi_j}\right|_S\bar{\psi}_i\psi_{Lj}-ig\sqrt{2}\left.\frac{\partial^2 \mathcal{K}}{\partial\phi_i^\dagger\partial\phi_j}\right|_S S^\dagger_i t_A\bar{\psi}_j\lambda_{AL}
        \\ 
        &
        +\frac{1}{2}\left.\frac{\partial^2 \mathcal{K}}{\partial\phi_i^\dagger\partial\phi_j}\right|_S\mathcal{F}^\dagger_i\mathcal{F}_j+\left.\frac{\partial \mathcal{W}}{\partial\phi_i}\right|_S\mathcal{F}_i-\frac{1}{2}\left.\frac{\partial^3 \mathcal{K}}{\partial\phi_k\partial\phi_i^\dagger\partial\phi_j^\dagger}\right|_S\bar{\psi}_i \psi_{Lj} \mathcal{F}_k-\frac{1}{4}\left.\frac{\partial \mathcal{H}_{AB}}{\partial\phi_i}\right|_S\bar{\lambda}_A\lambda_{BL} \mathcal{F}_i
        \\ 
        &+\frac{1}{4} \mathcal{H}_{AB}|_S \mathcal{D}_A\mathcal{D}_B-g\left.\frac{\partial \mathcal{K}}{\partial\phi_i}\right|_S  t_A S_i\mathcal{D}_A+\frac{i}{2\sqrt{2}}\left.\frac{\partial \mathcal{H}_{AB}}{\partial\phi_i}\right|_S \bar{\lambda}_B \psi_{Li} \mathcal{D}_A
        \\ & +\frac{1}{2}\left.\frac{\partial^3 \mathcal{K}}{\partial\phi_i\partial\phi_j\partial\phi_k^\dagger}\right|_S \bar{\psi}_j\gamma^\mu \psi_{Rk} D_\mu \phi_i-\frac{1}{4\sqrt{2}}\left.\frac{\partial \mathcal{H}_{AB}}{\partial\phi_i}\right|_S\bar{\lambda}_B\gamma^\mu\gamma^\nu \psi_{Li} F_{A\mu\nu}
        \\ &+\frac{1}{8}\left.\frac{\partial^4 \mathcal{K}}{\partial\phi_i\partial\phi_j\partial\phi_k^\dagger\partial\phi_l^\dagger}\right|_S(\bar{\psi}_i \psi_{Lj})(\bar{\psi}_l  \psi_{Rk})+\frac{1}{8}\left.\frac{\partial^2 \mathcal{H}_{AB}}{\partial\phi_i\partial\phi_j}\right|_S (\bar{\lambda}_A \lambda_{BL})(\bar{\psi}_{i}\psi_{Lj})\\
        &+{\rm h.c.} \,,
        \label{eq:fulla}
    \end{split}
\end{equation}
where the first and second line have the kinetic terms, the third line has terms with two fermions, which would generate fermion masses. The fourth line has the terms involving $\mathcal{F}$, the fifth line contains terms involving $\mathcal{D}$. These latter two lines would generate the scalar potential and further fermion masses after SUSY breaking. The sixth line has derivative couplings and the seventh one has four-fermion couplings. 

The auxiliary fields solve the equations
\begin{equation}
\begin{split}
    \left.\frac{\partial^2 \mathcal{K}}{\partial\phi_j^\dagger\partial\phi_i}\right|_S\mathcal{F}_i=&-\left.\frac{\partial \mathcal{W}}{\partial\phi_j^\dagger}\right|_S+\frac{1}{2}\left.\frac{\partial^3 \mathcal{K}}{\partial\phi_k\partial\phi_i\partial\phi_j^\dagger}\right|_S\bar{\psi}_k \psi_{Li} +\frac{1}{4}\left.\frac{\partial \mathcal{H}_{AB}}{\partial\phi_j^\dagger}\right|_S\bar{\lambda}_A\lambda_{LB},
    \\
   \mathcal{H}_{AB}|_S\  \mathcal{D}_B=&g\left.\frac{\partial \mathcal{K}}{\partial\phi_i}\right|_S  t_A S_i+g\left.\frac{\partial \mathcal{K}}{\partial\phi_i^\dagger}\right|_S  t_A S_i^\dagger-\frac{i}{2\sqrt{2}}\left.\frac{\partial \mathcal{H}_{AB}}{\partial\phi_i}\right|_S \bar{\lambda}_B \psi_L+\frac{i}{2\sqrt{2}}\left.\frac{\partial \mathcal{H}_{AB}}{\partial\phi_i^\dagger}\right|_S \bar{\psi}_i \lambda_{LB} \,.
  \end{split}
\end{equation}
Their VEVs, which would break SUSY, are then defined by
\begin{equation}
\begin{split}
    \left\langle\left.\frac{\partial^2 \mathcal{K}}{\partial\phi_j^\dagger\partial\phi_i}\right|_S\right\rangle\braket{\mathcal{F}_i}=-\left\langle\left.\frac{\partial \mathcal{W}}{\partial\phi_j^\dagger}\right|_S\right\rangle,\ \ \ \
   \left\langle\mathcal{H}_{AB}|_S\right\rangle\braket{ \mathcal{D}_B}=g\left\langle\left.\frac{\partial \mathcal{K}}{\partial\phi_i}\right|_S\right\rangle  t_A \braket{S_i}+g\left\langle\left.\frac{\partial \mathcal{K}}{\partial\phi_i^\dagger}\right|_S \right\rangle t_A \braket{S_i^\dagger} \,.
   \label{eq:susybreak}
  \end{split}
\end{equation}
Knowing that
\begin{equation}
\begin{split}
    \left.\frac{\partial^2 \mathcal{K}}{\partial\phi_j^\dagger\partial\phi_i}\right|_S=\ \delta^{ij}-\tilde{\kappa}^{ij}(S,S^\dagger)\,, \ \ \
     \mathcal{H}_{AB}|_S=\ \delta_{AB}-\tilde{\eta}_{AB}(S,S^\dagger)\,,
\end{split}
\end{equation}
one can use the geometric series expansion
\begin{equation}
    \begin{split}
        \frac{1}{\delta-\tilde{\kappa}}= \sum_{n=0}^\infty \tilde{\kappa}^n \,,\ \ \ \ \
        \frac{1}{\delta-\tilde{\eta}}=\sum_{n=0}^\infty \tilde{\eta}^n \,,
    \end{split}
\end{equation}
so that we have the polynomial solution
\begin{equation}
\begin{split}
    \mathcal{F}_k=&\sum_n(\tilde{\kappa}^n)_{kj}\left[-\left.\frac{\partial \mathcal{W}}{\partial\phi_j^\dagger}\right|_S+\frac{1}{2}\left.\frac{\partial^3 \mathcal{K}}{\partial\phi_s\partial\phi_i\partial\phi_j^\dagger}\right|_S\bar{\psi}_s \psi_{Li} +\frac{1}{4}\left.\frac{\partial \mathcal{H}_{AB}}{\partial\phi_j^\dagger}\right|_S\bar{\lambda}_A\lambda_{LB}\right] \,,\\
   \mathcal{D}_B=& \sum_n(\tilde{\eta}^n)_{AB}\left[g\left.\frac{\partial \mathcal{K}}{\partial\phi_i}\right|_S  t_A S_i+g\left.\frac{\partial \mathcal{K}}{\partial\phi_i^\dagger}\right|_S  t_A S_i^\dagger-\frac{i}{2\sqrt{2}}\left.\frac{\partial \mathcal{H}_{AC}}{\partial\phi_i}\right|_S \bar{\lambda}_C \psi_L+\frac{i}{2\sqrt{2}}\left.\frac{\partial \mathcal{H}_{AC}}{\partial\phi_i^\dagger}\right|_S \bar{\psi}_i \lambda_{LC}\right] \,,
   \label{eq:fdsol}
  \end{split}
\end{equation}
which can be plugged in \cref{eq:fulla} to obtain the full Lagrangian.

%%%%%%%%%%%%%%%%%%%%%%%%%%%%%%%%
 \section{$\E{8}$ generators}
 \label{app:e8gen}
 %%%%%%%%%%%%%%%%%%%%%%%%%%%%%%%%%%
 
The group $\E{8}$ can be fully described by the commutation algebra of its generators
The full list of 248 generators can be named just as the previously named  field representations $\textcolor{magenta}{\Delta_{C\ b}^{\ a}},
 \textcolor{magenta}{\Delta_{L\ b}^{\ a}},
 \textcolor{magenta}{\Delta_{R\ b}^{\ a}},
 \textcolor{magenta}{\Delta_{F\ b}^{\ a}},
 \textbf{X}^a_{\ bc\ },
 \overline{\textbf{X}}_a^{\ bc\ },
 \textcolor{ForestGreen}{\textbf{L}_{\ \ bc}^{\ a}},
 \textcolor{blue}{\textbf{Q}_{Lab\ c}},
 \textcolor{blue}{\textbf{Q}_{R\ \ c}^{\ a\ b}},
 \textcolor{orange}{\overline{\textbf{L}}^{\ \ bc}_{\ a}},
 \textcolor{red}{\overline{\textbf{Q}}^{\ ab\ c}_L},
 \textcolor{red}{\overline{\textbf{Q}}^{\ \ \ \ c}_{R a\ b}}$, noting that in this section, they are not 
 fields but $\E{8}$ generators. They have the following index convention
\begin{itemize}
 \item All indices run $a=1,2,3$ regardless of the corresponding $\SU{3}{}$.
 \item Upper index denotes antitriplet and lower index denotes triplet.
 \item The adjoint $\SU{3}{}$ generators $\textcolor{magenta}{\Delta}$ have an antitriplet-triplet index of the corresponding group.
 \item All the other generators have three of the four $\SU{3}{}$ indices always in the order $C,L,R,F$ and a blank space where a singlet is placed.
\end{itemize}
With these, the commutation relations can be written in Koca's convention
\cite{Koca:1981dr}
\begin{equation}
     \begin{array}{ll}
     [ \textcolor{magenta}{\Delta_{A\ j}^{\ i}},\textcolor{magenta}{\Delta_{B\ a}^{\ b}}]=(\delta^{b}_j\textcolor{magenta}{\Delta_{A\ a}^{\ i}}-\delta^{i}_a\textcolor{magenta}{\Delta_{A\ j}^{\ b}})\delta_{AB},\ \ \
     & 
      \ {\rm for}\ A,B=C,L,R,F, 
     \\ 
     \\
      {[}\textcolor{magenta}{\Delta_{C\ j}^{\ i}},\textbf{X}^a_{\ bc\ }{]}=\delta^a_j \textbf{X}^i_{\ bc\ }-\tfrac{1}{3}\delta_{ j}^{ i}\textbf{X}^a_{\ bc\ },
    & {[}\textcolor{magenta}{\Delta_{F\ j}^{\ i}},\textbf{X}^a_{\ bc\ }{]}=0,
    \\
    {[} \textcolor{magenta}{\Delta_{L\ j}^{\ i}},\textbf{X}^a_{\ bc\ }]=-\delta^i_b \textbf{X}^a_{\ jc\ }+\tfrac{1}{3}\delta_{ j}^{ i}\textbf{X}^a_{\ bc\ },
     &
     [ \textcolor{magenta}{\Delta_{R\ j}^{\ i}},\textbf{X}^a_{\ bc\ }]=-\delta^i_c \textbf{X}^a_{\ bj\ }+\tfrac{1}{3}\delta_{ j}^{ i}\textbf{X}^a_{\ bc\ },
     \\
     {[}\textbf{X}^a_{\ bc\ },\textbf{X}^i_{\ jk\ }]=-\epsilon_{dbj}\epsilon_{eck}\epsilon^{fai}\overline{\textbf{X}}_f^{\ de\ },
     &
     {[}\textbf{X}^a_{\ bc\ },\overline{\textbf{X}}_i^{\ jk\ }]=-\delta^a_i\delta^k_c \textcolor{magenta}{\Delta_{L\ b}^{\ j}}-\delta^a_i\delta_b^j\textcolor{magenta}{\Delta_{R\ c}^{\ k}}+\delta_b^j\delta_c^k\textcolor{magenta}{\Delta_{C\ i}^{\ a}},
     \\
     \\
      {[}\textcolor{magenta}{\Delta_{C\ j}^{\ i}},\textcolor{ForestGreen}{\textbf{L}_{\ \ bc}^{\ a}}]=0,
      &
       {[}\textcolor{magenta}{\Delta_{F\ j}^{\ i}},\textcolor{ForestGreen}{\textbf{L}_{\ \ bc}^{\ a}}]=-\delta^i_c \textcolor{ForestGreen}{\textbf{L}_{\ \ bj}^{\ a}}+\tfrac{1}{3}\delta_{ j}^{ i}\textcolor{ForestGreen}{\textbf{L}_{\ \ bc}^{\ a}},
       \\
      {[}\textcolor{magenta}{\Delta_{L\ j}^{\ i}},\textcolor{ForestGreen}{\textbf{L}_{\ \ bc}^{\ a}}]=\delta^a_j \textcolor{ForestGreen}{\textbf{L}_{\ \ bc}^{\ i}}-\tfrac{1}{3}\delta_{ j}^{ i}\textcolor{ForestGreen}{\textbf{L}_{\ \ bc}^{\ a}},
    & 
     {[}\textcolor{magenta}{\Delta_{R\ j}^{\ i}},\textcolor{ForestGreen}{\textbf{L}_{\ \ bc}^{\ a}}]=-\delta^i_b \textcolor{ForestGreen}{\textbf{L}_{\ \ jc}^{\ a}}+\tfrac{1}{3}\delta_{ j}^{ i}\textcolor{ForestGreen}{\textbf{L}_{\ \ bc}^{\ a}},
     \\
      {[}\textcolor{ForestGreen}{\textbf{L}_{\ \ bc}^{\ a}},\textcolor{ForestGreen}{\textbf{L}_{\ \ jk}^{\ i}}]=-\epsilon_{dbj}\epsilon_{eck}\epsilon^{fai}\textcolor{orange}{\overline{\textbf{L}}^{\ \ de}_{\ f}},
      &
      [\textcolor{ForestGreen}{\textbf{L}_{\ \ bc}^{\ a}},\textcolor{orange}{\overline{\textbf{L}}^{\ \ jk}_{\ i}}]=-\delta^a_i\delta^k_c \textcolor{magenta}{\Delta_{R\ b}^{\ j}}-\delta^a_i\delta_b^j\textcolor{magenta}{\Delta_{F\ c}^{\ k}}+\delta_b^j\delta_c^k\textcolor{magenta}{\Delta_{L\ i}^{\ a}},
      \\
      \\
       {[}\textcolor{magenta}{\Delta_{C\ j}^{\ i}},\textcolor{blue}{\textbf{Q}_{Lab\ c}}]=-\delta^i_a \textcolor{blue}{\textbf{Q}_{Ljb\ c}}+\tfrac{1}{3}\delta_{ j}^{ i}\textcolor{blue}{\textbf{Q}_{Lab\ c}},\ \
       &
        {[}\textcolor{magenta}{\Delta_{F\ j}^{\ i}},\textcolor{blue}{\textbf{Q}_{Lab\ c}}]=-\delta^i_c \textcolor{blue}{\textbf{Q}_{Lab\ j}}+\tfrac{1}{3}\delta_{ j}^{ i}\textcolor{blue}{\textbf{Q}_{Lab\ c}},
        \\
         {[}\textcolor{magenta}{\Delta_{L\ j}^{\ i}},\textcolor{blue}{\textbf{Q}_{Lab\ c}}]=-\delta^i_b \textcolor{blue}{\textbf{Q}_{Laj\ c}}+\tfrac{1}{3}\delta_{ j}^{ i}\textcolor{blue}{\textbf{Q}_{Lab\ c}},
         &
          {[}\textcolor{magenta}{\Delta_{R\ j}^{\ i}},\textcolor{blue}{\textbf{Q}_{Lab\ c}}]=0,
          \\
           {[}\textcolor{blue}{\textbf{Q}_{Lab\ c}},\textcolor{blue}{\textbf{Q}_{Lij\ k}}]=\epsilon_{dbj}\epsilon_{eck}\epsilon_{fai}\textcolor{red}{\overline{\textbf{Q}}_L^{\ de\ f}},
      &
      [\textcolor{blue}{\textbf{Q}_{Lab\ c}},\textcolor{red}{\overline{\textbf{Q}}_L^{\ ij\ k}}]=-\delta_a^i\delta^k_c \textcolor{magenta}{\Delta_{L\ b}^{\ j}}-\delta_a^i\delta_b^j\textcolor{magenta}{\Delta_{F\ c}^{\ k}}-\delta_b^j\delta_c^k\textcolor{magenta}{\Delta_{C\ a}^{\ i}},
      \\
      \\
       {[}\textcolor{magenta}{\Delta_{C\ j}^{\ i}},\textcolor{blue}{\textbf{Q}_{R\ \ c}^{\ a\ b}}]=\delta^a_j \textcolor{blue}{\textbf{Q}_{R\ \ c}^{\ i\ b}}-\tfrac{1}{3}\delta_{ j}^{ i}\textcolor{blue}{\textbf{Q}_{R\ \ c}^{\ a\ b}},\ \
       &
        {[}\textcolor{magenta}{\Delta_{F\ j}^{\ i}},\textcolor{blue}{\textbf{Q}_{R\ \ c}^{\ a\ b}}]=-\delta^i_c \textcolor{blue}{\textbf{Q}_{R\ \ c}^{\ a\ j}}+\tfrac{1}{3}\delta_{ j}^{ i}\textcolor{blue}{\textbf{Q}_{R\ \ c}^{\ a\ b}},
        \\
         {[}\textcolor{magenta}{\Delta_{L\ j}^{\ i}},\textcolor{blue}{\textbf{Q}_{R\ \ c}^{\ a\ b}}]=0,
         &
          {[}\textcolor{magenta}{\Delta_{R\ j}^{\ i}},\textcolor{blue}{\textbf{Q}_{R\ \ c}^{\ a\ b}}]=\delta^b_j\textcolor{blue}{\textbf{Q}_{R\ \ c}^{\ a\ i}}-\tfrac{1}{3}\delta^i_j\textcolor{blue}{\textbf{Q}_{R\ \ c}^{\ a\ b}},
          \\
           {[}\textcolor{blue}{\textbf{Q}_{R\ \ c}^{\ a\ b}},\textcolor{blue}{\textbf{Q}_{R\ \ k}^{\ i\ j}}]=\epsilon^{dbj}\epsilon^{eck}\epsilon_{fai}\textcolor{red}{\overline{\textbf{Q}}^{\ \ \ \ f}_{R d\ e}},
      &
      [\textcolor{blue}{\textbf{Q}_{R\ \ c}^{\ a\ b}},\textcolor{red}{\overline{\textbf{Q}}^{\ \ \ \ k}_{R i\ j}}]=\delta^a_i\delta^k_c \textcolor{magenta}{\Delta_{L\ j}^{\ b}}-\delta^a_i\delta^b_j\textcolor{magenta}{\Delta_{F\ c}^{\ k}}+\delta^b_j\delta_c^k\textcolor{magenta}{\Delta_{C\ i}^{\ a}},
      \\
      \\
      {[}\textcolor{ForestGreen}{\textbf{L}_{\ \ jk}^{\ i}},  \textcolor{blue}{\textbf{Q}_{Lab\ c}}]=\delta^i_b\epsilon_{kcd} \textcolor{red}{\overline{\textbf{Q}}^{\ \ \ \ d}_{R a\ j}},
      &
      [\textcolor{ForestGreen}{\textbf{L}_{\ \ jk}^{\ i}}, \textcolor{blue}{\textbf{Q}_{R\ \ c}^{\ a\ b}}]=\delta^b_j\epsilon_{kcd} \textcolor{red}{\overline{\textbf{Q}}^{\ ai\ d}_L},
      \\
      {[}\textcolor{ForestGreen}{\textbf{L}_{\ \ jk}^{\ i}},\textbf{X}^a_{\ bc\ }]=\delta^i_b\epsilon_{jcd}  \textcolor{blue}{\textbf{Q}_{R\ \ k}^{\ a\ d}},
      &
      {[}\textcolor{ForestGreen}{\textbf{L}_{\ \ jk}^{\ i}}, \textcolor{red}{\overline{\textbf{Q}}^{\ ab\ c}_L}]=-\delta^c_k\epsilon^{ibd}\textbf{X}^a_{\ dj\ },
      \\
      {[}\textcolor{ForestGreen}{\textbf{L}_{\ \ jk}^{\ i}},\textcolor{red}{\overline{\textbf{Q}}^{\ \ \ \ c}_{R a\ b}}=\delta^c_k\epsilon_{jbd} \overline{\textbf{X}}_a^{\ id\ },
      &
      {[}\textcolor{ForestGreen}{\textbf{L}_{\ \ jk}^{\ i}},\overline{\textbf{X}}_a^{\ bc\ }]=-\delta^c_j \epsilon^{ibd}\textcolor{blue}{\textbf{Q}_{Lad\ k}},
      \\
      {[}\textcolor{blue}{\textbf{Q}_{Lij\ k}}, \textcolor{blue}{\textbf{Q}_{R\ \ c}^{\ a\ b}}]=\delta^a_i\epsilon_{kcd} \textcolor{orange}{\overline{\textbf{L}}^{\ \ bd}_{\ j}},
      &
      {[}\textcolor{blue}{\textbf{Q}_{Lij\ k}},\textbf{X}^a_{\ bc\ }]=\delta^a_i\epsilon_{jbd} \textcolor{ForestGreen}{\textbf{L}_{\ \ ck}^{\ d}},
      \\
      {[}\textcolor{blue}{\textbf{Q}_{Lij\ k}},\textcolor{red}{\overline{\textbf{Q}}^{\ \ \ \ c}_{R a\ b}}]=\delta_k^c \epsilon_{iad}\textbf{X}^d_{\ jb\ },
      &
      {[}\textcolor{blue}{\textbf{Q}_{Lij\ k}},\overline{\textbf{X}}_a^{\ bc\ }]=\delta_j^b \epsilon_{iad} \textcolor{blue}{\textbf{Q}_{R\ \ k}^{\ d\ c}},
      \\
     {[}\textcolor{blue}{\textbf{Q}_{R\ \ k}^{\ i\ j}},\textbf{X}^a_{\ bc\ }]=-\delta^j_c\epsilon^{iad} \textcolor{blue}{\textbf{Q}_{Ldb\ k}},
     &
     {[}\textcolor{blue}{\textbf{Q}_{R\ \ k}^{\ i\ j}},\overline{\textbf{X}}_a^{\ bc\ }]=-\delta^i_a\epsilon^{jcd}\textcolor{ForestGreen}{\textbf{L}_{\ \ dk}^{\ b}},
     \label{eq:e8comm}
     \end{array}
     \end{equation}
which, together with their complex conjugate relations, determine completely the corresponding Lie algebra.

%%%%%%%%%%%%%%%%%%%%%%%%%%%%%%%%%%%%%%%%%%%
\section{Detailed Lagrangian of the model}
\label{app:modlag}
%%%%%%%%%%%%%%%%%%%%%%%%%%%%%%%%%%%%%%%%%%%%

The superpotential was described in \cref{eq:quyu}. Due to the non renormalization theorem, it does not receive any loop contributions. The SUSY breaking contributions from \cref{eq:dbreak} do affect the K\"ahler and $\mathcal{H}$ potential. In this section, these are worked out at the compactification scale.

One can define the tensors\footnote{The index conventions are: 
\begin{itemize}
    \item Color indices are not shown explicitly.
    \item Last index is always the family index.
    \item Where there are three sets of indices, the first one is left and the second one is right, the final one is family.
    \item Upper index denotes antitriplet and lower index denotes triplet.
\end{itemize}}
\begin{equation}\begin{split}
    \textbf{K}^{i\ \ q\ \ r}_{\ p\ \ m\ \ n}=&\textcolor{ForestGreen}{(\textbf{L}^\dagger)_a^{\ bc}} \textcolor{magenta}{(e^{-2V})^{a\ \ q\ \ r}_{\ p\ \ b\ \ c}}\textcolor{ForestGreen}{\textbf{L}^i_{\ mn}}\\
    &+
   \delta^q_m \textcolor{blue}{(\textbf{Q}^\dagger_{L})^{ab}}  \textcolor{magenta}{(e^{-2V})^{i\ \ r}_{\ a\ \ b}} \textcolor{blue}{\textbf{Q}_{Lpn}}/3\\
    &+
    \delta^i_p \textcolor{blue}{(\textbf{Q}_R^\dagger)^{\ a}_{ b}} \textcolor{magenta}{(e^{-2V})^{ b\ \ r}_{\ \ m\ \ a}} \textcolor{blue}{\textbf{Q}_{R\ n}^{\ q}}/3 \,,
    \\
    \widetilde{\textbf{K}}^{i\ \ q\ \ r}_{\ p\ \ m\ \ n}=&\textbf{K}^{i\ \ q\ \ r}_{\ p\ \ m\ \ n}+\textbf{K}^{a\ \ q\ \ r}_{\ a\ \ m\ \ n}\delta^i_p+\textbf{K}^{i\ \ a\ \ r}_{\ p\ \ a\ \ n}\delta^q_m+\textbf{K}^{i\ \ q\ \ a}_{\ p\ \ m\ \ a}\delta^r_n
    \\
    &+\textbf{K}^{i\ \ a\ \ b}_{\ p\ \ a\ \ b}\delta^q_m\delta^r_n+\textbf{K}^{a\ \ q\ \ b}_{\ a\ \ m\ \ b}\delta^i_p\delta^r_n+\textbf{K}^{a\ \ b\ \ r}_{\ a\ \ b\ \ n}\delta^i_p\delta^q_m
    +\textbf{K}^{a\ \ b\ \ c}_{\ a\ \ b\ \ c}\delta^i_p\delta^q_m\delta^r_n\,,
    \end{split}
\end{equation}
so that the full K\"ahler potential is
\begin{equation}
    \begin{split}
        \mathcal{K}\sim& \ \textbf{K}^{i\ \ m\ \ n}_{\ i\ \ m\ \ n}+\tfrac{1}{\Lambda^2}\widetilde{\textbf{K}}^{i\ \ q\ \ r}_{\ p\ \ m\ \ n}\widetilde{\textbf{K}}^{p\ \ m\ \ n}_{\ i\ \ q\ \ r}
        \\
        &
        +\tfrac{1}{\Lambda^3}\epsilon_{ijk}\epsilon^{lmn}\ \textcolor{ForestGreen}{\textbf{L}^i_{\ l(o-1)}} 
        \widetilde{\textbf{K}}^{j\ \ b\ \ p}_{\ a\ \ m\ \ o} \textcolor{ForestGreen}{\textbf{L}^a_{\ bp}}\textcolor{ForestGreen}{\textbf{L}^k_{\ n(o+1)} }
        \\
        &
        +\tfrac{1}{\Lambda^3}\epsilon_{ijk}\ \textcolor{blue}{\textbf{Q}_{R\ (o-1)}^{\ i}} \widetilde{\textbf{K}}^{s\ \ j\ \ p}_{\ s\ \ n\ \ o} \textcolor{blue}{\textbf{Q}_{R\ p}^{\ n}} \textcolor{blue}{\textbf{Q}_{R\ (o+1)}^{\ k}}
        \\
        &
        +\tfrac{1}{\Lambda^3}\epsilon^{ijk}\ \textcolor{blue}{\textbf{Q}_{Li(o-1)}}  \widetilde{\textbf{K}}^{a\ \ s\ \ p}_{\ j\ \ s\ \ o} \textcolor{blue}{\textbf{Q}_{Lap}}  \textcolor{blue}{\textbf{Q}_{Lk(o+1)}}
        \\
        &
        + \tfrac{1}{\Lambda^3}\textcolor{ForestGreen}{\textbf{L}^l_{\ m a}}\textcolor{blue}{\textbf{Q}_{R\ b}^{\ n}}\textcolor{blue}{\textbf{Q}_{Lpc}} \Big[\widetilde{\textbf{K}}^{p\ \ m\ \ a}_{\ l\ \ n\ \ (o+1)}\delta^b_{o-1}\delta^c_o+\widetilde{\textbf{K}}^{p\ \ m\ \ b}_{\ l\ \ n\ \ (o+1)}\delta^c_{o+1}\delta^a_o+\widetilde{\textbf{K}}^{p\ \ m\ \ c}_{\ l\ \ n\ \ (o+1)}\delta^a_{o-1} \delta^b_o
    \\ &+{\rm h.c.}+\mathcal{O}(1/\Lambda^4) \,,
    \label{eq:kah}
    \end{split}
\end{equation}
where the dimensionless constants have been ignored and the index $o=1,2,3$ is summed and $o=0\to 3,\ o=4\to 1$ . The possible $SU(3)_C$ adjoint tensor $\textbf{K}^{a\ \ b\ \ c}_{\ a\ \ b\ \ c}$ has also been ignored. The scale $\Lambda$ is the KK mass scale. The full superpotential was already written in \cref{eq:quyu}.

The K\"ahler potential in \cref{eq:kah} can be truncated to the relevant low energy terms. The first term contains the scalar and fermion kinetic terms. In any other term, the function $\textbf{K}$ can be approximated to only the $\textcolor{ForestGreen}{\textbf{L}}$ part, which involves Wilson line effective VEVs.
The next two terms would only generate subleading corrections and may be ignored. The second line contains lepton and Higgs interactions. The third and fourth line may contain proton decay terms of the form $QQQL$; they are studied in \cref{sec:prodec}. The last lines would bring corrections to the quark masses. Defining 
\begin{equation}
\begin{split}
\textcolor{ForestGreen}{(\tilde{\Delta}_L)^i_{\ j}}=\textcolor{ForestGreen}{\textbf{L}^i_{\ mn}(\textbf{L}^\dagger)_j^{\ mn}},\ \ \ 
\textcolor{ForestGreen}{(\tilde{\Delta}_R)^i_{\ j}}=\textcolor{ForestGreen}{(\textbf{L}^\dagger)_m^{\ in}\textbf{L}^m_{\ jn}},\ \ \ 
\textcolor{ForestGreen}{(\tilde{\Delta}_F)^i_{\ j}}=&\textcolor{ForestGreen}{(\textbf{L}^\dagger)_m^{\ ni}\textbf{L}^m_{\ nj}},
\end{split}
\end{equation}
one can define the truncated K\"ahler potential
\begin{equation}
    \begin{split}
        \mathcal{K}'=&  \ \textcolor{ForestGreen}{(\textbf{L}^\dagger)_a^{\ bc}} \textcolor{magenta}{(e^{-2V})^{a\ \ q\ \ r}_{\ p\ \ b\ \ c}}\textcolor{ForestGreen}{\textbf{L}^p_{\ qr}}+
  \textcolor{blue}{(\textbf{Q}^\dagger_{L})^{ab}}  \textcolor{magenta}{(e^{-2V})^{i\ \ r}_{\ a\ \ b}} \textcolor{blue}{\textbf{Q}_{Lir}}+
    \textcolor{blue}{(\textbf{Q}_R^\dagger)^{\ a}_{ b}} \textcolor{magenta}{(e^{-2V})^{ b\ \ r}_{\ \ m\ \ a}} \textcolor{blue}{\textbf{Q}_{R\ r}^{\ m}}
    \\ 
    &
    + \tfrac{1}{\Lambda^2}(\textcolor{ForestGreen}{\textbf{L}^a_{\ bc}} \textcolor{ForestGreen}{(\textbf{L}^\dagger)_a^{\ bc}})^2 
    \\
        &
        +\tfrac{1}{\Lambda^3}\epsilon_{ijk}\epsilon^{lmn}\ \textcolor{ForestGreen}{\textbf{L}^i_{\ l(o-1)}}
        \left[
       \textcolor{ForestGreen}{(\tilde{\Delta}_L)^j_{\ a}}  \textcolor{ForestGreen}{\textbf{L}^a_{\ mo}} 
       \textcolor{ForestGreen}{\textbf{L}^k_{\ n(o+1)} }+
        \textcolor{ForestGreen}{\textbf{L}^j_{\ mo}} \textcolor{ForestGreen}{(\tilde{\Delta}_F)^f_{\ (o+1)}}
        \textcolor{ForestGreen}{\textbf{L}^k_{\ nf} }+
        \textcolor{ForestGreen}{(\tilde{\Delta}_R)^b_{\ m}} \textcolor{ForestGreen}{\textbf{L}^j_{\ bo}} 
       \textcolor{ForestGreen}{\textbf{L}^k_{\ n(o+1)} }
        \right]
        \\
        &
        +\tfrac{1}{\Lambda^3} 
        \Big[
        \textcolor{ForestGreen}{\textbf{L}^l_{\ n (o-1)}}\textcolor{blue}{\textbf{Q}_{R\ o}^{\ n}}\textcolor{blue}{\textbf{Q}_{Lp(o+1)}}\textcolor{ForestGreen}{(\Delta_L)^p_{\ l}}
        +\textcolor{ForestGreen}{\textbf{L}^p_{\ m (o-1)}}\textcolor{blue}{\textbf{Q}_{R\ o}^{\ n}}\textcolor{blue}{\textbf{Q}_{Lp(o+1)}}\textcolor{ForestGreen}{(\Delta_R)^m_{\ n}}
        \\
        &\quad\quad +\textcolor{ForestGreen}{\textbf{L}^p_{\ n a}}\textcolor{blue}{\textbf{Q}_{R\ o}^{\ n}}\textcolor{blue}{\textbf{Q}_{Lp(o+1)}}\textcolor{ForestGreen}{(\Delta_F)^a_{\ (o-1)}}
        +
        \textcolor{ForestGreen}{\textbf{L}^p_{\ m i(o-1)}}\textcolor{blue}{\textbf{Q}_{R\ b}^{\ m}}\textcolor{blue}{\textbf{Q}_{Lp(o+1)}}\textcolor{ForestGreen}{(\Delta_F)^b_{\ o}}
        +
        \textcolor{ForestGreen}{\textbf{L}^l_{\ n (o-1)}}\textcolor{blue}{\textbf{Q}_{R\ o}^{\ n}}\textcolor{blue}{\textbf{Q}_{Llc}}\textcolor{ForestGreen}{(\Delta_F)^c_{\ (o+1)}}\Big]
    \\ &+{\rm h.c.} \,,
    \end{split}
    \label{eq:kappap}
\end{equation}
where the subleading terms may be ignored and the index $o=1,2,3$ is summed and $o=0\to 3,\ o=4\to 1$.
This potential generate the Lagrangian to be studied in this article.

The SUSY breaking contributions, which are proportional to $\braket{\mathcal{D}}$, can be written in terms of three different effective adjoint VEVs
\begin{equation}
\begin{split}
\braket{\textcolor{ForestGreen}{(\tilde{\Delta}_L)^i_{\ j}}}=&\braket{\textcolor{ForestGreen}{\textbf{L}^i_{\ mn}(\textbf{L}^\dagger)_j^{\ mn}}}-\braket{\textcolor{ForestGreen}{\textbf{L}^l_{\ mn}(\textbf{L}^\dagger)_l^{\ mn}}}\delta^i_j/3\\
=& \textcolor{Sepia}{\textbf{w}^2}\ (\delta^i_3\delta^3_j-\delta^i_j/3),\\
\braket{\textcolor{ForestGreen}{(\tilde{\Delta}_R)^i_{\ j}}}=&\braket{\textcolor{ForestGreen}{(\textbf{L}^\dagger)_m^{\ in}\textbf{L}^m_{\ jn}}}-\braket{\textcolor{ForestGreen}{\textbf{L}^l_{\ mn}(\textbf{L}^\dagger)_l^{\ mn}}}\delta^i_j/3\\
=&\delta^i_1\delta^1_j \braket{\textcolor{violet}{\tilde{\nu}^{c\dagger k}\tilde{\nu}^c_k}}
+\delta^i_1\delta^3_j\braket{\textcolor{violet}{\tilde{\nu}^{c\dagger k}}\textcolor{brown}{\varphi_k}}
+\delta^i_3\delta_j^1 \braket{\textcolor{brown}{\varphi^{\dagger k}}\textcolor{violet}{\tilde{\nu}^{c }_k}}
+\delta^i_3\delta_j^3 \braket{\textcolor{brown}{\varphi^{\dagger k}}\textcolor{brown}{\varphi_k}}-\textcolor{Sepia}{\textbf{w}^2}\delta^i_j/3,\\
\braket{\textcolor{ForestGreen}{(\tilde{\Delta}_F)^i_{\ j}}}=&\braket{\textcolor{ForestGreen}{(\textbf{L}^\dagger)_m^{\ ni}\textbf{L}^m_{\ nj}}}-\braket{\textcolor{ForestGreen}{\textbf{L}^l_{\ mn}(\textbf{L}^\dagger)_l^{\ mn}}}\delta^i_j/3\\
=&\textcolor{Sepia}{(\textbf{w}^2)^i_{\ j} }-\textcolor{Sepia}{\textbf{w}^2 }\delta^i_j/3 \,, \\
\label{eq:adj}
\end{split}
\end{equation}
The SUSY breaking Wilson lines also break the gauge symmetry. They generate the broken gaugino mass matrix
\begin{equation}
    \textcolor{magenta}{\lambda_A} M^\dagger_A M_B \textcolor{magenta}{\lambda_B}=\textcolor{magenta}{\lambda_A\lambda_B}\left(\braket{\textcolor{ForestGreen}{\textbf{L}}^\dagger}t_A t_B\braket{\textcolor{ForestGreen}{\textbf{L}}}\right) \,,
\end{equation}
where $t_{\tilde{a}}$ with $\tilde{a}=1,...,8$ are the Gell-Mann matrices which can belong to any $\SU{3}{C,L,R,F}$. Their product can be defined as
\begin{equation}
    [t_{\tilde{a}},t_{\tilde{b}}]=2if^{\ \ \tilde{c}}_{\tilde{a}\tilde{b}} t_{\tilde{c}},\ \ \ \{t_{\tilde{a}},t_{\tilde{b}}\}=\frac{4}{3}\delta_{\tilde{a}\tilde{b}}+2d^{\ \ \tilde{c}}_{\tilde{a}\tilde{b}} t_{\tilde{c}} \,.
\end{equation}
The squared broken gaugino mass matrix can then be written as
\begin{equation}
\begin{split}
       \textcolor{magenta}{\lambda_A}& M^\dagger_A M_B\textcolor{magenta}{\lambda_B}=
       \\
       &\textcolor{magenta}{\lambda^F_{\tilde{a}}\lambda^F_{\tilde{b}}}\left[\frac{2}{3}\textcolor{Sepia}{\textbf{w}^2}\delta_{\tilde{a}\tilde{b}}+d^{\ \ \tilde{c}}_{\tilde{a}\tilde{b}}(\braket{\textcolor{violet}{\tilde{\nu}^{c\dagger i}}}t_{\tilde{c}i}^{\ \ j}\braket{\textcolor{violet}{\tilde{\nu}^{c }_j}}+\braket{\textcolor{brown}{{\varphi}^{\dagger  i}}}t_{\tilde{c}i}^{\ \ j}\braket{\textcolor{brown}{\varphi_j}})\right]\Big|_{\tilde{a},\tilde{b},\tilde{c}=3,8}
       \\
       &
       +\textcolor{magenta}{\lambda^R_{\tilde{a}}\lambda^R_{\tilde{b}}}\left[\braket{\textcolor{brown}{\varphi^{\dagger i}\varphi_i}}\left(\delta^4_{\tilde{a}
       }\delta^4_{\tilde{b}}+\delta^5_{\tilde{a}}\delta^5_{\tilde{b}}+\delta^6_{\tilde{a}}\delta^6_{\tilde{b}}+\delta^7_{\tilde{a}}\delta^7_{\tilde{b}}+\frac{4}{3}\delta^8_{\tilde{a}}\delta^8_{\tilde{b}}\right)\right.
       \\
       &
       \quad \quad \quad +\braket{\textcolor{violet}{\tilde{\nu}^{c\dagger i}\tilde{\nu}^c_i}}\left(\delta^1_{\tilde{a}
       }\delta^1_{\tilde{b}}+\delta^2_{\tilde{a}}\delta^2_{\tilde{b}}+\delta^3_{\tilde{a}}\delta^3_{\tilde{b}}+\delta^4_{\tilde{a}}\delta^4_{\tilde{b}}+\delta^5_{\tilde{a}}\delta^5_{\tilde{b}}+\frac{2}{3}\delta^8_{\tilde{a}}\delta^8_{\tilde{b}}+\frac{2}{\sqrt{3}}\delta^{(3}_{\tilde{a}}\delta^{8)}_{\tilde{b}}\right)
       \\
       &
       \quad \quad \quad +\braket{\textcolor{violet}{\tilde{\nu}^{c\dagger i}\textcolor{brown}{\varphi_i}}}\left(\delta^{(1}_{\tilde{a}}\delta^{6)}_{\tilde{b}}+i\delta^{(1}_{\tilde{a}}\delta^{7)}_{\tilde{b}}+i\delta^{(2}_{\tilde{a}}\delta^{6)}_{\tilde{b}}+\delta^{(2}_{\tilde{a}}\delta^{7)}_{\tilde{b}}+\delta^{(3}_{\tilde{a}}\delta^{4)}_{\tilde{b}}+i\delta^{(3}_{\tilde{a}}\delta^{5)}_{\tilde{b}}-\frac{1}{\sqrt{3}}\delta^{(4}_{\tilde{a}}\delta^{8)}_{\tilde{b}}-i\frac{1}{\sqrt{3}}\delta^{(5}_{\tilde{a}}\delta^{8)}_{\tilde{b}}
       \right)
       \\
       &
       \quad \quad \quad \left.+\braket{\textcolor{brown}{\varphi^{\dagger i}}\textcolor{violet}{\tilde{\nu}^{c }_i}}\left(\delta^{(1}_{\tilde{a}}\delta^{6)}_{\tilde{b}}-i\delta^{(1}_{\tilde{a}}\delta^{7)}_{\tilde{b}}-i\delta^{(2}_{\tilde{a}}\delta^{6)}_{\tilde{b}}+\delta^{(2}_{\tilde{a}}\delta^{7)}_{\tilde{b}}+\delta^{(3}_{\tilde{a}}\delta^{4)}_{\tilde{b}}-i\delta^{(3}_{\tilde{a}}\delta^{5)}_{\tilde{b}}-\frac{1}{\sqrt{3}}\delta^{(4}_{\tilde{a}}\delta^{8)}_{\tilde{b}}+i\frac{1}{\sqrt{3}}\delta^{(5}_{\tilde{a}}\delta^{8)}_{\tilde{b}}
       \right)\right]
       \\
       &
       +\textcolor{magenta}{\lambda^L_{\tilde{a}}\lambda^L_{\tilde{b}}}\left[\textcolor{Sepia}{\textbf{w}^2}\left(\delta^4_{\tilde{a}
       }\delta^4_{\tilde{b}}+\delta^5_{\tilde{a}}\delta^5_{\tilde{b}}+\delta^6_{\tilde{a}}\delta^6_{\tilde{b}}+\delta^7_{\tilde{a}}\delta^7_{\tilde{b}}+\frac{4}{3}\delta^8_{\tilde{a}}\delta^8_{\tilde{b}}\right)\right]
       \\
       &
       +\textcolor{magenta}{\lambda^L_{\tilde{a}}\lambda^R_{\tilde{b}}}\left[
       \braket{\textcolor{violet}{\tilde{\nu}^{c\dagger i}}\textcolor{violet}{\tilde{\nu}^{c }_i}}\frac{4}{3}\delta^8_{\tilde{a}}\delta^8_{\tilde{b}}
       -\frac{2}{\sqrt{3}}\braket{\textcolor{brown}{{\varphi}^{\dagger  i}}\textcolor{brown}{\varphi_i}}\left( \delta^3_{\tilde{b}}+\frac{1}{\sqrt{3}}\delta^8_{\tilde{b}} \right)\right.
       \\
       &\quad\quad\quad\left.
       -\braket{\textcolor{brown}{{\varphi}^{\dagger  i}}\textcolor{violet}{\tilde{\nu}^{c }_i}}\frac{2}{\sqrt{3}}\delta^8_{\tilde{a}}\left(\delta^4_{\tilde{a}}+i\delta^5_{\tilde{a}}\right)-\braket{\textcolor{violet}{\tilde{\nu}^{c\dagger i}}\textcolor{brown}{\varphi_i}}\frac{2}{\sqrt{3}}\delta^8_{\tilde{a}}\left(\delta^4_{\tilde{a}}-i\delta^5_{\tilde{a}}\right)
       \right]
       \\
       &
       +\textcolor{magenta}{\lambda^L_{\tilde{a}}\lambda^F_{\tilde{b}}}\left[-\frac{2}{\sqrt{3}}\delta^8_{\tilde{a}}\left(\braket{\textcolor{violet}{\tilde{\nu}^{c\dagger i}}}t_{\tilde{b}i}^{\ \ j}\braket{\textcolor{violet}{\tilde{\nu}^{c }_j}}+\braket{\textcolor{brown}{{\varphi}^{\dagger  i}}}t_{\tilde{b}i}^{\ \ j}\braket{\textcolor{brown}{\varphi_j}}\right)\right]\Big|_{\tilde{b}=3,8}
       \\
       &
       +\textcolor{magenta}{\lambda^R_{\tilde{a}}\lambda^F_{\tilde{b}}}\left[\braket{\textcolor{violet}{\tilde{\nu}^{c\dagger i}}}\left( \delta^3_{\tilde{a}}+\frac{1}{\sqrt{3}}\delta^8_{\tilde{a}} \right)t_{\tilde{b}i}^{\ \ j}\braket{\textcolor{violet}{\tilde{\nu}^{c }_j}}-\frac{2}{\sqrt{3}}\braket{\textcolor{brown}{{\varphi}^{\dagger  i}}}\delta^8_{\tilde{a}}t_{\tilde{b}i}^{\ \ j}\braket{\textcolor{brown}{\varphi_j}}\right.
       \\
       &\quad\quad\quad\left.
       +\braket{\textcolor{brown}{{\varphi}^{\dagger  i}}}\left(\delta^4_{\tilde{a}}-i\delta^5_{\tilde{a}}\right)t_{\tilde{b}i}^{\ \ j}\braket{\textcolor{violet}{\tilde{\nu}^{c }_j}}+\braket{\textcolor{violet}{\tilde{\nu}^{c\dagger i}}}\left(\delta^4_{\tilde{a}}+i\delta^5_{\tilde{a}}\right)t_{\tilde{b}i}^{\ \ j}\braket{\textcolor{brown}{\varphi_j}}
       \right]\Big|_{\tilde{b}=3,8} \,.
\end{split}
\label{eq:gauginomass}
\end{equation}
This matrix $M_A^\dagger M_B$ is supersymmetric, therefore this is also the mass matrix for the broken gauge vectors.

%%%%%%%%%%%%%%%%%%%%%%%%%%%%%%%%%%%%%%%%%%%%%%%%%%%%%%%%%%%%%%%%%%%%%%
\section{Effective field content at each different scale}
\label{app:ensca}
%%%%%%%%%%%%%%%%%%%%%%%%%%%%%%%%%%%%%%%%%%%%%%%%%%%%%%%%%%%%%%%%%%%%%%

In what follows, the coefficients of the beta-functions for each of the energy regimes will be determined. 

%%%%%%%%%%%%%%%%%%%%%%%%%%%%%%%%%%%%%%%%%%%%%%%%%%%%%%%%%%%%%%%%%%%%%%
\subsection*{Running between $\Lambda_6$ and $\Lambda_5$ - Region I}
%%%%%%%%%%%%%%%%%%%%%%%%%%%%%%%%%%%%%%%%%%%%%%%%%%%%%%%%%%%%%%%%%%%%%%

The lowest-scale theory, denoted as region-I, is defined by $\Lambda_6$ which is the Electroweak scale. According to \cref{tab:smf}, the field content is exactly that of the SM one (plus one RHN) in the low-scale spectrum.
\begin{table}[h] 
    \centering
    \footnotesize
   \begin{tabular}{l r}
$\SO{1,3}{}$ & \hspace{6cm} $\SU{3}{C}\times \SU{2}{L}\times \U{Y}$\\
\hline
Massless complex scalars   & $ \textcolor{OliveGreen}{(\textbf{1},{\textbf{2}},3)}$\\
\hline
Massless Weyl fermions &   $	3\times\textcolor{blue}{(\textbf{1},\textbf{2},-3)}
+3\times\textcolor{blue}{(\textbf{3},\textbf{2},1)}
+3\times\textcolor{blue}{(\bar{\textbf{3}},\textbf{1},-4)}+3\times\textcolor{blue}{(\bar{\textbf{3}},\textbf{1},2)}+ 3\times\textcolor{blue}{(\textbf{1},\textbf{1},6)}+\textcolor{Sepia}{(\textbf{1},{\textbf{1}},0)}$\\
\hline
Massless real vectors  & $ \textcolor{magenta}{(\textbf{8},\textbf{1},0)}+ \textcolor{magenta}{(\textbf{1},\textbf{3},0)}+ \textcolor{magenta}{(\textbf{1},\textbf{1},0)}$ \\
\hline
\end{tabular}
    \caption{Field content in Region I}
    \label{tab:smf}
\end{table}
Knowing that for $\SU{3}{}$ 
    \begin{equation}
        T\left(\bm{8}\right) = C_2\left(\bm{8}\right) = 3\,, \qquad
        T\left(\bm{3}\right) = \frac12\,,
    \end{equation}
for $\SU{2}{}$
    \begin{equation}
        T\left(\bm{3}\right) = C_2\left(\bm{3}\right) = 2\,, \qquad
        T\left(\bm{2}\right) = \frac12\,,
    \end{equation}
and considering the Abelian charges and respective multiplicities in \cref{tab:smf}, one obtains
\begin{equation}
        b^\mathrm{I}_\mathrm{3} = -7\,, \qquad b^\mathrm{I}_\mathrm{2} = -\frac{19}{6}\,, \qquad
        b^{' \mathrm{I}}_\mathrm{1} = \frac{41}{10}\,.
        \label{coeff:I} 
    \end{equation}

%%%%%%%%%%%%%%%%%%%%%%%%%%%%%%%%%%%%%%%%%%%%%%%%%%%%%%%%%%%%%%%%%%%%%%%
\subsection*{Running between $\Lambda_5$ and $\Lambda_4$ - Region II}  
%%%%%%%%%%%%%%%%%%%%%%%%%%%%%%%%%%%%%%%%%%%%%%%%%%%%%%%%%%%%%%%%%%%%%%%

The next threshold scale,  introduces the fields that obtain a mass at the lowest Wilson line scale $\Lambda_5= g/R_3$ (with the gauge coupling evaluated at the corresponding scale). 

\begin{table}[h]
    \centering
    \footnotesize
   \begin{tabular}{l r}
$\SO{1,3}{}$ & \hspace{6cm} $\SU{3}{C}\times \SU{2}{L}\times \U{Y}$\\

\hline
Massless complex scalars   & $ \textcolor{OliveGreen}{(\textbf{1},{\textbf{2}},3)}+\textcolor{blue}{(\textbf{3},\textbf{2},1)}+\textcolor{blue}{(\bar{\textbf{3}},\textbf{1},-4)}+\textcolor{blue}{(\textbf{1},\textbf{1},6)}$\\
\hline
Massless Weyl fermions &  $ \textcolor{magenta}{(\textbf{8},\textbf{1},0)}+ \textcolor{magenta}{(\textbf{1},\textbf{3},0)}+  \textcolor{OliveGreen}{(\textbf{1},{\textbf{2}},3)}+4 \times \textcolor{ForestGreen}{(\textbf{1},{\textbf{2}},-3)}+5\times \textcolor{Sepia}{(\textbf{1},{\textbf{1}},0)}$ 
\\
  & $
+ 3\times\textcolor{blue}{(\textbf{1},\textbf{1},6)}+3\times\textcolor{blue}{(\textbf{3},\textbf{2},1)}
+3\times\textcolor{blue}{(\bar{\textbf{3}},\textbf{1},-4)}+4\times\textcolor{blue}{(\bar{\textbf{3}},\textbf{1},2)}+\textcolor{red}{({\textbf{3}},\textbf{1},2)}$\\
\hline
Massless real vectors  & $ \textcolor{magenta}{(\textbf{8},\textbf{1},0)}+ \textcolor{magenta}{(\textbf{1},\textbf{3},0)}+ 2\times\textcolor{magenta}{(\textbf{1},\textbf{1},0)}$ \\
\hline
\end{tabular}
    \caption{Field content in Region II}
    \label{tab:tri}
\end{table}
Taking into account the particle content in \cref{tab:tri}, one obtains the following coefficients
\begin{equation}
        b^\mathrm{II}_\mathrm{3} = -\frac{23}{6}\,, \qquad b^\mathrm{II}_\mathrm{2} = -\frac23\,, \qquad
        b^{\prime \mathrm{II}}_\mathrm{1} = \frac{79}{15}\,.
        \label{coeff:III}
    \end{equation}

%%%%%%%%%%%%%%%%%%%%%%%%%%%%%%%%%%%%%%%%%%%%%%%%%%%%%%%%%%%%%%
\subsection*{RG evolution between $\Lambda_4$ and $\Lambda_3$ - Region III}
%%%%%%%%%%%%%%%%%%%%%%%%%%%%%%%%%%%%%%%%%%%%%%%%%%%%%%%%%%%%%%

Region III contains the fields that obtain a mass through the intermediate Wilson line scale $\Lambda_4= g/R_2$ (with the gauge coupling evaluated at the corresponding scale).

%%%%%%%%%%%%%%%%%%%%%%%%%%%%%%%%%%%%%%%%%%%%%%%%%%%%%%%%%%%%%%
\begin{table}[h]
    \centering
    \footnotesize
   \begin{tabular}{l r}
$\SO{1,3}{}$ & \hspace{6cm} $\SU{3}{C}\times \SU{2}{L}\times \U{Y}$\\
\hline
Massless complex scalars   & $\textcolor{OliveGreen}{(\textbf{1},{\textbf{2}},3)}+2\times\textcolor{blue}{(\textbf{3},\textbf{2},1)}+2\times\textcolor{blue}{(\bar{\textbf{3}},\textbf{1},-4)}+2\times\textcolor{blue}{(\textbf{1},\textbf{1},6)}+\textcolor{blue}{(\bar{\textbf{3}},\textbf{1},2)}+(\textbf{3},\textbf{1},-2)$\\
\hline
Massless Weyl fermions &  $ \textcolor{magenta}{(\textbf{8},\textbf{1},0)}+ \textcolor{magenta}{(\textbf{1},\textbf{3},0)}+ 3\times \textcolor{OliveGreen}{(\textbf{1},{\textbf{2}},3)}+6 \times \textcolor{ForestGreen}{(\textbf{1},{\textbf{2}},-3)}+9\times \textcolor{Sepia}{(\textbf{1},{\textbf{1}},0)}$ 
\\
  & $
+ 4\times\textcolor{blue}{(\textbf{1},\textbf{1},6)}+\textcolor{red}{(\textbf{1},\textbf{1},-6)}+3\times\textcolor{blue}{(\textbf{3},\textbf{2},1)}
+3\times\textcolor{blue}{(\bar{\textbf{3}},\textbf{1},-4)}+5\times\textcolor{blue}{(\bar{\textbf{3}},\textbf{1},2)}+2\times\textcolor{red}{({\textbf{3}},\textbf{1},2)}$\\
\hline
Massless real vectors  & $ \textcolor{magenta}{(\textbf{8},\textbf{1},0)}+ \textcolor{magenta}{(\textbf{1},\textbf{3},0)}+ 4\times\textcolor{magenta}{(\textbf{1},\textbf{1},0)}$ \\
& $ +\textcolor{OliveGreen}{(\textbf{1},{\textbf{2}},3)}+\textcolor{ForestGreen}{(\textbf{1},{\textbf{2}},-3)}+\textcolor{blue}{(\textbf{1},\textbf{1},6)}+\textcolor{red}{(\textbf{1},\textbf{1},-6)}$\\
\hline
\end{tabular}
    \caption{Field content in Region III}
    \label{tab:trikksm}
\end{table}
%%%%%%%%%%%%%%%%%%%%%%%%%%%%%%%%%%%%%%%%%%%%%%%%%%%%%%%%%%%%%%
%
At this stage, and for the field content in \cref{tab:trikksm}, the coefficients governing the evolution of the gauge couplings read as
\begin{equation}
        b^\mathrm{III}_\mathrm{3} = -\frac73\,, \qquad b^\mathrm{III}_\mathrm{2} = -\frac52\,, \qquad
        b^{' \mathrm{III}}_\mathrm{1} = \frac76\,.
        \label{coeff:IV-1}
    \end{equation}

%%%%%%%%%%%%%%%%%%%%%%%%%%%%%%%%%%%%%%%%%%%%%%%%%%%%%%%%%%%%%%%%%%%%%%
\subsection*{Running between $\Lambda_3$ and $\Lambda_2$ - Region IV}
%%%%%%%%%%%%%%%%%%%%%%%%%%%%%%%%%%%%%%%%%%%%%%%%%%%%%%%%%%%%%%%%%%%%%%

Above the energy scale $\Lambda_3=(R_1 R_2 R_3)^{-1/3}$, denoted as region IV, the effect of the KK modes is introduced. This region ends up with the highest Wilson line scale $\Lambda_2=g/R_1$. Due to the steep running of the gauge coupling, this region is Its field content is described in \cref{tab:e6}.

\begin{table}[h]
    \centering
    \footnotesize
   \begin{tabular}{l r}
$\SO{1,3}{}$ & \hspace{6cm} $\SU{3}{C}\times \SU{2}{L}\times \U{Y}$\\
\hline
Complex Scalar KK tower
 &  $2\times \textcolor{magenta}{(\textbf{8},\textbf{1},0)}+2\times\textcolor{magenta}{(\textbf{1},\textbf{3},0)}+50\times \textcolor{Sepia}{(\textbf{1},{\textbf{1}},0)}+2\times(\textbf{3},\textbf{2},-5)+2\times(\bar{\textbf{3}},{\textbf{2}},5)$ 
\\
& $+20 \times \textcolor{ForestGreen}{(\textbf{1},{\textbf{2}},-3)}+10\times\textcolor{blue}{(\textbf{1},\textbf{1},6)}+10\times\textcolor{blue}{(\bar{\textbf{3}},\textbf{1},-4)}+10\times\textcolor{blue}{(\textbf{3},\textbf{2},1)}
+20\times\textcolor{blue}{(\bar{\textbf{3}},\textbf{1},2)}$
\\
& $+20\times \textcolor{OliveGreen}{(\textbf{1},{\textbf{2}},3)}+10\times\textcolor{red}{(\textbf{1},\textbf{1},-6)}+10\times\textcolor{red}{({\textbf{3}},\textbf{1},4)}+
10\times\textcolor{red}{(\bar{\textbf{3}},\textbf{2},-1)}
+20\times\textcolor{red}{({\textbf{3}},\textbf{1},-2)}$
\\
\hline
Real scalar KK tower  &  $ \textcolor{magenta}{(\textbf{8},\textbf{1},0)}+\textcolor{magenta}{(\textbf{1},\textbf{3},0)}+25\times \textcolor{Sepia}{(\textbf{1},{\textbf{1}},0)}+(\textbf{3},\textbf{2},-5)+(\bar{\textbf{3}},{\textbf{2}},5)$ 
\\
& $+10 \times \textcolor{ForestGreen}{(\textbf{1},{\textbf{2}},-3)}+5\times\textcolor{blue}{(\textbf{1},\textbf{1},6)}+5\times\textcolor{blue}{(\bar{\textbf{3}},\textbf{1},-4)}+5\times\textcolor{blue}{(\textbf{3},\textbf{2},1)}
+10\times\textcolor{blue}{(\bar{\textbf{3}},\textbf{1},2)}$
\\
& $+10\times \textcolor{OliveGreen}{(\textbf{1},{\textbf{2}},3)}+5\times\textcolor{red}{(\textbf{1},\textbf{1},-6)}+5\times\textcolor{red}{({\textbf{3}},\textbf{1},4)}+
5\times\textcolor{red}{(\bar{\textbf{3}},\textbf{2},-1)}
+10\times\textcolor{red}{({\textbf{3}},\textbf{1},-2)}$\\
\hline
Weyl fermion KK tower   &  $4\times \textcolor{magenta}{(\textbf{8},\textbf{1},0)}+4\times\textcolor{magenta}{(\textbf{1},\textbf{3},0)}+100\times \textcolor{Sepia}{(\textbf{1},{\textbf{1}},0)}+4\times(\textbf{3},\textbf{2},-5)+4\times(\bar{\textbf{3}},{\textbf{2}},5)$ 
\\
& $+40 \times \textcolor{ForestGreen}{(\textbf{1},{\textbf{2}},-3)}+20\times\textcolor{blue}{(\textbf{1},\textbf{1},6)}+20\times\textcolor{blue}{(\bar{\textbf{3}},\textbf{1},-4)}+20\times\textcolor{blue}{(\textbf{3},\textbf{2},1)}
+40\times\textcolor{blue}{(\bar{\textbf{3}},\textbf{1},2)}$
\\
& $+40\times \textcolor{OliveGreen}{(\textbf{1},{\textbf{2}},3)}+20\times\textcolor{red}{(\textbf{1},\textbf{1},-6)}+20\times\textcolor{red}{({\textbf{3}},\textbf{1},4)}+
20\times\textcolor{red}{(\bar{\textbf{3}},\textbf{2},-1)}
+40\times\textcolor{red}{({\textbf{3}},\textbf{1},-2)}$\\
\hline
Real vector KK tower   &  $ \textcolor{magenta}{(\textbf{8},\textbf{1},0)}+\textcolor{magenta}{(\textbf{1},\textbf{3},0)}+25\times \textcolor{Sepia}{(\textbf{1},{\textbf{1}},0)}+(\textbf{3},\textbf{2},-5)+(\bar{\textbf{3}},{\textbf{2}},5)$ 
\\
& $+10 \times \textcolor{ForestGreen}{(\textbf{1},{\textbf{2}},-3)}+5\times\textcolor{blue}{(\textbf{1},\textbf{1},6)}+5\times\textcolor{blue}{(\bar{\textbf{3}},\textbf{1},-4)}+5\times\textcolor{blue}{(\textbf{3},\textbf{2},1)}
+10\times\textcolor{blue}{(\bar{\textbf{3}},\textbf{1},2)}$
\\
& $+10\times \textcolor{OliveGreen}{(\textbf{1},{\textbf{2}},3)}+5\times\textcolor{red}{(\textbf{1},\textbf{1},-6)}+5\times\textcolor{red}{({\textbf{3}},\textbf{1},4)}+
5\times\textcolor{red}{(\bar{\textbf{3}},\textbf{2},-1)}
+10\times\textcolor{red}{({\textbf{3}},\textbf{1},-2)}$\\
\hline
Massless complex scalars   & $\textcolor{OliveGreen}{(\textbf{1},{\textbf{2}},3)}+2\times\textcolor{blue}{(\textbf{3},\textbf{2},1)}+2\times\textcolor{blue}{(\bar{\textbf{3}},\textbf{1},-4)}+2\times\textcolor{blue}{(\textbf{1},\textbf{1},6)}+\textcolor{blue}{(\bar{\textbf{3}},\textbf{1},2)}+(\textbf{3},\textbf{1},-2)$\\
\hline
Massless Weyl fermions &  $ \textcolor{magenta}{(\textbf{8},\textbf{1},0)}+ \textcolor{magenta}{(\textbf{1},\textbf{3},0)}+ 3\times \textcolor{OliveGreen}{(\textbf{1},{\textbf{2}},3)}+6 \times \textcolor{ForestGreen}{(\textbf{1},{\textbf{2}},-3)}+9\times \textcolor{Sepia}{(\textbf{1},{\textbf{1}},0)}$ 
\\
  & $
+ 4\times\textcolor{blue}{(\textbf{1},\textbf{1},6)}+\textcolor{red}{(\textbf{1},\textbf{1},-6)}+3\times\textcolor{blue}{(\textbf{3},\textbf{2},1)}
+3\times\textcolor{blue}{(\bar{\textbf{3}},\textbf{1},-4)}+5\times\textcolor{blue}{(\bar{\textbf{3}},\textbf{1},2)}+2\times\textcolor{red}{({\textbf{3}},\textbf{1},2)}$\\
\hline
Massless real vectors  & $ \textcolor{magenta}{(\textbf{8},\textbf{1},0)}+ \textcolor{magenta}{(\textbf{1},\textbf{3},0)}+ 4\times\textcolor{magenta}{(\textbf{1},\textbf{1},0)}$ \\
& $ +\textcolor{OliveGreen}{(\textbf{1},{\textbf{2}},3)}+\textcolor{ForestGreen}{(\textbf{1},{\textbf{2}},-3)}+\textcolor{blue}{(\textbf{1},\textbf{1},6)}+\textcolor{red}{(\textbf{1},\textbf{1},-6)}$\\
\hline
\end{tabular}
    \caption{Field content in Region IV}
    \label{tab:e6}
\end{table}
The coefficients governing both the logarithmic and power-law evolution of the gauge couplings are
\begin{equation}
       \tilde{b}^\mathrm{IV}_\mathrm{3} = b^\mathrm{IV}_\mathrm{3} = -\frac73\,, \qquad  
       \tilde{b}^\mathrm{IV}_\mathrm{2} = b^\mathrm{IV}_\mathrm{2} = -\frac52\,, \qquad
        \tilde{b}^{'\mathrm{IV}}_\mathrm{1} = b^{' \mathrm{IV}}_\mathrm{1} = \frac76\,.
        \label{coeff:IV-1}
    \end{equation}
%%%%%%%%%%%%%%%%%%%%%%%%%%%%%%%%%%%%%%%%%%%%%%%%%%%%%%%%%%%%%%%%%%%%%%
\subsection*{Running between $\Lambda_2$ and $\Lambda_1$ - Region V}    
%%%%%%%%%%%%%%%%%%%%%%%%%%%%%%%%%%%%%%%%%%%%%%%%%%%%%%%%%%%%%%%%%%%%%%

As one passes the $\Lambda_2$ scale, contributions from the KK modes  as well as all zero modes, incluiding the ones that obtain a mass at the highest Wilson line scale $\Lambda_2=g/R_1$ are considered.

\begin{table}[h]
    \centering
   \begin{tabular}{l r}
$\SO{1,3}{}$ & \hspace{6cm} $\SU{3}{C}\times \SU{2}{L}\times \U{Y}$\\
\hline
Complex Scalar KK tower
 &  $2\times \textcolor{magenta}{(\textbf{8},\textbf{1},0)}+2\times\textcolor{magenta}{(\textbf{1},\textbf{3},0)}+50\times \textcolor{Sepia}{(\textbf{1},{\textbf{1}},0)}+2\times(\textbf{3},\textbf{2},-5)+2\times(\bar{\textbf{3}},{\textbf{2}},5)$ 
\\
& $+20 \times \textcolor{ForestGreen}{(\textbf{1},{\textbf{2}},-3)}+10\times\textcolor{blue}{(\textbf{1},\textbf{1},6)}+10\times\textcolor{blue}{(\bar{\textbf{3}},\textbf{1},-4)}+10\times\textcolor{blue}{(\textbf{3},\textbf{2},1)}
+20\times\textcolor{blue}{(\bar{\textbf{3}},\textbf{1},2)}$
\\
& $+20\times \textcolor{OliveGreen}{(\textbf{1},{\textbf{2}},3)}+10\times\textcolor{red}{(\textbf{1},\textbf{1},-6)}+10\times\textcolor{red}{({\textbf{3}},\textbf{1},4)}+
10\times\textcolor{red}{(\bar{\textbf{3}},\textbf{2},-1)}
+20\times\textcolor{red}{({\textbf{3}},\textbf{1},-2)}$
\\
\hline
Real scalar KK tower  &  $ \textcolor{magenta}{(\textbf{8},\textbf{1},0)}+\textcolor{magenta}{(\textbf{1},\textbf{3},0)}+25\times \textcolor{Sepia}{(\textbf{1},{\textbf{1}},0)}+(\textbf{3},\textbf{2},-5)+(\bar{\textbf{3}},{\textbf{2}},5)$ 
\\
& $+10 \times \textcolor{ForestGreen}{(\textbf{1},{\textbf{2}},-3)}+5\times\textcolor{blue}{(\textbf{1},\textbf{1},6)}+5\times\textcolor{blue}{(\bar{\textbf{3}},\textbf{1},-4)}+5\times\textcolor{blue}{(\textbf{3},\textbf{2},1)}
+10\times\textcolor{blue}{(\bar{\textbf{3}},\textbf{1},2)}$
\\
& $+10\times \textcolor{OliveGreen}{(\textbf{1},{\textbf{2}},3)}+5\times\textcolor{red}{(\textbf{1},\textbf{1},-6)}+5\times\textcolor{red}{({\textbf{3}},\textbf{1},4)}+
5\times\textcolor{red}{(\bar{\textbf{3}},\textbf{2},-1)}
+10\times\textcolor{red}{({\textbf{3}},\textbf{1},-2)}$\\
\hline
Weyl fermion KK tower   &  $4\times \textcolor{magenta}{(\textbf{8},\textbf{1},0)}+4\times\textcolor{magenta}{(\textbf{1},\textbf{3},0)}+100\times \textcolor{Sepia}{(\textbf{1},{\textbf{1}},0)}+4\times(\textbf{3},\textbf{2},-5)+4\times(\bar{\textbf{3}},{\textbf{2}},5)$ 
\\
& $+40 \times \textcolor{ForestGreen}{(\textbf{1},{\textbf{2}},-3)}+20\times\textcolor{blue}{(\textbf{1},\textbf{1},6)}+20\times\textcolor{blue}{(\bar{\textbf{3}},\textbf{1},-4)}+20\times\textcolor{blue}{(\textbf{3},\textbf{2},1)}
+40\times\textcolor{blue}{(\bar{\textbf{3}},\textbf{1},2)}$
\\
& $+40\times \textcolor{OliveGreen}{(\textbf{1},{\textbf{2}},3)}+20\times\textcolor{red}{(\textbf{1},\textbf{1},-6)}+20\times\textcolor{red}{({\textbf{3}},\textbf{1},4)}+
20\times\textcolor{red}{(\bar{\textbf{3}},\textbf{2},-1)}
+40\times\textcolor{red}{({\textbf{3}},\textbf{1},-2)}$\\
\hline
Real vector KK tower   &  $ \textcolor{magenta}{(\textbf{8},\textbf{1},0)}+\textcolor{magenta}{(\textbf{1},\textbf{3},0)}+25\times \textcolor{Sepia}{(\textbf{1},{\textbf{1}},0)}+(\textbf{3},\textbf{2},-5)+(\bar{\textbf{3}},{\textbf{2}},5)$ 
\\
& $+10 \times \textcolor{ForestGreen}{(\textbf{1},{\textbf{2}},-3)}+5\times\textcolor{blue}{(\textbf{1},\textbf{1},6)}+5\times\textcolor{blue}{(\bar{\textbf{3}},\textbf{1},-4)}+5\times\textcolor{blue}{(\textbf{3},\textbf{2},1)}
+10\times\textcolor{blue}{(\bar{\textbf{3}},\textbf{1},2)}$
\\
& $+10\times \textcolor{OliveGreen}{(\textbf{1},{\textbf{2}},3)}+5\times\textcolor{red}{(\textbf{1},\textbf{1},-6)}+5\times\textcolor{red}{({\textbf{3}},\textbf{1},4)}+
5\times\textcolor{red}{(\bar{\textbf{3}},\textbf{2},-1)}
+10\times\textcolor{red}{({\textbf{3}},\textbf{1},-2)}$\\
\hline
Msssless Complex Scalars
 & $ 3\times \textcolor{OliveGreen}{(\textbf{1},{\textbf{2}},3)}+6\times \textcolor{ForestGreen}{(\textbf{1},{\textbf{2}},-3)}+6\times \textcolor{Sepia}{(\textbf{1},{\textbf{1}},0)} +3\times\textcolor{blue}{(\textbf{1},\textbf{1},6)}$\\
&
$+3\times\textcolor{blue}{(\textbf{3},\textbf{2},1)}+3\times\textcolor{blue}{(\bar{\textbf{3}},\textbf{1},-4)}+6\times\textcolor{blue}{(\bar{\textbf{3}},\textbf{1},2)}+3\times\textcolor{red}{({\textbf{3}},\textbf{1},-2)}$
\\
\hline
Massless Weyl fermions  & $\textcolor{magenta}{(\textbf{8},\textbf{1},0)}+\textcolor{magenta}{(\textbf{1},\textbf{3},0)}+ 4\times \textcolor{OliveGreen}{(\textbf{1},{\textbf{2}},3)}+7\times \textcolor{ForestGreen}{(\textbf{1},{\textbf{2}},-3)}+19\times \textcolor{Sepia}{(\textbf{1},{\textbf{1}},0)} +4\times\textcolor{blue}{(\textbf{1},\textbf{1},6)}$\\
&
$+3\times\textcolor{blue}{(\textbf{3},\textbf{2},1)}+3\times\textcolor{blue}{(\bar{\textbf{3}},\textbf{1},-4)}+6\times\textcolor{blue}{(\bar{\textbf{3}},\textbf{1},2)}+3\times\textcolor{red}{({\textbf{3}},\textbf{1},-2)}+\textcolor{red}{(\textbf{1},\textbf{1},-6)}$
\\
\hline
Massless Real vectors  & $ \textcolor{magenta}{(\textbf{8},\textbf{1},0)}+\textcolor{magenta}{(\textbf{1},\textbf{3},0)}+13\times \textcolor{Sepia}{(\textbf{1},{\textbf{1}},0)}$ 
\\
 & $\textcolor{OliveGreen}{(\textbf{1},{\textbf{2}},3)}+ \textcolor{ForestGreen}{(\textbf{1},{\textbf{2}},-3)}+\textcolor{blue}{(\textbf{1},\textbf{1},6)}+\textcolor{red}{(\textbf{1},\textbf{1},-6)}$\\
\hline
\end{tabular}
    \caption{Field content in Region V}
    \label{tab:e6kk}
\end{table}
Using the field content in \cref{tab:e6kk}, one obtains the following coefficients for what we denote as region V
\begin{equation}
        b^\mathrm{V}_\mathrm{3} = 0\,, \qquad b^\mathrm{V}_\mathrm{2} = 0\,, \qquad
        b^{' \mathrm{V}}_\mathrm{1} = \frac{18}{5}\,.
        \label{coeff:V-1}
    \end{equation}
and
\begin{equation}
        \tilde{b}^\mathrm{V}_\mathrm{3} = -10\,, \qquad 
        \tilde{b}^\mathrm{V}_\mathrm{2} = -5\,, \qquad
        \tilde{b}^{' \mathrm{V}}_\mathrm{1} = -13\,.
        \label{coeff:V-2}
    \end{equation}
Note the sign inversion for the $\U{Y}$ beta-functions. This is due to the presence of several charged vector fields that contribute with a large negative factor to the beta function coefficients, implying an asymptotic unification of the gauge couplings as we discuss in the main text.

%%%%%%%%%%%%%%%%%%%%%%%%%%%%%%%%%%%%%%%%%%%%%%%%%%%%%%%%%%%%%%%%%%%%%%
\subsection*{Running above $\Lambda_1$ - Region VII}
%%%%%%%%%%%%%%%%%%%%%%%%%%%%%%%%%%%%%%%%%%%%%%%%%%%%%%%%%%%%%%%%%%%%%%

Finally above the largest energy $\Lambda_1=1/R_1$, the unification scale, there is only a single $\E{8}$ vector superfield, which is decomposed into KK modes as in \cref{tab:e8field}.
\begin{table}[h]
    \centering
   \begin{tabular}{l r}
$\SO{1,3}{}$ & \hspace{8cm} $\mathrm{E}_8$\\
\hline
Complex Scalar KK tower
 & $3\times \textcolor{magenta}{\textbf{248}}$\\
\hline
Weyl fermion KK tower  & $4\times \textcolor{magenta}{\textbf{248}}$\\
\hline
Real vector KK tower  & $\textcolor{magenta}{\textbf{248}}$
\end{tabular}
    \caption{Field content in Region VII}
    \label{tab:e8field}
\end{table}
%%%%%%
The field content has an effective $\mathcal{N}=4$ SUSY and the single gauge coupling beta-function vanishes \cite{Sohnius:1985qm} i.e., $\beta^{(n)}_{g_8} = 0$, thus no longer evolving with the energy-scale.

\end{document}